\documentclass[apj,twocolumn,numberedappendix,appendixfloats]{emulateapj}
\usepackage{apjfonts}
\usepackage[pagebackref=false,colorlinks=true,citecolor=blue,linkcolor=blue,breaklinks=true,bookmarks=true]{hyperref}
\usepackage{amsmath}

\shorttitle{SMG Cold Stream} 
\shortauthors{Fu et al.}
\journalinfo{To appear in the Astrophysical Journal}
\submitted{Submitted on 2020 Dec 23, accepted on 2021 Jan 11}

\newcommand{\rmxaa}{RMxAA}
\newcommand{\kms}{{km s$^{-1}$}}
\newcommand{\msun}{$M_\odot$}
\newcommand{\msunyr}{$M_\odot\,{\rm yr}^{-1}$}
\newcommand{\lsun}{$L_\odot$}
\newcommand{\um}{$\mu$m}
\newcommand{\uJy}{$\mu$Jy}
\newcommand{\cc}{cm$^{-3}$}
\newcommand{\pcm}{cm$^{-2}$}

\newcommand{\logNHI}{\log N_{\rm HI}}
\newcommand{\zabs}{z_{\rm abs}}
\newcommand{\zsmg}{z_{\rm SMG}}
\newcommand{\zsys}{z_{\rm sys}}
\newcommand{\Ha}{H$\alpha$}

\newcommand{\Lya}{Ly$\alpha$}
\newcommand{\Lyb}{Ly$\beta$}
\newcommand{\Lyg}{Ly$\gamma$}
\newcommand{\Lyd}{Ly$\delta$}
\newcommand{\HI}{H\,{\sc i}}
\newcommand{\HII}{H\,{\sc ii}}

\newcommand{\NII}{[N\,{\sc ii}]}
\newcommand{\MgII}{Mg\,{\sc ii}}
\newcommand{\CIV}{C\,{\sc iv}}
\newcommand{\cothree}{CO\,(3$-$2)}
\newcommand{\coone}{CO\,(1$-$0)}

\newcommand{\mh}{M_{\rm halo}}
\newcommand{\Herschel}{{\it Herschel}}
\newcommand{\system}{GAMA\,J0913$-$0107} 
\newcommand{\SMGfull}{ALMA\,J091339.55$-$010656.4} 
\newcommand{\COMfull}{ALMA\,J091338.28$-$010643.8} 
\newcommand{\Chostfull}{ALMA\,J091338.49$-$010705.5} 
\newcommand{\bgQSOfull}{SDSS\,J091338.97$-$010704.6} 
\newcommand{\fgQSOfull}{SDSS\,J091338.30$-$010708.6} 
\newcommand{\SMG}{SMM\,J0913} 
\newcommand{\COM}{Comp a} 
\newcommand{\Chost}{Comp b} 
\newcommand{\bgQSO}{QSO1} 
\newcommand{\fgQSO}{QSO2} 

\begin{document}

\title{A Long Stream of Metal-Poor Cool Gas around a Massive Starburst Galaxy at z = 2.67}
\author{
Hai~Fu\altaffilmark{1,2}, R.~Xue\altaffilmark{1,3}, J.~X.~Prochaska\altaffilmark{4}, A.~Stockton\altaffilmark{2}, S.~Ponnada\altaffilmark{1,5}, M.~W.~Lau\altaffilmark{6}, A.~Cooray\altaffilmark{7}, and D.~Narayanan\altaffilmark{8}
}
\altaffiltext{1}{Department of Physics \& Astronomy, University of Iowa, Iowa City, IA 52242}
\altaffiltext{2}{Institute for Astronomy, University of Hawaii, Honolulu, HI 96822}
\altaffiltext{3}{National Radio Astronomy Observatory, Charlottesville, VA, 22903}
\altaffiltext{4}{Department of Astronomy and Astrophysics, UCO/Lick Observatory, University of California, Santa Cruz, CA 95064}
\altaffiltext{5}{Astronomy Department, California Institute of Technology, Pasadena, CA 51125}
\altaffiltext{6}{Department of Physics and Astronomy, University of California, Riverside, CA 92521}
\altaffiltext{7}{Department of Physics and Astronomy, University of California, Irvine, CA 92697}
\altaffiltext{8}{Department of Astronomy, University of Florida, Gainesville, FL, 32611}

\begin{abstract}

We present the first detailed dissection of the circumgalactic medium (CGM) of massive starburst galaxies at $z > 2$. Our target is a submillimeter galaxy (SMG) at $z=2.674$ that has a star formation rate of 1200\,\msunyr\ and a molecular gas reservoir of $1.3\times10^{11}$\,\msun. We characterize its CGM with two background QSOs at impact parameters of 93\,kpc and 176\,kpc. We detect strong \HI\ and metal-line absorption near the redshift of the SMG toward both QSOs, each consisting of three main subsystems spanning over 1500\,\kms. The absorbers show remarkable kinematic and metallicity coherence across a separation of $\sim$86\,kpc. In particular, the cool gas in the CGM of the SMG exhibits high \HI\ column densities ($\logNHI/{\rm cm}^{-2} = 20.2, 18.6$), low metallicities (${\rm [M/H]} \approx -2.0$), and similar radial velocities ($\delta v \sim -300$\,\kms). While the \HI\ column densities match previous results on the CGM around QSOs at $z > 2$, the metallicities are lower by more than an order of magnitude, making it an outlier in the line width$-$metallicity relation of damped \Lya\ absorbers. The large physical extent, the velocity coherence, the high surface density, and the low metallicity are all consistent with the cool, inflowing, and near-pristine gas streams predicted to penetrate hot massive halos at $z > 1.5$. We estimate a total gas accretion rate of $\sim$100\,\msunyr\ from three such streams, which falls short of the star formation rate but is consistent with simulations. At this rate, it takes about a gigayear to acquire the molecular gas reservoir of the central starburst. 

\end{abstract}


\section{Introduction} \label{sec:intro}

The global gas supply for in situ star formation is a central question in galaxy formation and evolution, because star formation and merging are the two primary channels through which galaxies grow \citep{Oser10}. According to spherical hydrodynamical models \citep{Birnboim03} and cosmological simulations \citep{Keres05}, stable accretion shocks are established near the virial radius when a dark matter (DM) halo grows to a mass threshold of $M_{\rm shock} = 2-3\times10^{11}$\,\msun. So in massive halos, a significant fraction of the accreted gas is expected to be shock-heated to the virial temperature ($T_{\rm vir} = 8\times10^6~(\mh/10^{13}~M_\odot)^{2/3}$\,K) and develops an atmosphere of hot diffuse gas. The virial shock effectively cuts off the fuel supply for star formation, because of the inefficient radiative cooling of the hot gas even in the denser inner regions \citep{Keres09}. But at high redshift, narrow filaments of cool gas ($T \lesssim 10^5$\,K) from the cosmic web may penetrate the hot atmospheres of rare, massive halos without ever being shock-heated to the virial temperature, thanks to the lower masses of typical halos at higher redshifts that define the width of the filaments \citep{Dekel06,Dekel09}. In fact, this {\it cold mode} accretion may dominate over the {\it hot mode} accretion (radiative cooling of shock-heated virialized gas) at all halo masses at $z > 2$ \citep{Keres09}. 

In emission, the predicted cold-mode accretion streams (or ``cold streams'' in short) feeding high-redshift massive galaxies may appear as giant filamentary \Lya\ nebulae around QSOs \citep{Weidinger04,Cantalupo14,Martin15,Martin19a} and in dense protocluster environments \citep{Moller01,Hennawi15,Umehata19,Li19a,Daddi20}. However, it has been difficult to rule out outflows as the alternative interpretation, especially when QSO photoionization contributes to the \Lya\ emission and the chemical abundance of the nebulae cannot be easily measured.
In absorption, cold streams can be detected and distinguished from other gaseous components based on neutral hydrogen (\HI) column density, kinematics, and particularly chemical abundance \citep{Fumagalli11a,Theuns21}. 

In this project, we have selected a sample of massive starbursts at high redshifts in the vicinity of background QSOs to trace the cool gas supply in these early massive halos. There, the problem of gas supply is the most acute because of the extremely short gas exhaustion timescale. We then utilize the absorption-line spectra of background QSOs to characterize the physical state of their circumgalactic medium (CGM) -- the gas between the inner regions of galaxies and the diffuse intergalactic medium (IGM) -- and to search for large-scale cool gas reservoirs. In this section, we review our knowledge of the target galaxy sample and QSO absorption-line systems in the literature. These earlier studies have motivated this project and will provide valuable reference samples that can be compared with the system dissected in this work. 

\subsection{Submillimeter Galaxies}

Heated dust in the interstellar medium cools by emitting a modified blackbody spectrum (MBB; $S_\nu \propto (1-e^{-\tau_\nu})B_\nu(T)$) with temperatures in the range $10~{\rm K} \lesssim T \lesssim 100~{\rm K}$, forming the far-infrared (IR) hump in the spectral energy distribution (SED) of a galaxy. At any given frequency along the Rayleigh-Jeans tail where the dust should be optically thin, the observed flux density ($S_{\nu,\rm obs}$) of the MBB is proportional to the dust mass ($M_{\rm dust}$), the dust temperature ($T$), and the redshift ($z$) --- 
\begin{align}
S_{\nu,{\rm obs}} &\propto M_{\rm dust}~T~(1+z)^{\beta-1}~\nu_{\rm obs}^{2+\beta}/d_A(z)^2 \nonumber \\ \
&\propto M_{\rm dust}~T~(1+z)^{\beta-1},
\end{align}
where $d_A(z)$ is the angular diameter distance at redshift $z$ (which varies by only 22\% between $z = 1$ and $z = 4$) and $\beta \approx 2$ is the dust emissivity parameter ($\kappa_\nu \propto \nu^\beta$). Therefore, galaxies selected at long wavelengths, such as the (sub)millimeter regime, preferentially have higher dust mass, higher dust temperature, and are at higher redshift than galaxies selected at shorter wavelengths. Furthermore, holding the metallicity ($Z_{\rm gas}$) constant, high dust mass together with high temperature implies that the galaxies are gas-rich ($M_{\rm gas} = M_{\rm dust}/Z_{\rm gas}$) and have high star-formation efficiency (${\rm SFE} = {\rm SFR}/M_{\rm gas}$, where SFR is the star formation rate), because 
\begin{equation}
T^4 \propto L_{\rm bol}/M_{\rm dust} \propto {\rm SFR}/(Z_{\rm gas} M_{\rm gas}),
\end{equation}
a result from the Stefan-Boltzmann law. 

Indeed, follow-up observations of the brightest galaxies selected at 850~\um\ ($S_{850} \gtrsim 3$\,mJy), the submillimeter galaxies (SMGs; \citealt{Smail97,Barger98,Blain02}), have revealed a significant population of gas-rich starburst galaxies that contribute almost as much to the cosmic SFR density as UV-selected Lyman break galaxies at $z = 2-3$ \citep{Chapman05,Casey14}. The SMGs are mature \citep[$\langle M_{\rm star} \rangle \sim 10^{11}$~\msun;][]{Hainline11, Michalowski12, Targett13}, metal-rich \citep[$\langle Z \rangle \sim Z_\odot$;][]{Swinbank04}, gas-rich \citep[$\langle M_{\rm mol} \rangle \sim 3\times10^{10}$~\msun;][]{Greve05, Tacconi08, Ivison11, Bothwell13}, extreme star-forming systems ($\overline{\rm SFR} \sim 500$~\msunyr) with a broad redshift distribution that peaks at $\langle z \rangle \sim 2.5$ \citep{Chapman05, Wardlow11}. The molecular gas reservoirs are turbulent, likely due to starburst-driven galactic outflows \citep{Falgarone17}. Notably, the nearly linear relation between CO and IR luminosities implies an almost constant gas depletion timescale of $\tau_{\rm dep} \equiv M_{\rm mol}/{\rm SFR} \sim 0.1~{\rm Gyr}$ \citep{Bothwell13}, which is far shorter than that of normal star-forming galaxies on the main sequence \citep[$\tau_{\rm dep} \sim 0.6~{\rm Gyr}$ at $z = 2.5$;][]{Tacconi18}, justifying the usage of ``starburst'' in describing SMGs\footnote{Although most of the difference in $\tau_{\rm dep}$ is driven by the conversion factor from CO to molecular gas ($\alpha_{\rm CO} \equiv M_{\rm mol}/L_{\rm CO}^\prime$ in units of $M_\odot/({\rm K~km~s^{-1}~pc^2})$), constraints from dynamical masses and dust masses have shown that SMGs indeed have lower $\alpha_{\rm CO}$ than that of normal star-forming galaxies \citep[e.g.,][]{Hodge12,Magnelli12b,Xue18} and that the value adopted from local ultraluminous IR galaxies \citep[ULIRGs, $\alpha_{\rm CO} = 1.0$;][]{Downes98,Papadopoulos12b} is more appropriate for SMGs than the Galactic value \citep[$\alpha_{\rm CO} = 4.3$;][]{Bolatto13}.}.

On the other hand, the autocorrelation length for SMGs of $\sim$11~Mpc at $z = 1-3$ implies a characteristic dark matter halo mass of $\mh \sim 9\times10^{12}$~\msun\ for $h=0.7$ \citep{Hickox12}. The high halo mass is consistent with the high maximum rotation velocities ($V_{\rm circ} \gtrsim 500$\,\kms) observed in several bright SMGs with spatially resolved kinematics \citep[e.g.,][]{Hodge12,Xue18}. For Navarro–Frenk–White (NFW) halos \citep{Navarro96} at $z = 2.5$, the halo mass is directly related to the maximum circular velocity by a power law: $\mh = 10^{13}~M_\odot~(V_{\rm circ}/500~{\rm km~s}^{-1})^3$ \citep{Bullock01, Klypin11}. Such a mass is well above the threshold mass for stable virial shocks ($M_{\rm shock}$), and atmospheres of hot gas at the virial temperature ($\sim8\times10^6$\,K) are expected to fill the halo. But as previously discussed, at the early epoch of the SMGs, cool gas filaments can penetrate their halos, which could potentially deliver enough gas to build the molecular gas reservoir that supports the ongoing intense star formation.

\subsection{QSO Absorption-line Systems}

Ever since the discovery of multiple absorption redshifts in QSO spectra \citep{Burbidge68}, quasar absorption-line spectroscopy has become a powerful tool to study diffuse gas at various phases in the IGM and the CGM, which account for the majority of the baryonic mass in the universe (see \citealt{Peroux20} for a recent review). The optical depth at the Lyman limit ($\lambda_{\rm rest} = 912$\,\AA) reaches unity when the \HI\ column density reaches $\logNHI = 17.2$\footnote{Column densities are given in units of \pcm\ throughout the paper.} 
Accumulating evidence suggests that these optically thick absorbers trace material in virialized structures 
\citep[i.e., the CGM,][]{Fumagalli16,Lehner16}, 
while the optically thin absorbers in the \Lya\ forest (LYAF) likely trace the IGM \citep{Rauch98}. Due to their distinct physical properties, the optically thick absorbers are empirically subdivided into three categories based on their \HI\ column densities: the Lyman limit systems (LLSs, $17.2 \leq \logNHI < 19$) that are mostly ionized, the damped \Lya\ absorbers (DLAs; $\logNHI \geq 20.3$) that are mostly neutral, and lastly the super-LLSs or sub-DLAs\footnote{The two terms have been used interchangeably.} for the intermediate category of absorbers with $19 \leq \logNHI < 20.3$. Unlike absorbers at lower column densities, gas in the DLAs is mostly neutral. In fact, at all epochs since $z \sim 5$, the DLAs have contained most of the neutral gas that is poised to fuel star formation in galaxies \citep{Wolfe05}. 

\subsection{Emission$-$Absorption Connection} \label{sec:intro-connection}

Because the \HI\ column density threshold of DLAs was set by the observed limit of 21\,cm emission at the cutoff boundaries of nearby spiral disks \citep{Wolfe86}, the DLAs were expected to arise from gas-rich galactic disks even at high redshifts. However, the emission counterparts (i.e., the DLA galaxies) of most DLAs have eluded detection. Among the limited detections in optical searches, it is found that the DLA galaxies are very faint ($r \gtrsim 24$) and close ($\sim$2\arcsec) to the QSOs \citep[e.g.,][]{Steidel92,Fynbo08}, making it difficult to measure their redshifts. To improve efficiency, searches of the emission counterparts of DLAs have focused on DLAs that are clearly chemically enriched ([M/H] $> -0.7$) \citep[e.g.,][]{Fynbo13,Jorgenson14} or sightlines that pass through multiple DLAs \citep[e.g.,][]{Srianand16}. But still, only 16 $z > 1.9$ DLA host galaxies have been identified via emission lines in the optical \citep[see][for compilations]{Krogager17,Moller20} over an extensive search period of nearly three decades. The advent of (sub)millimeter interferometers such as the Atacama Large Millimeter/submillimeter Array (ALMA) and the Very Large Array (VLA) have significantly improved the success rate of identifying absorption-selected galaxies, because (1) the contrast between DLA host galaxies and the QSO is more favorable at longer wavelengths, and (2) the interferometers have an unattenuated view over a wide FoV (thus does not require lucky slit placements). In only a few years, there have been four $z \sim 4$ DLA galaxies identified in [C\,{\sc ii}]\,158\,\um\ \citep{Neeleman17,Neeleman19} and five $z \sim 2$ DLA in CO(4-3) and CO(3-2) \citep{Kanekar20}. Interestingly, the DLA galaxies previously identified in the optical/near-IR are not detected in CO and {\it vice versa} \citep{Kanekar20}, indicating that observations at different wavelength are complementary to one another and that \HI-absorption-selection tags gas-rich galaxies of all types. Some DLAs also have multiple emission counterparts that are consistent with the absorption redshifts, suggesting a group/cluster environment \citep[e.g.,][]{Fynbo18}. 
 
The opposite approach from the searches of DLA galaxies is to start from an emission-selected galaxy sample and look for corresponding absorption lines in the spectra of nearby background QSOs. This approach requires chance alignments of foreground galaxies and background QSOs, thus requires large samples of both populations. The implicit assumption is that the emission-selected galaxies have similar CGM properties, and therefore, the absorption signals obtained from different galaxy-QSO pairs can be combined to provide meaningful average properties of a typical halo in the studied galaxy population. The searches for absorbers are no longer limited to DLAs, but to all optically thick absorbers (i.e., LLSs and sub-DLAs). At $z \gtrsim 2$, the targeted galaxy populations have included Lyman Break Galaxies (LBGs) \citep[e.g.,][]{Simcoe06,Rudie12,Rudie13,Crighton13,Crighton15} and QSOs \citep[e.g.,][]{Hennawi06a,Prochaska13,Lau16}. In addition, using a sample of projected QSO pairs where one of the QSOs intercepts a DLA, \citet{Rubin15} have probed the CGM of the DLA galaxy without identifying the DLA galaxy in emission. These studies have mapped out the \HI\ column density, the ion ratios, and the metallicity as a function of impact parameters ($R_\bot$). The covering fraction of optically thick \HI\ absorbers increases from $\sim$30\% around LBGs \citep{Rudie12} and DLAs \citep{Rubin15} to $\gtrsim$60\% around QSOs \citep{Prochaska13} for sightlines out to $R_\bot = 100-200$\,kpc (comparable to the virial radius of DM halos with $\mh = 10^{12.5}$\,\msun\ at $z = 2$: $R_{\rm vir} = 154$\,kpc). The abundance of neutral gas in the halos of QSOs is particularly puzzling. Simulations predict that such massive halos are dominated by a hot $T\sim 10^7\,{\rm K}$ virialized plasma and a significantly lower covering factor of optically thick \HI\ absorbers \citep[e.g.,][]{Faucher-Giguere15}. 

The unexpectedly large covering factor of LLSs around $z \sim 2$ QSOs and the difficulty of reproducing the SMG population in galaxy formation models, could both be symptoms of the same problem. Attempting to reduce this tension between observations and theory, more recent cosmological zoom-in simulations have implemented recipes of stronger and presumably more realistic stellar feedback, which manages to preserve cool gas reservoirs in the accreted sub-halos during earlier phases of star formation before their infall into the massive halo. The presence of these gas-rich sub-halos increases the cool gas covering factor around QSOs \citep{Faucher-Giguere16} and their prolonged bombardment to the central galaxy leads to a rising star formation history that eventually produces SMGs between $2 < z < 3$ \citep{Narayanan15,Lovell21}. 

\subsection{Organization} \label{sec:org}

QSO absorption-line spectroscopy combined with efficient emission-line mapping provides a powerful method to link star-forming galaxies with the neutral gas reservoir that may fuel future star formation. Our understanding of the formation and evolution of massive galaxies is severely limited by the lack of observational constraints of the CGM of SMGs. The advent of \Herschel\ large-scale far-infrared surveys have provided an opportunity to use projected SMG$-$QSO pairs to probe the CGM of SMGs. In this paper, we focus on one particularly interesting system -- \system\ -- where two background QSOs have revealed an unusually \HI-rich CGM around a luminous SMG.

The main text of the paper is organized as follows. We first provide an orientation of the system in \S~\ref{sec:system}, then proceed with a detailed study of the emission sources (\S~\ref{sec:smg}) and the absorption-line systems (\S~\ref{sec:qsos}), before finally drawing connections between the absorbers and their emission counterparts in \S~\ref{sec:connection}. We conclude the paper with a summary of the main results and a discussion of the implications in \S\,\ref{sec:summary}. To keep the main text focused on the SMG$-$DLA system at $z \approx 2.67$, we move additional material to the {\it Appendices}. We analyze the nearby optical source to the SMG and its potential lensing effect in Appendix\,\ref{sec:fakecp}, present the methodology and result of our blind search of line emitters in the ALMA band-3 data in Appendix\,\ref{sec:blindCO}, give an inventory of the line-of-sight contaminating absorbers at other redshifts in Appendix\,\ref{sec:otherabs}, provide tables of detailed ionic column density and metallicity measurements in Appendix\,\ref{sec:aodmtables}, and describe our attempt to detect CO emission from faint optical sources near \bgQSO\ and the identification of \Chost\ in Appendix\,\ref{sec:optCO}.

Throughout this paper, we adopt a model optimization method that combines a heuristic $\chi^2$ minimization algorithm with a Markov Chain Monte Carlo (MCMC) algorithm (hereafter, the ``{\sc amoeba + mcmc}'' method). It begins with using the downhill simplex method {\sc amoeba} \citep{Press92} with simulated annealing {\sc amoeba\_sa} to find the solution that minimizes the residual. Although computationally more expensive than other least-$\chi^2$ solvers (e.g., the Levenberg-Marquardt technique), {\sc amoeba\_sa} has the advantage of avoiding being trapped in local minima in a multidimensional parameter space. This advantage is particularly important in more complex problems such as fitting the \HI\ absorption profiles with many Voigt profiles (\S~\ref{sec:voigt}). Next, starting from the minimum-$\chi^2$ solution of {\sc amoeba\_sa}, we use the Differential Evolution MCMC algorithm \citep{Ter-Braak06} implemented in {\sc exofast\_demc} \citep{Eastman13,Eastman19} to obtain the final solution and the statistical uncertainties of the parameters. The {\sc exofast\_demc} routine first determines the stepping scale of each parameter by varying it from the minimum $\chi^2$ solution until the $\chi^2$ increases by one. It then starts the chains from positions that are randomly offset from the minimum $\chi^2$ solution. The routine stops when the chains are considered well-mixed and the steps in the initial ``burn-in'' phase are removed. The marginalized 1$\sigma$ confidence interval of each parameter is determined from the values at 15.8 and 84.1 percentiles of the concatenated chains, and the {\it median} values are adopted as the {\it formal} solution. Parameters derived from the model parameters are treated likewise: their formal values and uncertainties are calculated from the 50, 15.8, and 84.1 percentiles of the array directly calculated from the chains of model parameters.

We assume the $\Lambda$CDM cosmology with $\Omega_{\rm m}=0.3$, $\Omega_\Lambda=0.7$, and $h \equiv H_0/(100~{\rm km~s}^{-1}~{\rm Mpc}^{-1}) = 0.7$ and quote proper/physical distances.


\begin{figure*}[!tb]
\epsscale{1.19}
\plotone{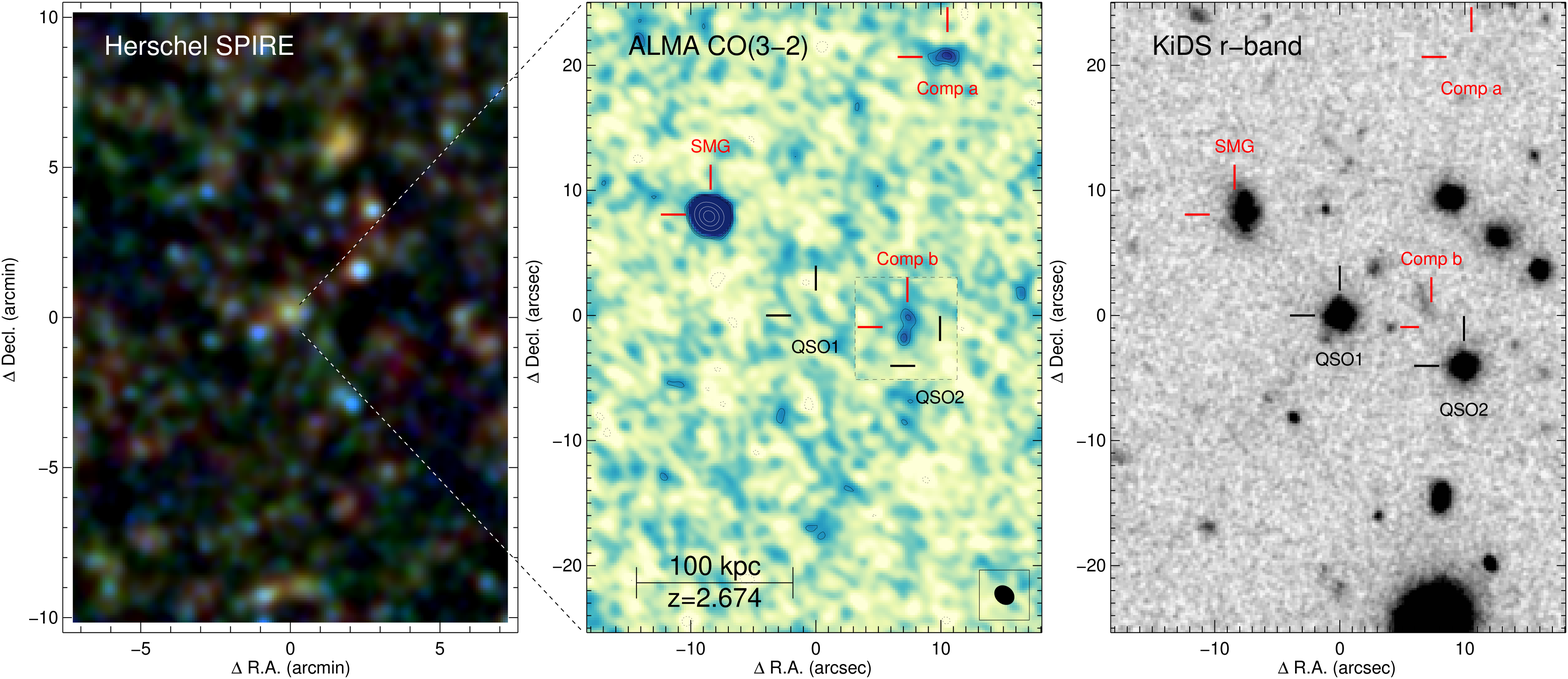}
\caption{Multi-wavelength images of the \system\ system. 
{\it Left} - A wide-field {\it Herschel} pseudo-color image combining 250~\um\ (blue), 350~\um\ (green) and 500~\um\ (red) images. The SMG is the bright source near the center of the 15.2\arcmin$\times$21.0\arcmin\ region. 
{\it Middle} - An ALMA map zoomed in onto the SMG showing \cothree\ emission between $2.67 < z < 2.70$. This 36.5\arcsec$\times$50.4\arcsec\ region encloses the SMG and its CO companion galaxies ({\it red} tickmarks), and the two QSOs in the background of the system ({\it black} tickmarks). This {\it composite} CO image is formed by combining the 11 channels within $\pm$140\,\kms\ of $z = 2.674$ (i.e., $\nu_{\rm obs} = 94.120 \pm 0.039$~GHz). To show the CO emission from \Chost, the 8\arcsec$\times$8\arcsec\ region centered on \Chost\ ({\it dotted box}) is formed by combining the two channels where CO emission is detected ($\nu_{\rm obs} = 93.7535$, and 93.6676\,GHz, corresponding to $z = 2.6884$ and 2.6917). The contours are drawn at $-3$ ({\it black dotted}), 3, 4, 5({\it black solid}), 20, 40, and 60$\sigma$ ({\it white solid}). The synthesized beam of 1.6\arcsec$\times$1.3\arcsec\ is shown at the lower right corner. 
{\it Right} - A deep $r$-band image of the same region from KiDS (5$\sigma$ detection limit at $\sim$25 mag).
In all images, the position of \bgQSO\ sets the origin of the coordinates.
\label{fig:fchart}} 
\epsscale{1.0}
\end{figure*} 

\section{System Overview} \label{sec:system}

\begin{table*}
\begin{center}
\caption{Major Components of the \system\ system and Impact Parameters}
\label{tab:coords}
\begin{tabular}{llcccccccc}
\hline
\hline
Designation & Short Name & R.A. (J2000) & Decl. (J2000) & $z$ & $L^\prime_{\rm CO3-2}$ & $\theta_1$ & $\theta_2$ & $R_{\bot,1}$ & $R_{\bot,2}$ \\
& & (deg) & (deg) & & (K\,\kms\,pc$^2$) & (arcsec) & (arcsec) & (kpc) & (kpc) \\
\hline
\SMGfull   & \SMG   & 138.4147767 & $-$1.1156772 & 2.674 & $6.5\times10^{10}$ & 11.7 & 22.1 & 93.1  & 175.5 \\
\COMfull   & \COM   & 138.4094849 & $-$1.1121639 & 2.6747 & $6.3\times10^9$ & 23.3 & 24.8 & 185.1 & 197.1 \\
\Chostfull & \Chost & 138.4103803 & $-$1.1181869 & 2.6884 & $6.8\times10^8$ & 7.4 &  4.1 & 58.9  &  32.2 \\
--- & --- & --- & --- & 2.6917 & $5.3\times10^8$ & --- & --- & --- & --- \\
\bgQSOfull & \bgQSO & 138.4124260 & $-$1.1179280 & 2.9161 & \nodata & \nodata & \nodata & \nodata  &  \nodata \\
\fgQSOfull & \fgQSO & 138.4096520 & $-$1.1190520 & 2.7488 & $5.6\times10^9$ & 10.8 & \nodata & 85.0  &  \nodata \\
\hline
\end{tabular}
\end{center}
\tablecomments{QSO coordinates are from the KiDS DR4 catalog, the SMG coordinates are from the higher-resolution ALMA band-6 image, and the coordinates of \COM\ and \Chost\ are from their CO emission detected in the ALMA band-3 datacube. \Chost\ has two rows because it is a superposition of two CO emitters likely involved in a merger. The quoted redshift of \fgQSO\ is from the \cothree\ line, which is slightly off from its optical redshift ($z_{\rm opt} = 2.7498$, $\delta v = 80$\,\kms). The column $L^\prime_{\rm CO3-2}$ lists the \cothree\ line luminosities from ALMA. The CO emission of \bgQSO\ is outside of our spectral coverage. The columns $\theta_1$ and $\theta_2$ list the angular separations from \bgQSO\ and \fgQSO, respectively, and the corresponding transverse proper distances at the source redshift (i.e., the impact parameters) are listed as $R_{\bot,1}$ and $R_{\bot,2}$.}
\end{table*}

Fig.~\ref{fig:fchart} illustrates the \system\ system, where we label its major components: the SMG, its CO companions, and the two background QSOs. Table~\ref{tab:coords} lists their coordinates, redshifts, \cothree\ luminosities, along with the impact parameters of the QSO sightlines. 

The \system\ system is one of the 163 SMG$-$QSO pairs with {\it apparent} separations between 5\arcsec\ and 30\arcsec, which were selected by cross-matching \Herschel-selected SMGs with optically-selected QSOs from a compilation of spectroscopic surveys \citep{Fu16,Fu17}. Located in the R.A.\,=\,9\,hr equatorial field of the \Herschel\ Astrophysical Terahertz Large Area Survey (H-ATLAS) survey \citep{Eales10}, the \Herschel\ source at R.A. = $09^{\rm h}13^{\rm m}39^{\rm s}$, Decl. = $-01^\circ06\arcmin59\arcsec$ is detected at high S/N by SPIRE \citep[Spectral and Photometric Imaging Receiver;][]{Griffin10} at 250, 350, and 500~\um, with de-boosted flux densities of $S_{250} = 52.5\pm7.4$\,mJy, $S_{350} = 69.4\pm8.8$\,mJy, and $S_{500} = 48.4\pm9.2$\,mJy \citep{Valiante16,Fu17}. The far-IR SED clearly peaks around 350~\um\ (i.e., ``350~\um\ peakers''), giving a rough photometric redshift of $\sim$2.5 assuming a dust temperature of $\sim$50\,K, following Wien's displacement law for $S_\nu$ over $\lambda$: $\lambda_{\rm peak} = 102~\mu{\rm m}~(50~{\rm K}/T)~(1+z)$. 

Our ALMA 345~GHz imaging pinpointed the position of the \Herschel\ source \citep{Fu17}, and Gemini near-IR and ALMA 94~GHz spectroscopy jointly determined a spectroscopic redshift of 2.674 (\S~\ref{sec:specID} \& \ref{sec:almaB3}). We designate the SMG as \SMGfull\ or \SMG\ in short. In addition, our ALMA 94~GHz observations detected companion galaxies in \cothree\ near the redshift of the SMG: \COM\ at $z = 2.6747$, and \Chost\ at $z = 2.6884, 2.6917$. Both are within 23\arcsec\ of the SMG position. \Chost\ has two redshifts because it is a superposition of two galaxies (see \S~\ref{sec:comps}). Because \Herschel\ has FWHM resolutions of 18\arcsec, 25\arcsec, and 35\arcsec\ at 250, 350, and 500\,\um, respectively, the SMG and its companions are blended in the \Herschel\ images. But the contribution of the companions to the \Herschel\ fluxes should be negligible given their orders-of-magnitude lower CO line luminosities. 

There are {\it two} bright QSOs within 22\arcsec\ of the SMG, \bgQSO\ (\bgQSOfull, $g = 20.78$, $r = 20.38$) at $z = 2.9161$ and \fgQSO\ (\fgQSOfull, $g = 20.71$, $r = 20.44$) at $z = 2.7488$. Both QSOs are in the background of the SMG, allowing us to probe its CGM at impact parameters of $R_\bot = 93.1$\,kpc and $R_\bot = 175.5$\,kpc, or approximately 0.5$\times$ and 0.9$\times$ the virial radius of a $10^{13}$\,\msun\ halo at $z = 2.674$ ($R_{\rm vir} = 186$\,kpc). Coincidentally, strong \HI\ and metal absorption lines near the SMG redshift have been previously detected in the QSO spectra \citep{Finley14}. 

When comparing the CO map with the KiDS $r$-band image in Fig.~\ref{fig:fchart}, we notice an $r = 21.6$ optical source just 0.8\arcsec\ from the SMG. As we will show in Appendix\,\ref{sec:fakecp}, it is a foreground galaxy with a spec-$z$ of $z = 0.055$ and its gravitational lensing effect on the SMG is negligible.  
We also find that \Chost\ is just $\sim$3\arcsec\ from an elongated optical source to the NNE. In the KiDS DR4 catalog, the optical source has a designation of J091338.527$-$010703.60 and magnitudes of $r = 23.8\pm0.1$ and $H = 22.0\pm0.3$. It has a photo-$z$ of $z_{\rm p} = 0.79^{+0.45}_{-0.06}$. Its SED shows a clear 1-magnitude drop-off between $Y$-band and $Z$-band, corresponding to a 4000~\AA-break at $z \sim 1.2-1.5$, which is consistent with the maximum-likelihood photo-$z$ of $z_{\rm p} = 1.37$ in the catalog. We thus conclude that the optical source is most likely a foreground galaxy, although the extraction of an ALMA spectrum near the position of the optical source led to the identification of \Chost\ (Appendix\,\ref{sec:optCO}).

\section{The Submillimeter Galaxy and Its Companions} \label{sec:smg}

\subsection{ALMA Position and Near-IR Spectroscopy} \label{sec:specID}

\Herschel/SPIRE positions have large uncertainties. A comparison between ALMA and \Herschel\ positions showed a 1$\sigma$ positional offset of $\sim$4.2\arcsec\ for sources with S/N $\sim$ 6 at 250\,\um\ \citep[Eq. 5 of][]{Fu17}. Near-IR slit spectroscopy requires sub-arcsec positions, so we carried out 0.5\arcsec-resolution ALMA band-7 (345~GHz/870~\um) observations of \system\ as part of our Cycle-3 project 2015.1.00131.S \citep[see][for details]{Fu17}. \system\ shared an hour-long observing session with nine other \Herschel\ SMGs in the same H-ATLAS field. With four scans and 44 antennas, we accumulated a total on-source integration time of 189.5~s. A single high S/N source is detected in the $\sim$17\arcsec\ Full Width at Half Power (FWHP) of the primary beam. It has a 870\,\um\ flux density of $S_{870} = 7.4\pm0.5$~mJy, an offset from the \Herschel\ position by 4.1\arcsec, and a beam-deconvolved FWHM of $0.46\arcsec\pm0.04\arcsec$ along the major axis (which corresponds to $\sim$4\,kpc at $z = 2.5$).  

\begin{figure}[!tb]
\epsscale{1.15}
\plotone{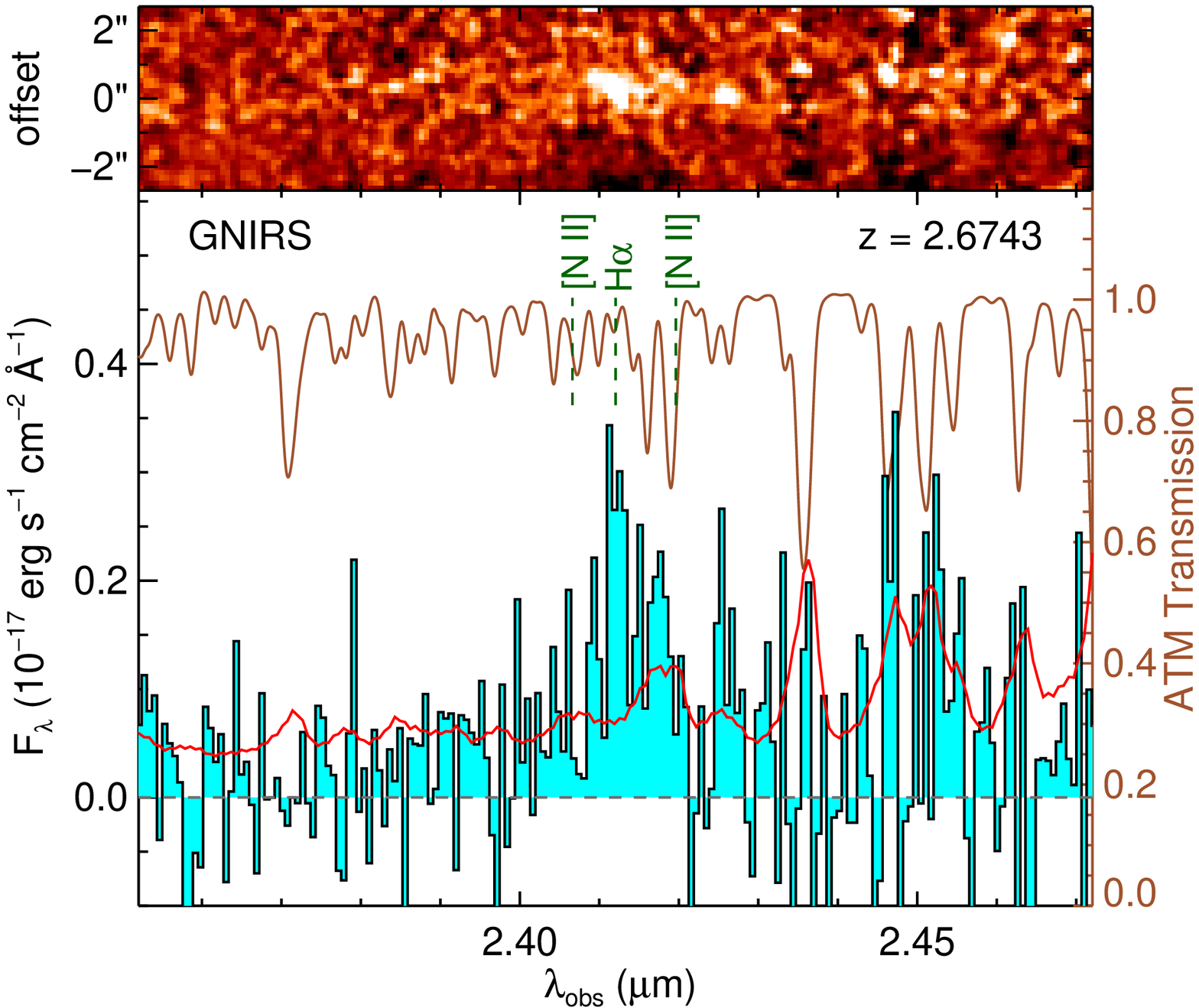}
\caption{The GNIRS near-IR spectrum of \SMG. The top panel shows the coadded 2D spectrum. The ordinate is the positional offset along the spatial direction, and is centered on the SMG location. The bottom panel shows the flux-calibrated 1D spectrum ({\it black}) and its 1$\sigma$ uncertainty ({\it red}). Wavelengths affected by strong sky lines show large errors. The dashed lines indicate the redshifted \Ha\ and \NII\,$\lambda\lambda$6550,6585 lines. The brown curve shows the atmosphere transmission curve, using the right-side ordinate. 
\label{fig:gnirs}} 
\epsscale{1.0}
\end{figure}

The accurate ALMA position enabled our follow-up near-IR slit spectroscopy. Observations of \SMG\ were carried out with Gemini near-infrared spectrograph \citep[GNIRS;][]{Elias06} on 2017 Feb 19 as part of our queue program GN-2017A-Q-31. The A0-type star HIP49125 ($V$ = 7.19, $K$ = 6.553 Vega mag) was observed right after the science target to provide telluric correction and flux calibration. Because our goal was to measure the redshift, we used the cross-dispersed mode with the 32 l/mm grating to achieve a continuous wavelength coverage between 0.85 and 2.5\,\um\ (orders 3 to 8). After applying a 30\arcsec\ offset from an offset star to the NE, we placed the 1\arcsec-wide 7\arcsec-long slit on the ALMA 870\,\um\ position at a position angle (PA) of 73 deg (E of N; the PA was chosen to reach a guide star). The expected spectral resolution of this configuration is $R = 510$ (FWHM = 590\,\kms), but the actual spectral resolution may be higher depending on the source size and the seeing. We took 24 exposures of 136\,s with a 3\arcsec-step ABBA dithering pattern. Data reduction was performed with a modified version of Spextool \citep{Cushing04} for GNIRS by K. Allers. The final coadded spectrum in Fig.\,\ref{fig:gnirs} includes all 24 frames and has a total on-source time of 54.4\,min. We detected an emission line at $\sim$4$\sigma$-level at 2.4121\,\um\ (heliocentric corrected, vacuum wavelength), which we identified as \Ha\ ($\lambda_{\rm rest} = 6564.63\,\AA$) at $z_{\rm H\alpha} = 2.6743\pm0.0003$. The \NII\,$\lambda$6585.28 line is undetected, likely due to the elevated background noise at its wavelength and its lower flux. Our best-fit Gaussian model yields a line ${\rm FWHM} = 230\pm60$\,\kms\ (i.e., the line is unresolved) and a line flux of $F_{\rm H\alpha} = (6.1\pm1.0)\times10^{-17}$\,erg\,s$^{-1}$\,cm$^{-2}$, which are comparable to those of the SMGs identified by the VLA \citep{Alaghband-Zadeh12,Fu16}.

\subsection{ALMA \cothree\ Spectral Line Imaging} \label{sec:almaB3}

To detect the molecular gas reservoir that fuels the intense star formation in the SMG, we carried out deep ALMA band-3 (100~GHz/3~mm) spectral line observations of \SMG\ on 2018 December 10 and 15 with our Cycle 6 project 2018.1.00548.S. We tuned the four 1.875~GHz-bandwidth spectral windows to center on 92.2 (BB3), 94.0 (BB4), 104.2 (BB1), and 106.0~GHz (BB2) in dual linear polarization mode ({\it XX} and {\it YY}). We chose a spectral averaging factor of 16 to bin the Frequency Division Mode (FDM)'s input channel spacing of 0.488\,MHz to an output channel spacing of 7.8125\,MHz. The spectral averaging significantly reduces the output data rate and essentially eliminates the correlation between adjacent channels introduced by the Hanning window function applied to the correlation functions (see \S~5.5 in the {\it ALMA Technical Handbook}). The resulting spectral response function is basically a top-hat function with a width of 7.8125\,MHz (i.e., the output channel spacing), which corresponds to a spectral resolution of 22$-$25\,\kms. The two lower frequency spectral windows provide a continuous frequency range between 91.27 and 94.94\,GHz, which covers the \cothree\ line ($\nu_{\rm rest} = 345.79599$\,GHz and $\lambda_{\rm rest} = 866.96337$\,\um) between $2.642 \leq z \leq 2.789$, encompassing the SMG at $z = 2.674$ and \fgQSO\ at $z = 2.7498$. The frequency range covers a velocity window between $-2620$ and $+9240$\,\kms\ relative to the SMG redshift. The two higher frequency spectral windows provide a continuous frequency coverage between 103.27 and 106.94\,GHz, which covers the \cothree\ line between $2.234 \leq z \leq 2.348$ and traces the continuum emission at rest-frame frequencies around $\nu_{\rm rest} = 386$\,GHz ($\lambda_{\rm rest} = 776$\,\um) at $z = 2.674$. 
 
The primary beam of the ALMA 12-m antennas has an FWHP of 62\arcsec\ at 94~GHz. We set the field center at R.A. = $09^{\rm h}13^{\rm m}38.89^{\rm s}$, Decl. = $-01^\circ07\arcmin03.6\arcsec$, which is near the position of \bgQSO\ but $\sim$12\arcsec\ offset from the SMG. Three of the four planned observing sessions were executed, accumulating a total on-source time of 143.8\,min with 6.05\,s integrations. Either 43 or 46 12-m antennas were operational, with baselines ranging between 15.1\,m and 740.5\,m. The BL Lac object J0854$+$2006 served as the amplitude, bandpass, and pointing calibrator, and the flat-spectrum radio quasar J0909$+$0121 as the phase calibrator \citep{Bonato19}. 

\begin{figure}[!tb]
\epsscale{1.15}
\plotone{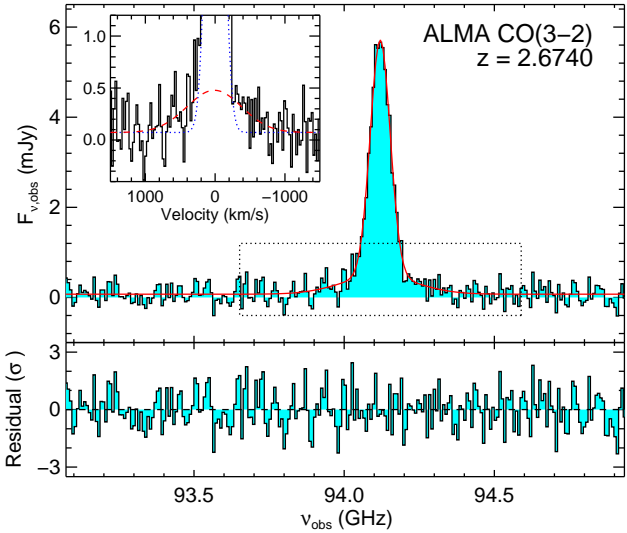}
\caption{ALMA \cothree\ spectrum of \SMG\ and the best-fit double-Gaussian model ({\it red solid curve}). The inset shows a zoomed-in version of the spectrum to highlight the broad component ({\it red dashed curve}) beneath the narrow component ({\it blue dotted curve}). The dotted rectangle shows the portion of the spectrum shown in the inset. The bottom panel shows the residual (data - model) in units of 1$\sigma$ error.
\label{fig:cospec}} 
\epsscale{1.0}
\end{figure}

\subsection{ALMA Data Processing}

The raw visibility data were flagged and calibrated by the ALMA pipeline (Pipeline ver. 42030M, CASA ver. 5.4.0-68). The calibrated visibilities of the three observing sessions were then combined to form the final calibrated measurement set (MS). The pipeline worked very well. After inspecting the amplitudes of the calibrated visibilities, we found that additional flagging was only necessary for a tiny fraction of data. We used the CASA task {\tt flagcmd} to flag the cross-correlation data of the antenna pair DA62 and DA65 in the 94.0\,GHz spectral window between channels 188 and 191.  

We use the CASA task {\tt tclean} to image the calibrated visibilities of each spectral window into spectral data cubes. When visibilities are gridded into regularized $uv$-cells, we adopt natural weighting to maximize the sensitivity. The synthesized beams are on average 1.7\arcsec$\times$1.3\arcsec\ in FWHM, so we set the imaging pixel size to 0.2\arcsec. In the spectral dimension, we retain the original channel spacing of 7.8125\,MHz. The data were recorded in the Topocentric (TOPO) reference frame. Due to the motion of the Earth, every observing scan has a slightly different sampling in sky frequency. We image the data to the solar System Barycenter (BARY) reference frame to be consistent with the velocities measured in the heliocentric-corrected optical and near-IR spectra.  

Significant continuum emission and a strong emission line at $\sim$94.1\,GHz (in BB4) is detected at the ALMA 870\,\um\ position of \SMG. To minimize the sidelobes from this bright source, we used Clark {\sc clean} deconvolution algorithm with a mask consisting of a single 2\arcsec-radius circle centered on the SMG. The {\sc clean} depth is set to be 2$\times$ the rms listed below. 

For each spectral window, we generate two datacubes: one avoids interpolation in the spectral dimension by setting {\tt interpolation = nearest} and is uncorrected for the primary beam, and the other uses linear interpolation and is corrected for the primary beam. The former is better suited for blind line searches because maps in adjacent channels remain uncorrelated, while the latter is better suited for measuring line parameters such as central frequency, width, and integrated flux. The resulting spectral cubes have a dimension of 540 pixels by 540 pixels by 240 channels. At the phase center, the sensitivities of the datacubes reach ${\rm rms} = 0.165, 0.171, 0.158, 0.155$\,mJy\,beam$^{-1}$\,channel$^{-1}$ for BB1, BB2, BB3, and BB4, respectively. The rms values are consistent with the visibility noise that we measured with {\tt visstat} ($\sigma \sim 250$\,mJy\,visibility$^{-1}$\,channel$^{-1}$), because ${\rm rms} \sim \sigma/\sqrt{n_{\rm ch} n_{\rm pol} n_{\rm baseline} n_{\rm int}}$, where $n_{\rm ch} = 1$ (1 channel binning), $n_{\rm pol} = 2$ (2 polarizations), $n_{\rm baseline} = n_{\rm ant} (n_{\rm ant}-1)/2 = 903$ for $n_{\rm ant} = 43$ (903 baselines for 43 antennae), and $n_{\rm int} = t_{\rm on\_source}/t_{\rm int} \sim 1426$ (total number of integrations).

In addition, we generate one continuum image from the two higher frequency spectral windows (BB1 and BB2). We used the Multi-term Multi-Frequency Synthesis ({\tt mtmfs}) {\sc clean} algorithm with a linear spectral model (i.e., {\tt nterms} = 2). Again, we used natural weighting, 0.2\arcsec\ pixel size, and the same {\sc clean} mask centered on the SMG. The sensitivity of the continuum image reaches ${\rm rms} = 7.4$\,\uJy\,beam$^{-1}$, consistent with the rms of the spectral cubes divided by $\sqrt{n_{\rm ch}} \simeq 22$. 

\subsection{SMG Properties} \label{sec:smg_prop}

\begin{figure*}[!tb]
\epsscale{0.38}
\plotone{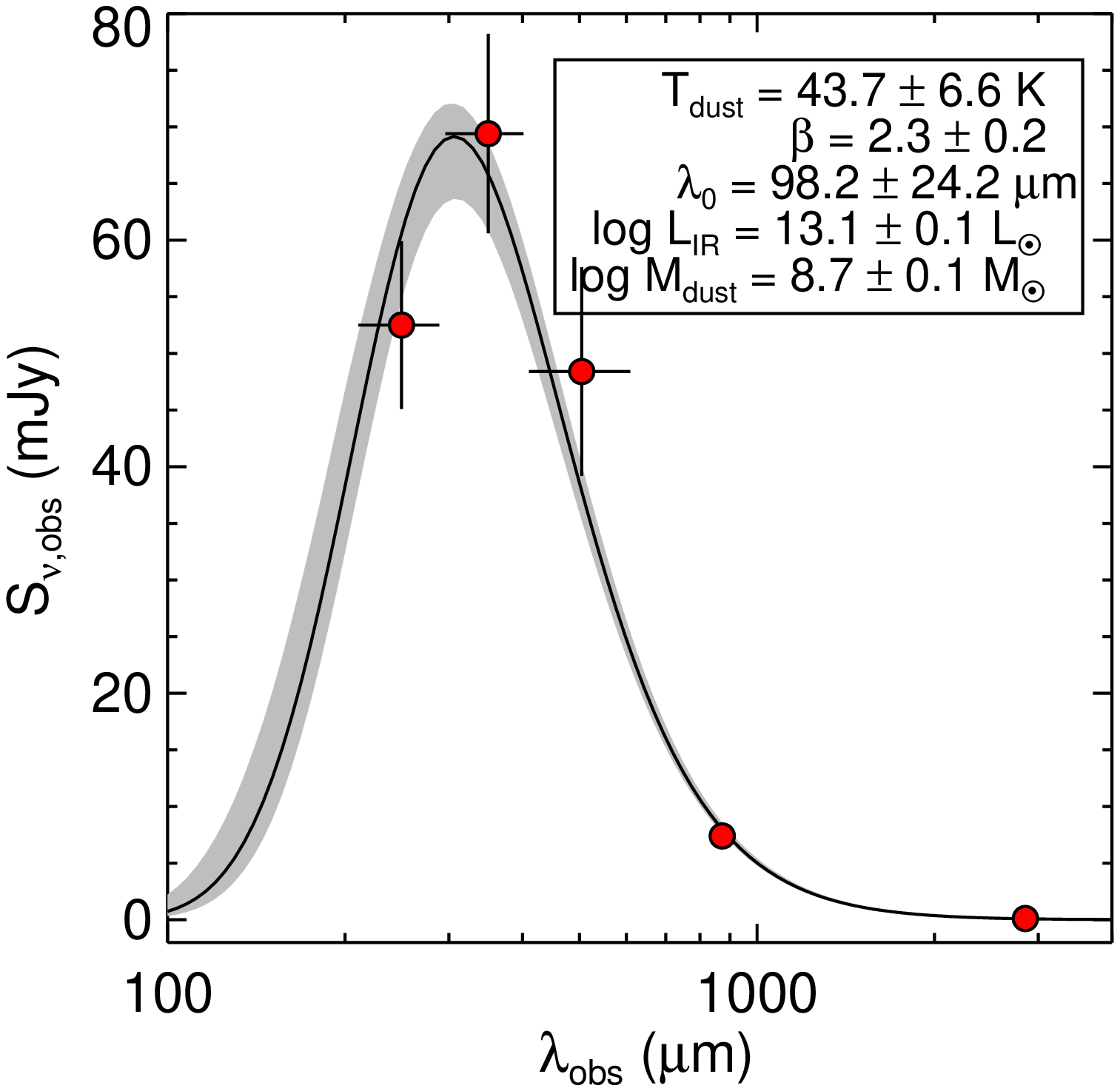}
\plotone{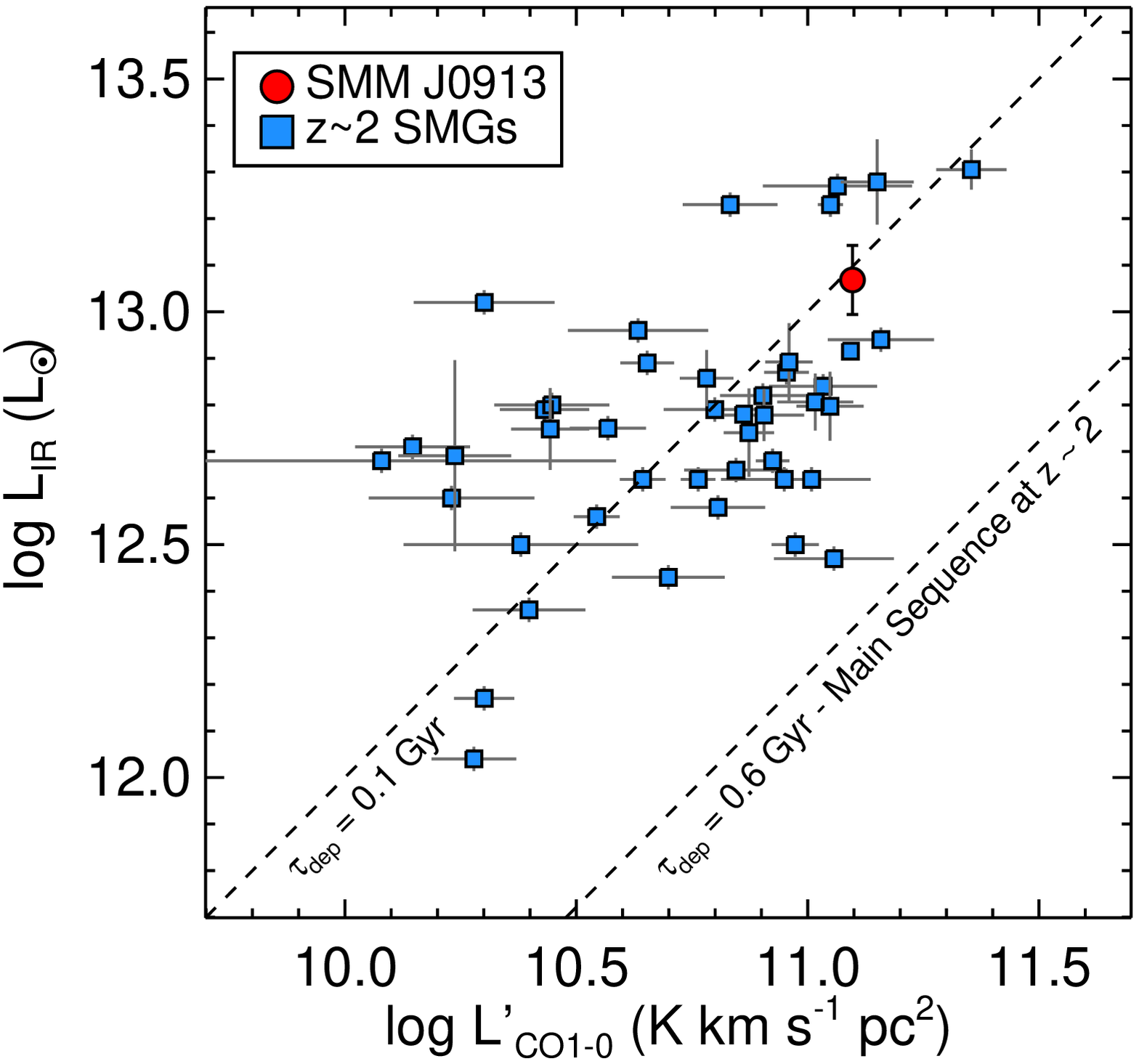}
\plotone{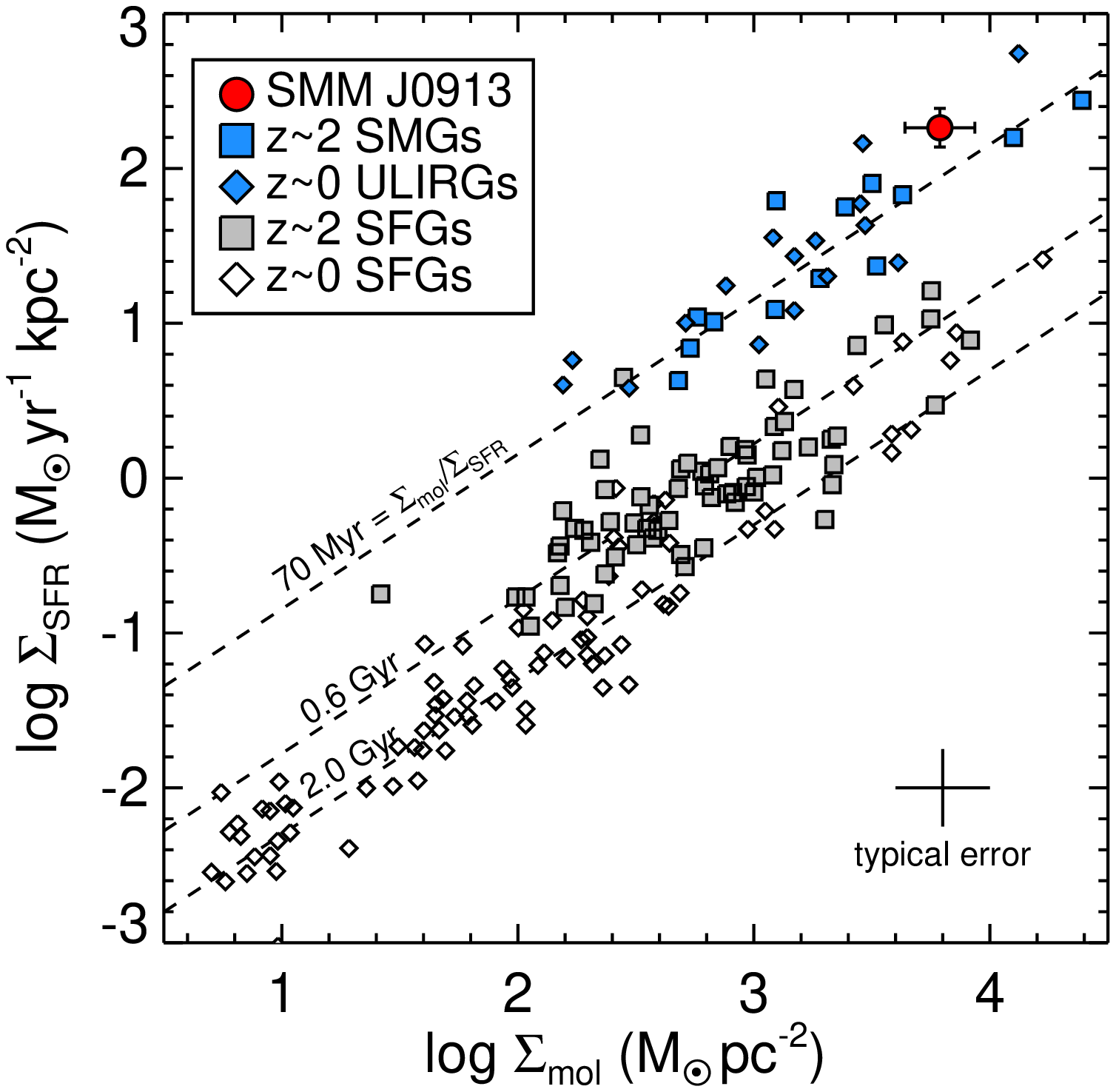}
\caption{Physical properties of \SMG\ ({\it red} data points). {\it Left}: the Far-IR SED from \Herschel\ and ALMA and the formal MBB solution. The shaded area shows the 1$\sigma$ spread of the models in the MCMC chains. {\it Middle}: \coone\ luminosity vs. IR luminosity. Blue squares show SMGs from the literature \citep{Harris10,Riechers11b,Ivison11,Bothwell13,Sharon13}. The dashed lines show contours of constant gas depletion timescales. {\it Right}: SFR surface density vs. molecular gas mass surface density (i.e., the Kennicutt-Schmidt relation). Other data points show local ULIRGs \citep{Kennicutt98b}, high-redshift SMGs \citep{Daddi09,Genzel10,Fu13}, normal star-forming galaxies at $z \sim 2$ and $z \sim 0$ \citep{Tacconi13}. The dashed lines indicate constant ratios of surface densities, which are another indicator of the gas depletion timescale. 
\label{fig:smg}} 
\epsscale{1.0}
\end{figure*}

\begin{table}
\begin{center}
\caption{Spectroscopy of the SMG}
\label{tab:smg_spec}
\begin{tabular}{lcc}
\hline
\hline
Quantity & Measurement & Unit \\
\hline
\multicolumn{3}{c}{H$\alpha$ from Gemini/GNIRS} \\
\hline
$z_{\rm H\alpha}$        & $2.6743(3)$ & \nodata \\ 
$\Delta V_{\rm FWHM}$    & $<300$ & \kms \\
$F_{\rm H\alpha}$        & $(6.1\pm1.0)\times10^{-17}$ & erg\,s$^{-1}$\,cm$^{-2}$ \\
$L_{\rm H\alpha}$        & $(2.9\pm0.5)\times10^{41}$ & erg\,s$^{-1}$ \\
SFR$_{\rm H\alpha}$      & $1.6\pm0.3$ & \msunyr \\
\hline
\multicolumn{3}{c}{\cothree\ from ALMA Band-3} \\
\hline
\multicolumn{3}{c}{{\it Narrow Component}} \\
$z_{\rm CO3-2}$          & 2.67399(3) & \nodata \\
$\Delta V_{\rm FWHM}$    & $249\pm8$        & \kms \\
$S_{\rm CO} \Delta V$    & $1.38\pm0.06$   & Jy\,\kms \\
$L'_{\rm CO3-2}$         & $(5.0\pm0.2)\times10^{10}$ & K\,\kms\,pc$^2$ \\
\multicolumn{3}{c}{{\it Broad Component}} \\
$z_{\rm CO3-2}$          & 2.6741(7)       & \nodata \\
$\Delta V_{\rm FWHM}$    & $906\pm206$     & \kms \\
$S_{\rm CO} \Delta V$    & $0.41\pm0.07$   & Jy\,\kms \\
$L'_{\rm CO3-2}$         & $(1.5\pm0.3)\times10^{10}$ & K\,\kms\,pc$^2$ \\
\multicolumn{3}{c}{{\it Total Emission}} \\
$L'_{\rm CO1-0}$         & $(1.25\pm0.07)\times10^{11}$ & K\,\kms\,pc$^2$ \\
$M_{\rm mol}$            & $(1.25\pm0.07)\times10^{11}$ & \msun \\
\multicolumn{3}{c}{{\it Intrinsic Source Size}} \\
Deconv. Maj.             & $0.76\pm0.09$ & arcsec \\
Deconv. Min.             & $0.54\pm0.17$ & arcsec \\
\hline
\end{tabular}
\end{center}
\tablecomments{We have adopted a \cothree/\coone\ brightness temperature ratio of $r_{31} = 0.52$ and a CO to molecular gas conversion factor of $\alpha_{\rm CO} \equiv M_{\rm mol}/L'_{\rm CO1-0} = 1.0~M_\odot/({\rm K~km~s^{-1}~pc^2})$. Here and in Table\,\ref{tab:smg_phot}, the intrinsic source sizes are given by the beam-deconvolved major- and minor-axis FWHMs ($a$ \& $b$), measured by CASA task \texttt{imfit}.}
\end{table}

\begin{table}
\begin{center}
\caption{Photometry of the SMG}
\label{tab:smg_phot}
\begin{tabular}{lcc}
\hline
\hline
Quantity & Measurement & Unit \\
\hline
\multicolumn{3}{c}{Herschel/SPIRE} \\
\hline
R.A.       & 09:13:39.32   & hms \\
Decl.      & $-$01:06:58.6 & dms \\
$S_{250}$ & $52.5\pm7.4$ & mJy \\
$S_{350}$ & $69.4\pm8.8$ & mJy \\
$S_{500}$ & $48.4\pm9.2$ & mJy \\
\hline
\multicolumn{3}{c}{ALMA Band-6 Continuum} \\
\hline
R.A.                & 09:13:39.55   & hms \\
Decl.               & $-$01:06:56.4 & dms \\
$S_{\rm 343.5GHz}$  & $7.4\pm0.5$ & mJy \\
Deconv. Maj.        & $0.46\pm0.04$ & arcsec \\
                    & $3.7\pm0.3$ & kpc \\
Deconv. Min.        & $0.28\pm0.06$ & arcsec \\
                    & $2.2\pm0.5$ & kpc \\
\hline
\multicolumn{3}{c}{ALMA Band-3 Continuum} \\
\hline
$S_{\rm 93.1GHz}$  & $106\pm22$ & $\mu$Jy \\
$S_{\rm 105.1GHz}$ & $111\pm20$ & $\mu$Jy \\
Deconv. Maj.       & $0.68\pm0.47$ & arcsec \\
Deconv. Min.       & $0.47\pm0.23$ & arcsec \\
\hline
\multicolumn{3}{c}{Modified Blackbody Fit} \\
\hline
$T$  & $44\pm7$ & K \\
$\beta$         & $2.3\pm0.2$ & \nodata \\
$\lambda_0$     & $98\pm24$  & \um \\
$\pi r_s^2$    & $9^{+10}_{-5}$ & kpc$^2$ \\
$M_{\rm dust}$  & $(5.4\pm1.2)\times10^8$    & \msun \\
$L_{\rm IR}$    & $1.17^{+0.17}_{-0.11}\times10^{13}$ & \lsun \\ 
SFR$_{\rm IR}$  & $1170^{+170}_{-110}$ & \msunyr \\ 
\hline
\end{tabular}
\end{center}
\tablecomments{The errors of the {\it Herschel} photometry include confusion noises of 5.3, 6.4 and 6.7~mJy~beam$^{-1}$ at 250, 350, and 500~\um, respectively \citep{Pascale11,Rigby11}. The IR luminosity, $L_{\rm IR}$, is defined as the integrated luminosity between 8 and 1000~\um\ at rest-frame. The dust-obscured SFR is estimated using the commonly used conversion: ${\rm SFR}/M_\odot~{\rm yr}^{-1} = 10^{-10} L_{\rm IR}/L_\odot$ \citep[e.g.,][]{Daddi10b}.}
\end{table}

Fig.~\ref{fig:cospec} shows the ALMA band-3 spectrum of the SMG. The spectrum is extracted with an elliptical aperture matching the beam-convolved source size. A prominent emission line peaks at $\nu_{\rm obs} = 94.12$\,GHz, which we identify as the \cothree\ line at $z_{\rm CO} = 2.67399\pm0.00003$. The CO detection thus confirms the H$\alpha$ redshift from the GNIRS spectrum ($z_{\rm H\alpha} = 2.6743\pm0.0003$, see \S~\ref{sec:specID}). Because the CO line is detected at a higher S/N and is less affected by dust extinction, we adopt the CO redshift for the SMG throughout the paper, i.e., $z_{\rm SMG} = z_{\rm CO} = 2.674$ (a slightly rounded-up value for simplicity).

A closer inspection of the \cothree\ spectrum reveals a broad (FWHM $\sim$ 1000\,\kms) emission-line component with a peak flux of $\sim$0.4\,mJy underneath the prominent narrow component (FWHM $\sim$ 250\,\kms). We thus model the CO spectrum with two Gaussians and compare its result with a single-Gaussian model.

We find that the improvement of the double-Gaussian model over the single-Gaussian model is highly significant. The formal double-Gaussian solution achieves a $\chi^2 = 220.2$ for a degree-of-freedom (DOF) of 209. For comparison, the formal single-Gaussian solution achieves a $\chi^2 = 251.8$ for DOF = 212. According to the $F$-test, such a difference rejects the null hypothesis, that the double-Gaussian model does not provide a significantly better fit, at a confidence level of 99.99964\% or 4.6$\sigma$.
The result of the double-Gaussian fit from the ``{\sc amoeba + mcmc}'' method is listed in Table\,~\ref{tab:smg_spec}. The broad component has an FWHM of $\sim$900\,\kms\ and accounts for almost a quarter of the total emission-line flux. The existence of a narrow CO line on top of a broad CO line with essentially no velocity offset indicates that the SMG's intense star-forming nucleus (the narrow component) is either embedded in a fast-rotating disk or driving a bipolar outflow (the broad component). Unfortunately, the spatial resolution of the ALMA data is inadequate to distinguish between the two scenarios. We note that such a broad CO component would not have been detected in shallower spectroscopic data that are more generally available to SMGs, so this feature may not be unique to \SMG. 

With the redshift determined, we fit the SED between 250~\um\ and 3~mm from {\it Herschel}/SPIRE and ALMA with a modified blackbody curve. We adopt the general solution of the radiative transfer equation assuming local thermal equilibrium at a constant temperature $T$: 
\begin{equation}
S_\nu = (1-e^{-\tau_\nu})~B_\nu(T)~\pi r_s^2/d_L^2,
\label{eq:mbb}
\end{equation}
where $B_\nu(T)$ is the Planck function at a temperature of $T$ and a rest-frame frequency $\nu$, $\pi r_s^2$ the effective size of the dust emitting region, and $d_L$ the luminosity distance. Assuming that the dust opacity follows a power-law with a negative slope of $-\beta$ at wavelengths greater than the dust size ($\sim$10~\um), the optical depth should follow the same power-law: 
\begin{equation}
\tau_\nu = (\nu/\nu_0)^\beta = (\lambda/\lambda_0)^{-\beta},
\label{eq:tau}
\end{equation}
where $\nu_0$ ($\lambda_0$) is the rest-frame frequency (wavelength) at which the dust becomes optically thick. Given the dust mass-absorption coefficient of $\kappa = 0.07$\,m$^2$~kg$^{-1}$ at 850~\um\ for Galactic dust \citep{Dunne00,James02}, it can be shown that the dust mass is:
\begin{equation}
M_{\rm dust} = 9.0\times10^9~M_\odot~(\pi r_s^2/{\rm kpc}^2)~(\lambda_0/850\,\mu {\rm m})^\beta.
\label{eq:mdust}
\end{equation}
The result of the SED fit gives us a measure of the dust temperature, the dust-obscured SFR, the dust mass, and the effective size of the dust photosphere. Fig.\,\ref{fig:smg} {\it left} shows the multi-band photometry, the median MCMC model, and the 1$\sigma$ spread of the models. Table\,\ref{tab:smg_phot} lists the formal parameters and their uncertainties. The dust is relatively warm ($T = 44\pm7$\,K) and the dust photosphere has an effective size of $9_{-5}^{+10}$\,kpc$^2$, comparable to the intrinsic source size measured at 343.5\,GHz -- $\pi a b/4 = 6.4\pm1.5$\,kpc$^2$.

The ALMA \cothree\ line luminosity offers an estimate of the mass of the molecular gas reservoir. Because \cothree\ traces the warm and moderately dense ($n_{\rm eff} \sim 10^4$\,\cc) component \citep[e.g.,][]{Juneau09}, we first convert the \cothree\ to \coone\ luminosity using the average brightness temperature ratio of $r_{31} \equiv L'_{\rm CO3-2}/L'_{\rm CO1-0} = 0.52$ observed in SMGs \citep{Bothwell13}. We then convert the \coone\ luminosity to the total molecular gas mass with a CO-to-Molecular-Gas conversion factor of $\alpha_{\rm CO} = 1.0$, a value found appropriate for high-redshift dusty starbursts \citep[e.g.,][]{Hodge12,Magnelli12,Xue18}. The result is a total molecular gas mass of $M_{\rm mol} = (1.25\pm0.07)\times10^{11}~(r_{31}/0.52)^{-1}~(\alpha_{\rm CO}/1.0)$\,\msun, near the high end of the molecular gas masses measured in SMGs (see Fig.\,\ref{fig:smg} {\it middle}). 

Combining the results from the SED fit and the \cothree\ spectroscopy, we found a gas depletion timescale of $\sim$0.1~Gyr, which is similar to other SMGs but 6$\times$ shorter than co-eval main-sequence galaxies (Fig.\,\ref{fig:smg} {\it middle}). The SMG's dust emission is resolved by the ALMA band-6 data with a beam-deconvolved size of 0.46\arcsec$\times$0.28\arcsec. Its \cothree\ emission is resolved by the ALMA band-3 data with a beam-deconvolved size of 0.76\arcsec$\times$0.54\arcsec. In both cases, we have measured the intrinsic source sizes from {\sc clean}'ed images using the CASA task \texttt{imfit}. These size measurements allow us to place the SMG on the Kennicutt-Schmit relation (Fig.\,\ref{fig:smg} {\it right}). It is characterized by high surface densities of SFR and molecular gas even compared to the SMG population. But like other SMGs, it features a high star-formation efficiency that is distinct from normal star-forming galaxies at $z \sim 2$ \citep[e.g.,][]{Daddi10b,Genzel10}.

\subsection{Companion Galaxies} \label{sec:comps}

We carried out a blind search of line emitters in the ALMA datacubes with a matched-filter algorithm and tested the fidelity of the detections with simulated noise-only interferometer data (see Appendix \S~\ref{sec:blindCO}). In the spectral window centered at 94\,GHz (i.e., BB4), we found two robust line emitters: the SMG at $z = 2.674$ (S/N = 67) and \COM\ at $z = 2.6747$ (S/N = 6.4). The other companion, \Chost, was detected at high significance when we combine the two ALMA channels closest in velocity to the metal-line-detected absorbing clouds C1 and C2 toward \bgQSO\ (see \S~\ref{sec:aodm}). The source has a peak S/N of 4.5 when combining the two channels, while its peak S/N is only 3.6 in the two individual channels, making it confused with noise spikes. The matched-filter algorithm fails to identify \Chost\ because it assumes that only adjacent channels can boost the S/N above the detection limit. In other words, the detection of \Chost\ is possible only because (1) we have utilized the prior knowledge of the redshifts of the absorption lines (with the implicit assumption that the emission counterparts have similar redshifts), and (2) the emission counterparts of the two clouds are superimposed on the sky (increasing the S/N when they are combined). 

Our blind search detected four additional high-fidelity line emitters: \fgQSO\ at $z = 2.7488$ (S/N = 7.0), its companion at $z = 2.7392$ ($\delta v = -770$\,\kms; S/N = 6.1) located $\sim$30\arcsec\ to the NE of \fgQSO, and two additional sources at $z = 2.3452$ (S/N = 5.5) and $z = 2.3324$ (S/N = 5.2) that may correspond to the $\zabs = 2.345$ \HI\ and C\,{\sc iv} absorbers that appear toward both QSOs (see Fig.~\ref{fig:absorbers} in Appendix~\ref{sec:otherabs}). But on the other hand, only the SMG is detected in the continuum image of the two higher-frequency spectral windows (i.e., BB1 and BB2).

While the detection of CO in the SMG and \fgQSO\ is expected, the detection of their companion CO emitters is not. We can use the ALMA Spectroscopic Survey in the Hubble Ultra Deep Field \citep[ASPECS;][]{Decarli19} to estimate a baseline level of source-detection probability in normal field environments. The ASPECS covers an area of 4.6\,arcmin$^2$ (${\rm PB} \geq 0.5$, where PB is a correction factor for the primary beam pattern) with 17 pointings in band 3 and a spectral range of 21\,GHz (84$-$105\,GHz) with five tunings. The rms sensitivity varies with frequency with a range between 0.12 and 0.4\,mJy~beam$^{-1}$~channel$^{-1}$ for a channel spacing of 7.8\,MHz (the same as ours). To provide a conservative estimate, we only count the 7 sources detected at ${\rm S/N} > 6$ between 96 and 103\,GHz \citep{Gonzalez-Lopez19}, where ${\rm rms} \simeq 0.135$ mJy~beam$^{-1}$~channel$^{-1}$. Only within this spectral range is ASPECS more sensitive than our data (${\rm rms} \simeq 0.16$\,mJy~beam$^{-1}$~channel$^{-1}$). This gives a source density of $0.22\pm0.08$\,arcmin$^{-2}$~GHz$^{-1}$ in the field. Given that our ALMA observations cover an area of 0.88\,arcmin$^2$ where ${\rm PB} \geq 0.5$ and are $\sim$20\% shallower, one would expect to identify less than $0.36\pm0.14$ sources at ${\rm S/N} > 6$ over the 1.875\,GHz bandwidth of a baseband, and only a third of these (i.e., $<0.12\pm0.05$ sources) are expected to fall within $\pm$0.3\,GHz (1000\,\kms) of the main galaxies to be considered as companions. In other words, one would need to increase our survey area by $>$8$\times$ to detect a chance ``companion'' galaxy of the SMG or \fgQSO\ in the field. Yet we have detected one ${\rm S/N} > 6$ companion within 30\arcsec\ of each main galaxy. Our result thus indicates that both the SMG and \fgQSO\ inhabit overdense environments, which is consistent with their purported large halo masses \citep{Hickox12}. 

In addition to the companion galaxies detected in CO emission, both the SMG and \fgQSO\ are also associated with absorbers of high \HI\ column density in the spectrum of a common background QSO (\bgQSO), as we will show in the next section. 

\section{Absorption-line Systems} \label{sec:qsos}

The two QSOs in the \system\ system were first identified in the Sloan Digital Sky Survey (SDSS) DR9 quasar catalog \citep{Paris12}. The pair has a separation of only 10.8\arcsec, and more importantly, two closely separated DLAs were immediately identified in the low-resolution SDSS spectrum of \bgQSO\ \citep{Noterdaeme12}. The stronger DLA ($\logNHI \simeq 21.3$) at $\zabs \approx 2.75$ is associated with \fgQSO\ at $z = 2.7488$, providing an important probe of the CGM around QSOs at an impact parameter of $R_\bot = 85$\,kpc (see Appendix~\ref{sec:otherabs}). The other DLA ($\logNHI \simeq 20.5$) at $\zabs \approx 2.68$ provides a window to probe the CGM of the SMG at $z = 2.674$, which is just 11.7\arcsec\ from the QSO. 

We searched the spectral databases with {\tt specdb}\footnote{\url{https://github.com/specdb/specdb}} and found that the QSO pair had accumulated an excellent set of spectroscopic data from Gemini Multiobject Spectrograph \citep[GMOS;][]{Prochaska13a}, VLT/X-shooter \citep{Finley14}, and Magellan Echellette Spectrograph \citep[MagE;][]{Rubin15}. \citet{Finley14} noticed the strong coincident absorption at $\zabs \approx 2.68$ in the SDSS spectra of both QSOs, which motivated them to obtain the higher resolution X-shooter spectra for a detailed analysis. The absorption structure toward both QSOs is resolved into three major subsystems of variable metallicities and with a total velocity span of $>$1700\,\kms. The observed kinematic and metallicity coherence across sightlines is remarkable, given the 86\,kpc separation between the QSOs. The authors interpreted the system as a gaseous overdensity extended by six Mpc along the line-of-sight, which is suggestive of a clumpy filamentary structure that may eventually collapse and form a proto-cluster. They attributed the two main subsystems at lower velocities (A and B) as part of the IGM because of their low metallicity (${\rm [Fe/H]} < -1.9$) and suspected that the third main subsystem (C) with ${\rm [Fe/H]} = -1.1$ is likely associated with a galaxy. Now with the detection of the SMG and its companion galaxies, we will use the coincident absorption-line system to characterize the CGM of these galaxies in \S~\ref{sec:connection}. We will show that subsystems A and B are cool gas streams in the CGM of the SMG, and subsystem C is indeed associated with a galaxy (\Chost).  

In this section, we present a re-analysis of the $\zabs \approx 2.68$ absorption system using a new reduction of the X-shooter spectra (\S~\ref{sec:xshooter}). \citet{Finley14} used {\tt vpfit}\footnote{\url{https://people.ast.cam.ac.uk/~rfc/vpfit.html}} to fit Voigt profiles to the entire spectrum. Our approach is complementary to the {\tt vpfit} analysis and our results show a good agreement with those presented in \citet{Finley14}. The main differences between the two analyses are:

\begin{enumerate}

\item We fit Voigt profiles to the \HI\ Lyman series after masking out contaminating LYAF lines and quantify the statistical and systematic uncertainties of the model parameters using an MCMC algorithm (\S~\ref{sec:voigt}). 

\item We measure the ionic column densities of metals with the apparent optical depth method (AODM; \S~\ref{sec:aodm}). This is a more direct and conservative technique compared with Voigt profile fitting using {\tt vpfit}, because it relies only on equivalent width measurements and uses a straightforward method to detect line saturation.   

\item We use ionic column density ratios to constrain the photoionization model for each cloud, which in turn provides the ionization correction factors necessary for metallicity estimates (\S\,\ref{sec:MH}).

\item We use the SMG to define the systemic redshift and adopt the solar abundance scale of \citet{Asplund09}.

\end{enumerate}
  
\subsection{VLT X-shooter Spectroscopy} \label{sec:xshooter}

The X-shooter observations of the QSOs took place between 2013 March 31 and 2013 May 1 on the 8.2\,m VLT/UT2 telescope (program ESO\,089.A-0855; \citealt{Finley14}). X-shooter uses three individual echelle spectrographs to cover a wide wavelength range between 0.3 and 2.5\,\um\ simultaneously \citep{Vernet11}. \citet{Finley14} estimated spectral resolutions of $R \sim 6400$ (FWHM = 47\,\kms) in the UVB arm (3000$-$5600\AA), $R \sim 11000$ (27\,\kms) in the VIS arm (5500\AA$-$1\um), and $R \sim 6600$ (45\,\kms) in the NIR arm (1$-$2.5\,\um). The total exposure times are 100\,min for \fgQSO\ and 310\,min for \bgQSO. We downloaded the raw data from the ESO archive and reduced the data with the spectroscopy data reduction pipeline developed by George Becker\footnote{\url{ftp://ftp.ast.cam.ac.uk/pub/gdb/}}. The final 1D spectra were corrected for the 0.2\AA\ (0.5 pixel) wavelength redshift of the spectra in the VIS arm noticed by \citet{Noterdaeme12a}, which is likely produced by uncompensated instrumental flexure. 

We fit the QSO continuum using the Python software package {\tt linetools}\footnote{\url{https://github.com/linetools/linetools}}. First, the spectrum is divided into a number of wavelength intervals, which are $\sim$50~\AA\ wide shortward of the \Lya\
emission, narrower across strong emission lines, and wider in regions free of
emission lines and longward of \Lya. Next, a spline is fit through the central wavelength and median flux of each interval (i.e., the spline ``knots''). Finally, these ``knots'' are iteratively added, deleted, or moved until a satisfactory continuum fit is obtained.

\begin{figure*}[!tb]
\epsscale{0.58}
\plotone{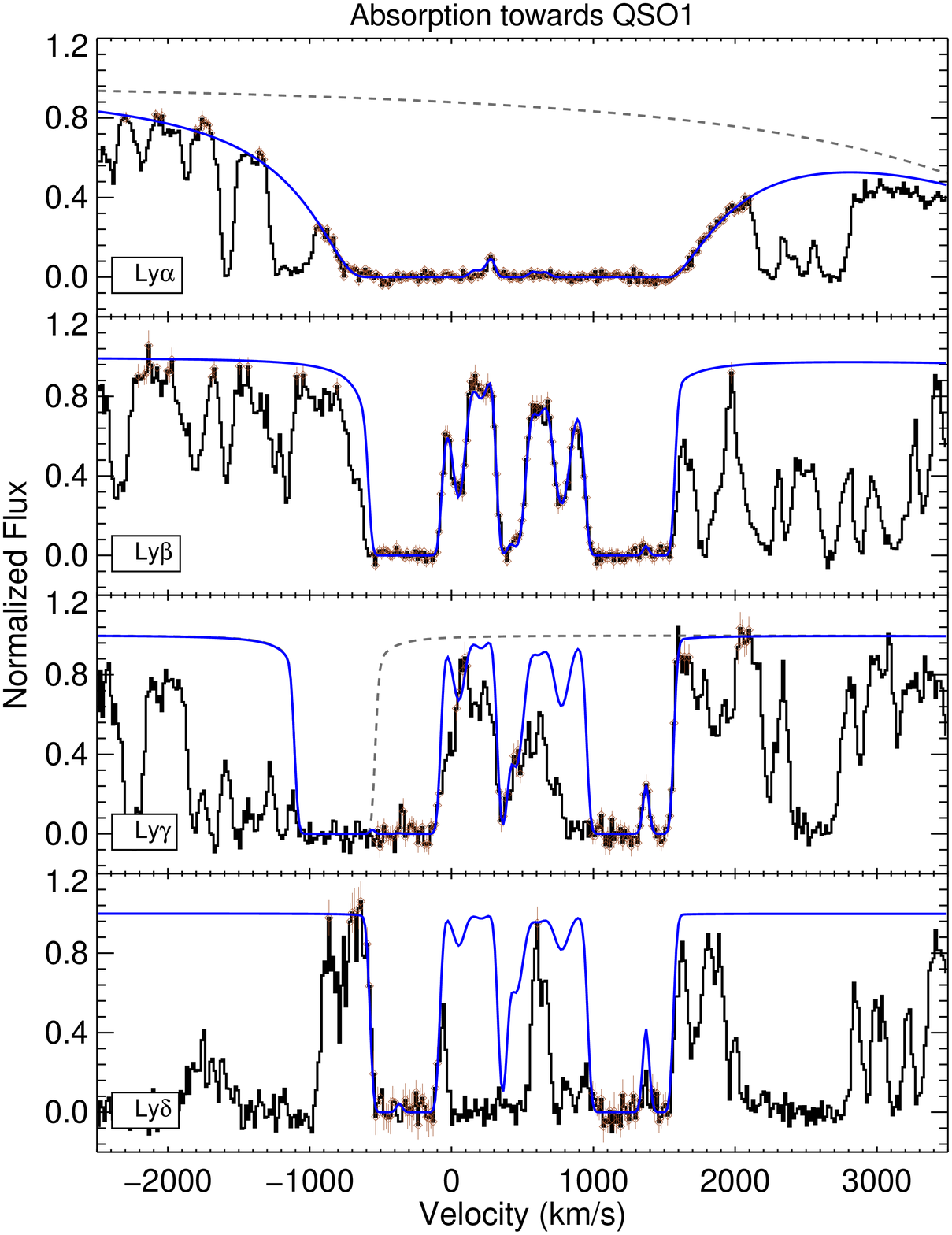}
\plotone{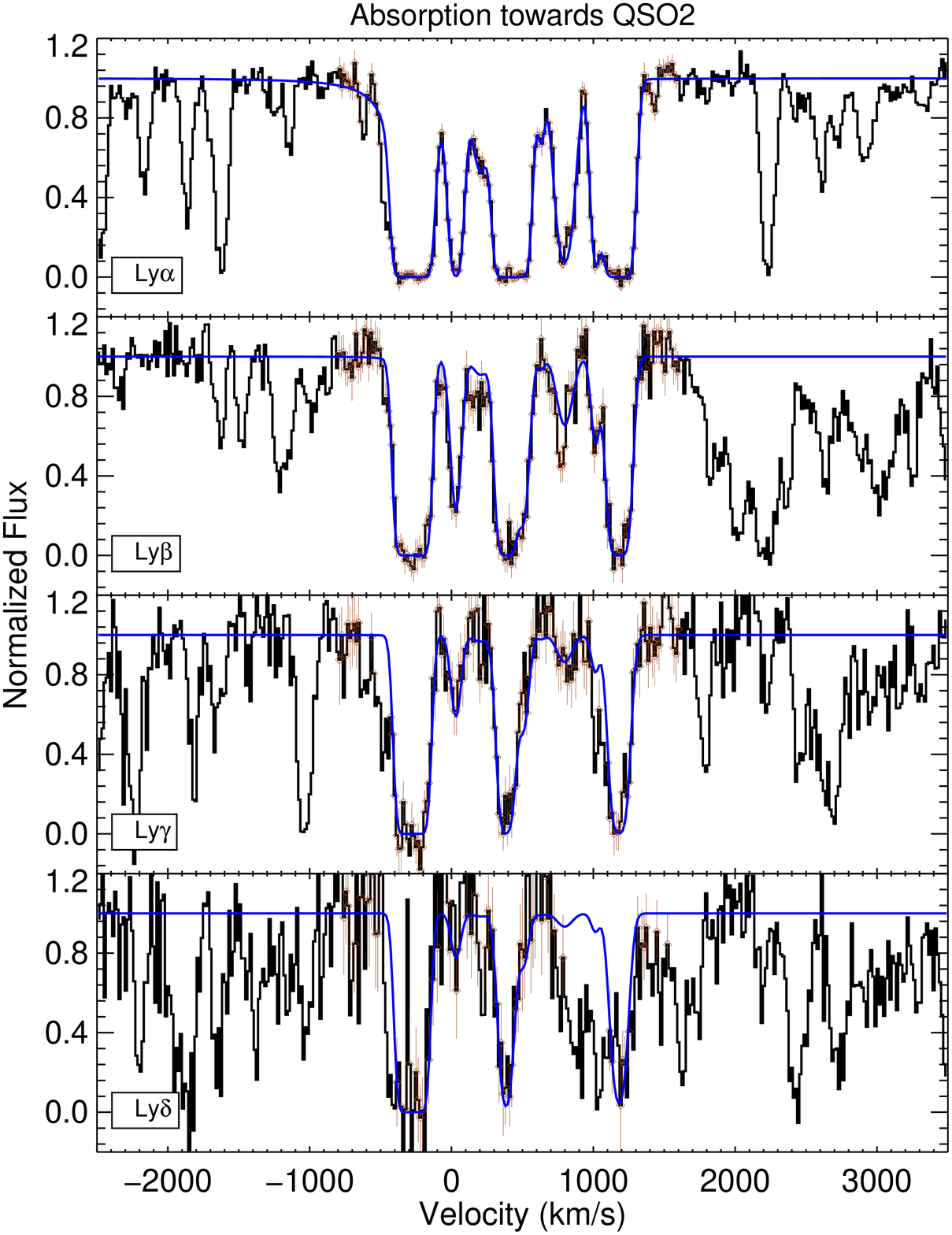}
\caption{Velocity profiles of \HI\ absorption near the SMG redshift toward \bgQSO\ and \fgQSO, overlaid with best-fit Voigt profiles ({\it blue curves}). We indicate data points in contamination-free regions with brown diamonds with error bars. The gray dashed lines in the left panels show the \HI\ \Lya\ and \Lyd\ absorption from the DLA at $\zabs = 2.751$; note how significantly they affect the \Lya\ and \Lyg\ profiles of the absorption at $\zabs \approx 2.68$. All velocities are relative to $\zsmg = 2.674$.
\label{fig:voigt}}
\epsscale{1.0}
\end{figure*}

\begin{figure*}[!tb]
\epsscale{0.58}
\plotone{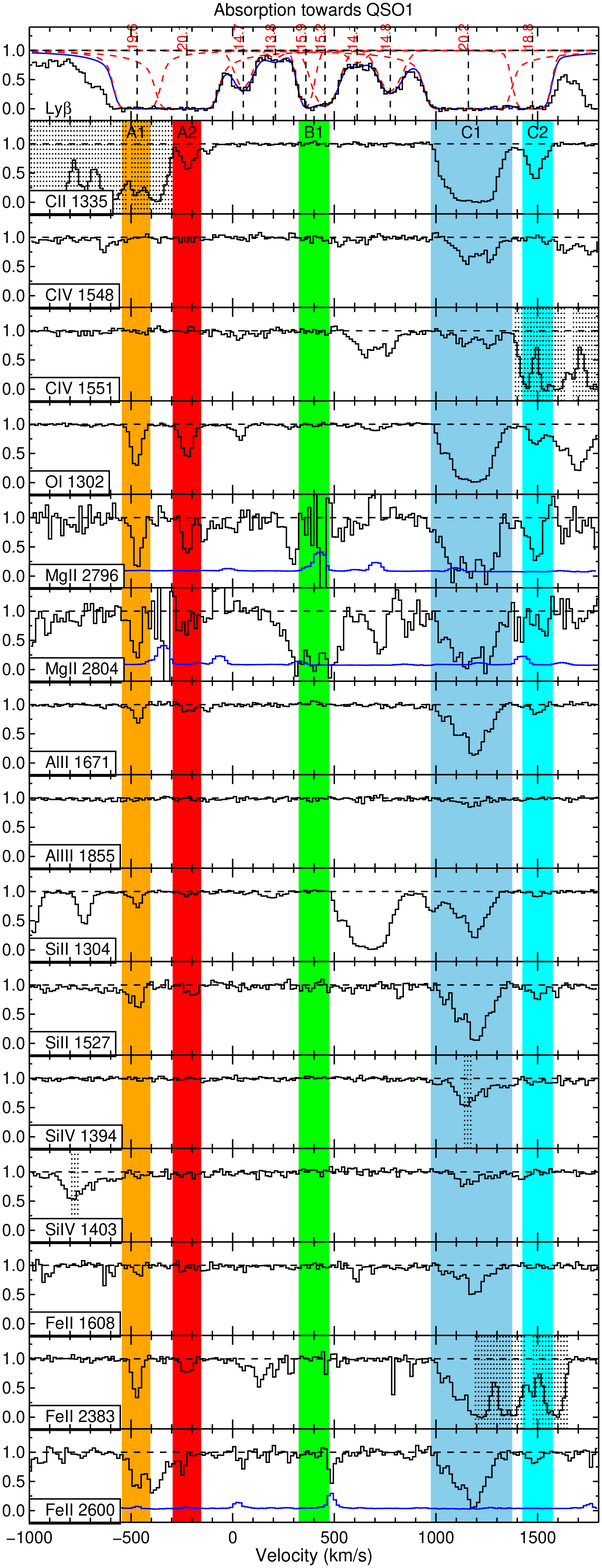}
\plotone{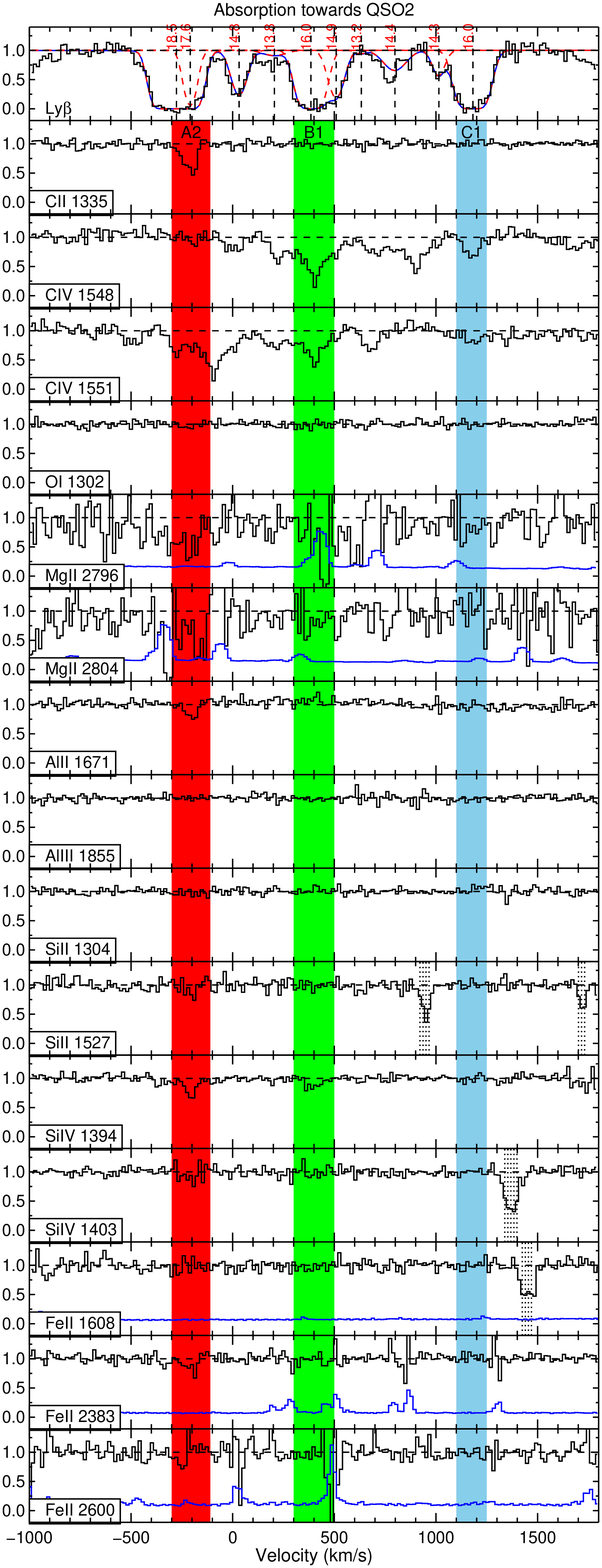}
\caption{Velocity profiles of \HI\ Ly$\beta$ and selected metal lines toward \bgQSO\ ({\it left}) and \fgQSO\ ({\it right}). All velocities are relative to $\zsmg = 2.674$. The velocity integration ranges of the clouds defined in Table\,\ref{tab:aodm} are highlighted in color. Vertical dotted lines strike out regions that are blended with lines from absorbers at other redshifts (see Appendix~\ref{sec:otherabs}). The error spectrum is plotted ({\it blue}) when it shows significant structures. 
\label{fig:aodm}}
\epsscale{1.0}
\end{figure*}

\subsection{Voigt Profile Fitting of Neutral Hydrogen} \label{sec:voigt}

Fig.~\ref{fig:voigt} shows the \HI\ absorption profiles (\Lya\ through \Lyd) of the absorbers at $\zabs \approx 2.68$ toward the two QSOs. Although the two QSOs are separated by 10.8\arcsec\ (86\,kpc at $z = 2.674$), their \HI\ absorption profiles show strikingly similar velocity structures spanning over 1800\,\kms, as first noted by \citet{Finley14}. The kinematic coherence indicates that the medium responsible for the absorption is extended at least 86\,kpc across the sky plane. 

Line blending from other absorbers is evident, as indicated by the disagreement in velocity profile among the Lyman series. To measure \HI\ column densities in such a complex situation, it is beneficial to first identify a guessed solution by iteratively varying the Voigt profiles (convolved to $R = 6400$) until an acceptable fit to the data is obtained. The guessed solution not only provides a good starting point for the formal minimum $\chi^2$ approach below, but also helps to identify regions contaminated by line blending (which thus should be flagged out). During this procedure, we find that a minimum of 10 clouds are needed to adequately fit the Lyman series in each sightline. Because each cloud is described by three parameters ($v, b, \logNHI$), our model for each QSO spectrum has a total of 30 free parameters.

To model the absorption toward \bgQSO, the \HI\ Lyman lines of the DLA at $\zabs \approx 2.751$ (see Fig.\,\ref{fig:absorbers}$a$) must be included in the model because its \Lya\ and \Lyd\ blend with the \Lya\ and \Lyg\ profiles of the absorber at $\zabs \approx 2.68$. We find that the DLA's \HI\ absorption is adequately modeled as two clouds separated by 290\,\kms\ ($\zabs = 2.7502,2.7538$), each with $\logNHI = 21.0$ and $b = 40$\,\kms\ (see Fig.\,\ref{fig:cgm_qso2}). These parameters for the DLA at $\zabs = 2.751$ are fixed in the fitting process.

The $\chi^2$ minimization is focused on the velocity range between $-1500$ and 2100\,\kms\ for \bgQSO\ and between $-600$ and 1350\,\kms\ for \fgQSO. With the guessed solution, we also mask out the pixels that are clearly contaminated by line blending within the fitting ranges. The surviving ``good'' pixels are indicated by diamond symbols with error bars in Fig.\,\ref{fig:voigt} and the Voigt models are optimized using the {\sc amoeba + mcmc} method described in \S~\ref{sec:org}. The priors of central velocities and column densities are centered around the guessed solution, with bounds of $\pm$100\,\kms\ for $v_{\rm HI}$ and $\pm$0.8\,dex for $\log N_{\rm HI}$. On the other hand, the Doppler parameter, $b_{\rm HI}$, is allowed to vary between 5 and 70\,\kms. 
For three \HI\ components, we found it necessary to fix their velocities to those measured from low-ion metal lines, because the \HI\ series alone do not constrain their velocities well. Specifically, these components are at $-470$ and $-225$\,\kms\ toward \bgQSO\ and at $-209$\,\kms\ toward \fgQSO.  
The optimized models are plotted against the data as blue curves in Fig.\,\ref{fig:voigt} and the formal parameters and their statistical uncertainties are tabulated in Table\,\ref{tab:voigt}. 

Because of the empirical nature of our placement of the unabsorbed QSO continuum, the Voigt parameters suffer from significant systematic uncertainties. In particular, we are interested in the systematic uncertainties of $\logNHI$, which depends on (1) the column density due to the varying gradient of the curve of growth, (2) the quality of the spectrum, and (3) the significance of line blending. To quantify this, we run the same modeling procedure as above but vary the QSO continuum model by $\pm$10\% and use the resulting offsets between the three formal solutions to estimate systematic uncertainties. All subsequent errors in $\logNHI$ and metallicities ([X/H]) include both statistical and systematic uncertainties.

Fig.\,\ref{fig:voigt} reveals that there are three separate kinematic clumps centered around $\delta v \approx -300, +400, +1200$\,\kms\ relative to $\zsmg = 2.674$ (i.e., $\zabs \approx 2.6703, 2.6789, 2.6887$), which we designate as subsystems A, B, and C, respectively, following the nomenclature of \citet{Finley14}. The same clumps appear toward both QSOs, although their column densities vary between sightlines. Half of the six subsystems are optically thick (i.e., $\logNHI > 17.2$), including two sub-DLAs (\bgQSO-A and \bgQSO-C) and one LLS (\fgQSO-A). Metal absorption lines from these subsystems are thus expected, as we will show in the next subsection. 

\begin{table}[!tb]
\begin{center}
\caption{Voigt Solution for \HI\ Lyman Lines}
\label{tab:voigt}
\begin{tabular}{cccccc}
\hline
\hline
\multicolumn{3}{c}{toward \bgQSO} & \multicolumn{3}{c}{toward \fgQSO} \\
$v_{\rm HI}$ & $b_{\rm HI}$ & $\log N_{\rm HI}$ & $v_{\rm HI}$ & $b_{\rm HI}$ & $\log N_{\rm HI}$ \\ 
(\kms)       & (\kms) & ($\log {\rm cm}^{-2}$) & (\kms)       & (\kms) & ($\log {\rm cm}^{-2}$) \\
\hline
$-470.1_{-0.0}^{+0.0}$&$32.2_{-1.2}^{+1.0}$&$19.64_{-0.05}^{+0.04}$&	$-276.3_{-1.3}^{+1.2}$&$47.2_{-0.7}^{+0.8}$&$18.56_{-0.09}^{+0.06}$\\
$-224.7_{-0.0}^{+0.0}$&$42.2_{-0.5}^{+0.6}$&$20.06_{-0.04}^{+0.04}$&	$-208.5_{-0.0}^{+0.0}$&$14.2_{-6.3}^{+6.6}$&$17.48_{-0.47}^{+0.41}$\\
$52.5_{-2.4}^{+2.3}$&$45.8_{-3.6}^{+3.7}$&$14.66_{-0.03}^{+0.03}$&	$31.6_{-1.3}^{+1.3}$&$38.2_{-1.4}^{+1.4}$&$14.79_{-0.04}^{+0.04}$\\
$210.1_{-8.2}^{+7.8}$&$43.3_{-12.0}^{+13.6}$&$13.75_{-0.11}^{+0.10}$&	$205.1_{-5.9}^{+5.9}$&$67.0_{-4.2}^{+2.2}$&$13.75_{-0.03}^{+0.03}$\\
$360.1_{-3.4}^{+4.3}$&$22.2_{-3.5}^{+3.6}$&$15.90_{-0.25}^{+0.41}$&	$384.7_{-4.2}^{+4.0}$&$46.2_{-3.6}^{+3.4}$&$16.03_{-0.13}^{+0.18}$\\
$454.5_{-8.1}^{+9.0}$&$58.3_{-8.3}^{+7.0}$&$15.22_{-0.07}^{+0.06}$&	$509.1_{-8.4}^{+7.0}$&$37.9_{-4.6}^{+5.3}$&$14.85_{-0.08}^{+0.08}$\\
$612.6_{-9.4}^{+8.9}$&$51.9_{-11.4}^{+11.5}$&$14.08_{-0.08}^{+0.08}$&	$632.9_{-4.8}^{+4.6}$&$18.4_{-7.9}^{+10.0}$&$13.24_{-0.07}^{+0.08}$\\
$775.4_{-2.7}^{+2.7}$&$62.1_{-3.9}^{+3.9}$&$14.83_{-0.02}^{+0.02}$&	$798.1_{-1.5}^{+1.5}$&$67.4_{-2.1}^{+1.7}$&$14.41_{-0.02}^{+0.02}$\\
$1159.5_{-2.0}^{+2.0}$&$59.0_{-0.7}^{+0.7}$&$20.23_{-0.02}^{+0.02}$&	$1014.4_{-2.3}^{+2.4}$&$29.4_{-2.6}^{+2.8}$&$14.30_{-0.05}^{+0.06}$\\
$1474.8_{-2.2}^{+1.6}$&$30.2_{-0.8}^{+1.0}$&$18.79_{-0.19}^{+0.14}$&	$1181.8_{-1.9}^{+1.9}$&$54.1_{-2.0}^{+2.0}$&$16.00_{-0.09}^{+0.10}$\\
\hline
\end{tabular} 
\end{center}
\tablecomments{All velocities are relative to the SMG redshift ($\zsmg = 2.674$). Reported parameters are the MCMC median values and their offsets from the 15.8 and 84.1 percentiles. Some velocities have zero uncertainties because they are fixed to those of the low-ion metal lines.}
\end{table}

\subsection{Ionic Column Densities from the AODM Method} \label{sec:aodm}

\begin{table}[!tb]
\begin{center}
\caption{Selected Metal Transitions}
\label{tab:metallines}
\begin{tabular}{lcrrr}
\hline
\hline
Ion & $\lambda_{\rm rest}$ & $\log f$ & IP$_0$ & IP$_1$ \\
    & (\AA)                &          & (eV)   & (eV) \\
\hline
 C~\sc{ii}&1334.5323&$-0.8935$&11.26&24.38\\
 C~\sc{iv}&1548.2040&$-0.7215$&47.89&64.49\\
 \nodata  &1550.7776&$-1.0234$&\nodata&\nodata\\
 O~\sc{i}&1302.1685&$-1.3188$& 0.00&13.62\\
Mg~\sc{ii}&2796.3543&$-0.2108$& 7.65&15.04\\
 \nodata  &2803.5315&$-0.5146$&\nodata&\nodata\\
Al~\sc{ii}&1670.7886&$ 0.2405$& 5.99&18.83\\
Al~\sc{iii}&1854.7183&$-0.2526$&18.83&28.45\\
Si~\sc{ii}&1304.3702&$-1.0640$& 8.15&16.35\\
 \nodata  &1526.7070&$-0.8761$&\nodata&\nodata\\
Si~\sc{iv}&1393.7602&$-0.2899$&33.49&45.14\\
 \nodata  &1402.7729&$-0.5952$&\nodata&\nodata\\
Fe~\sc{ii}&1608.4508&$-1.2388$& 7.90&16.20\\
 \nodata  &2382.7642&$-0.4949$&\nodata&\nodata\\
 \nodata  &2600.1725&$-0.6209$&\nodata&\nodata\\
\hline
\end{tabular}
\end{center}
\tablecomments{The columns $\lambda_{\rm rest}$ and $\log f$ list the rest-frame wavelengths and the oscillator strengths \citep{Morton03}. The columns IP$_0$ and IP$_1$ list the ionization potentials (IPs) to create the ion from the immediate lower state and to ionize it to the immediate higher state, respectively.} 
\end{table}

Fig.\,\ref{fig:aodm} compares the velocity profiles of \HI\ \Lyb\ and a selection of metal line transitions commonly observed in LLSs and DLAs (see Table\,\ref{tab:metallines} for transition data). We find that five of the six \HI\ subsystems are detected in at least one metal transition; the only exception is \bgQSO-B. Similar to their \HI\ absorption, the metal-line absorptions of subsystems \bgQSO-A and C, the two sub-DLAs, are resolved into multiple components. Note that the X-shooter spectrum has higher resolution for metal lines in the range $1500 < \lambda_{\rm rest} < 2700$\,\AA\ ($R = 11000$ or FWHM = 27\,\kms) than for the \HI\ Lyman lines at $\lambda_{\rm rest} < 1217$\,\AA\ ($R = 6400$ or FWHM = 47\,\kms). 
For each distinct metal-line cloud, we define a velocity integration window (highlighted in Fig.\,\ref{fig:aodm}) and name it by adding a number suffix to designate its associated subsystem. 
The cloud \bgQSO-C1 shows the strongest metal absorption with at least four blended components within $\sim$200\,\kms. We treat it as a single entity here, because for the purpose of measuring the gas metallicity it is unnecessary to deblend these components with Voigt profile fitting. 
Because the absorption toward \fgQSO\ spans a narrower velocity range than that toward \bgQSO, the former is missing the most blueshifted cloud ``A1'' and the most redshifted cloud ``C2''. For completeness, we defined \bgQSO-B1 based on its \HI\ absorption because no metal lines are detected there. As a result, there are a total of eight metal-line clouds.
The top section of Table~\ref{tab:aodm} lists the velocity integration windows of the clouds and their \HI\ column densities by summing $\logNHI$ of the Voigt components within the velocity windows. Each cloud contains only one \HI\ Voigt component except \bgQSO-B1 and \fgQSO-A2, both of which contain two closely separated components.

\begin{table*}[!tb]
\begin{center}
\caption{Properties of Metal-line-defined Clouds}
\label{tab:aodm}
\begin{tabular}{l rrrrrrrr}
\hline
\hline
Quantity   & QSO1-A1 & QSO1-A2 & QSO1-B1 & QSO1-C1 & QSO1-C2 & QSO2-A2 & QSO2-B1 & QSO2-C1 \\
\hline
$\delta v/{\rm km\,s}^{-1}$ &[$ -545, -405$]&[$ -295, -155$]&[$  325,  475$]&[$  975, 1375$]&[$ 1425, 1575$]&[$ -300, -110$]&[$  300,  500$]&[$ 1100, 1250$]\\
$\logNHI$ & $19.64_{-0.06}^{+0.04}$&$20.06_{-0.05}^{+0.08}$&$15.98_{-0.23}^{+0.39}$&$20.23_{-0.08}^{+0.07}$&$18.79_{-0.25}^{+0.33}$&$18.59_{-0.43}^{+0.23}$&$16.03_{-0.16}^{+0.21}$&$16.00_{-0.09}^{+0.10}$\\
\hline
log C\,{\sc iv}/C\,{\sc ii}   & \nodata & $<-1.06$ & \nodata & $<-1.10$ & $<-1.23$ & $<-0.80$ & $>0.93$ & $>0.36$ \\ 
log Al\,{\sc iii}/Al\,{\sc ii}& $<-0.20$ & $<0.17$ & \nodata & $-0.58$ & $<0.00$ & $<0.22$ & \nodata & \nodata \\ 
log Si\,{\sc iv}/Si\,{\sc ii} & $<-1.47$ & $<-0.92$ & \nodata & $-1.22$ & $<-0.86$ & $>-0.37$ & \nodata & \nodata \\ 
$\log U$ & $<-3.4$ & $<-2.9$ & \nodata & $-3.0$ & $<-3.2$ & $-2.9$ & $> -2.1$& $> -2.5$ \\
\hline
             $\log U$, adopted	&$-3.5$	&$-3.0$	&$-2.0$	&$-3.0$	&$-3.5$	&$-3.0$	&$-2.0$	&$-2.0$	\\
   $\log n_{\rm H}/{\rm cm}^3$	&$-1.4$	&$-1.9$	&$-2.9$	&$-1.9$	&$-1.4$	&$-1.9$	&$-2.9$	&$-2.9$	\\
   $\log N_{\rm H}/{\rm cm}^2$	&$19.9$	&$20.4$	&$19.5$	&$20.5$	&$19.7$	&$20.1$	&$19.5$	&$19.5$	\\
             $\log f_{\rm HI}$	&$-0.3$	&$-0.3$	&$-3.5$	&$-0.3$	&$-0.9$	&$-1.5$	&$-3.5$	&$-3.5$	\\
             $\log l/{\rm pc}$	&$ 2.8$	&$ 3.9$	&$ 3.9$	&$ 3.9$	&$ 2.6$	&$ 3.5$	&$ 3.9$	&$ 3.9$	\\
\hline
$[\alpha/{\rm H}]$&$-1.77\pm0.06$&$-2.28\pm0.08$&\nodata&$-1.08\pm0.08$&$-1.15\pm0.30$&$-1.91\pm0.34$&$-1.02\pm0.19$&$-1.58\pm0.15$\\
ion & O\,{\sc i}&O\,{\sc i}&\nodata&Si\,{\sc ii}& O\,{\sc i}&    C\,{\sc ii}& C\,{\sc iv}& C\,{\sc iv}\\
IC & $-0.01$ & $-0.01$ & \nodata & $-0.12$ & $-0.04$ & $-0.90$ & $-2.82$ & $-2.82$ \\
$[{\rm Fe/H}]$ & $-1.92\pm0.07$ & $-2.62\pm0.14$ & \nodata & $-1.27\pm0.09$ & $-1.76\pm0.34$ & \nodata & \nodata & \nodata \\
$[\alpha/{\rm Fe}]$ & $+0.15\pm0.09$ & $+0.34\pm0.16$ & \nodata & $+0.19\pm0.12$ & $+0.61\pm0.45$ & \nodata & \nodata & \nodata \\ 
\hline
\end{tabular}
\end{center}
\tablecomments{The first section gives the AODM velocity integration windows (relative to $\zsys = 2.674$) and the total \HI\ column densities of the Voigt components within these windows (the errors include systematic uncertainties due to continuum placement). The second section lists the observed ionic column density ratios, and their joint constraints on the ionization parameter ($\log U$). The third section lists the quantities implied by the {\sc cloudy} model for the adopted $\log U$ and $\log N_{\rm HI}$: total H volume density ($\log n_{\rm H}$), total H column density ($\log N_{\rm H}$), neutral H fraction ($\log f_{\rm HI}$), and characteristic line-of-sight depth ($\log l$). The last section lists the adopted $\alpha$-element metallicity ([$\alpha$/H]), the preferred ion, the ionization correction, the iron metallicity ([Fe/H]), and the $\alpha$-enhancement relative to iron ([$\alpha$/Fe]).}
\end{table*}

In Appendix \ref{sec:aodmtables}, we provide an overview of the AODM method and our measurements of ionic column densities from all of the selected transitions (Table\,\ref{tab:logN}). The listed uncertainties of the unsaturated and unblended detections in the Table include both statistical and systematic errors. Column densities from the AODM method are taken as lower limits for lines with more than one saturated pixels (which we define as $I_{\rm obs}/I_0 \leq 0.05$ or $\tau \geq 3$) and are taken as upper limits for lines that are blended with transitions from absorbers at other redshifts. Lastly, for undetected transitions, we quote 3$\sigma_{\rm sta}$ upper limits on the column densities. Systematic errors are not used here because it's not meaningful to adjust the QSO continuum around an undetected transition.

\subsection{Ionization Correction and Metallicities} \label{sec:MH}

A relative metallicity measurement of the intervening gas requires (1) \HI\ column density, (2) ionic column density of a metal element, (3) the reference solar abundances, and (4) the ionization correction. The definition of the relative metallicity makes this explicit: 
\begin{align}
{\rm [X/H]} &\equiv \log (N_{\rm X}/N_{\rm H}) - \log (N_{\rm X}/N_{\rm H})_\odot \nonumber \\
            &= [\log (N_{\rm X_i}/N_{\rm HI}) - \log (N_{\rm X}/N_{\rm H})_\odot] + (\log f_{\rm HI} - \log f_{\rm X_i}) \nonumber \\
            &\equiv {\rm [X/H]}^\prime + {\rm IC} \label{eq:XH}
\end{align}
where X$_i$ denotes the ionic state $i$ of element X, $f_{\rm X_i} \equiv N_{\rm X_i}/N_{\rm X}$ is the fraction of the element in the ionic state $i$, $f_{\rm HI} \equiv N_{\rm HI}/N_{\rm H}$ is the neutral fraction of Hydrogen, ${\rm [X/H]}^\prime \equiv \log (N_{\rm X_i}/N_{\rm HI}) - \log (N_{\rm X}/N_{\rm H})_\odot $ is the {\it raw} metallicity, and ${\rm IC} \equiv \log f_{\rm HI} - \log f_{\rm X_i}$ is the ionization correction. 

We have obtained the first two items ($\logNHI$ and $\log N_{\rm X_i}$) in \S~\ref{sec:voigt} and \S~\ref{sec:aodm} and the results are listed in Tables\,\ref{tab:voigt} and \ref{tab:logN}. Combined with the elemental abundances of the present-day solar photosphere from \citealt{Asplund09}, we are ready to calculate the {\it raw} metallicity ${\rm [X/H]}^\prime$. Next, we calculate the ionization correction (IC) using {\sc cloudy} photoionization models \citep{Ferland17}.

The IC factors are sensitive to both the \HI\ column density ($\logNHI$) and the ionization parameter $\log U = \log \Phi_{\rm H}/(n_{\rm H} c)$, where $\Phi_{\rm H}$ is the surface flux of ionizing photons with $h\nu > 1$\,Ryd at the illuminated face. The former has been measured, but the latter needs to be constrained by comparing the column density ratios of different ionic states of the same elements and predictions from photoionization models. Therefore, for each cloud listed in Table\,\ref{tab:aodm}, we calculate a set of photoionization models, with a termination condition set to meet the observed \HI\ column density. For the ionizing source, we use the \citet{Haardt12} radiation background interpolated to $z = 2.67$ with contributions from both galaxies and quasars. For the cloud, we assume plane-parallel geometry, the solar relative abundance pattern, a metallicity of ${\rm [M/H]} = -1.5$\footnote{The derived ICs are insensitive to the assumed metallicity.}, and a range of hydrogen volume densities ($1.09 \geq \log n_{\rm H}/{\rm cm}^{-3} \geq -4.91$) to cover ionization parameters between $-6 \leq \log U \leq 0$. 
For each cloud, we compare the observed ionic column ratios (C\,{\sc iv}/C\,{\sc ii} and Si\,{\sc iv}/Si\,{\sc ii}) and the model-predicted ratios to constrain the ionization parameter. We list the constraints on $\log U$ in Table\,\ref{tab:aodm}. It is rare to have detections of both high ions and low ions in the same cloud, leading to many upper or lower limits of $\log U$. But we found that the most plausible ionization parameters lie between $-4 \lesssim \log U \lesssim -2$, comparable to other published LLSs \citep[e.g.,][]{Prochaska99,Lehner16}. Depending on the data constraint, we adopt $\log U$ values of $-3.5, -3.0,$ or $-2.0$ for each cloud. Finally, once both $\logNHI$ and $\log U$ are fixed, we use the {\sc cloudy} model of the same parameters to calculate the ICs for all of the ions (Table\,\ref{tab:IC}), which are then used to obtain the ionization-corrected metallicity measurements for all of the transitions (Table\,\ref{tab:XH}). 
Notice that the ionization correction only becomes important for (1) low ions in clouds with $\logNHI \lesssim 19$ and (2) intermediate or high ions (e.g., C\,{\sc iv} and Si\,{\sc iv}) at all column densities. For low ions in sub-DLAs, the ICs are fairly small ($\pm$0.15 dex).
We also use the model to infer the total H column density ($\log N_{\rm H}$), the H neutral fraction ($\log f_{\rm HI}$), the H volume density ($\log n_{\rm H}$), and the characteristic line-of-sight depth of the cloud ($\log l = \log N_{\rm H} - \log n_{\rm H}$). The results are listed in Table\,\ref{tab:aodm}.

The best metallicity measurement is provided by O\,{\sc i}, because O\,{\sc i} has the smallest ionization correction factors due to its charge-exchange reactions with hydrogen \citep{Field71} and its hydrogen-like ionization potential. Unfortunately, O\,{\sc i}\,$\lambda1302$ is saturated in C1 toward \bgQSO\ (a common issue of the transition for DLAs) and is undetected in cloud B toward \bgQSO\ and the clouds toward \fgQSO. As a result, transitions from other ions need to be used. The bottom section of Table\,\ref{tab:aodm} lists the final adopted metallicities from our preferred $\alpha$-element transitions and Fe\,{\sc ii}. Note that the ICs for \fgQSO-B1 and \fgQSO-C1 are large because only the C\,{\sc iv} lines are detected in these clouds. We fail to obtain a reliable metallicity measurement for \bgQSO-B1 because of the absence of metal absorption. For the four clouds where we have both [$\alpha$/H] and [Fe/H], we found a moderate level of $\alpha$-enhancement ([$\alpha$/Fe]) between 0.15 and 0.61 (with an inverse-variance-weighted mean of 0.2), comparable to those previously measured in $z > 2$ DLAs ([$\alpha$/Fe] = $0.30\pm0.16$; \citealt{Rafelski12}).

We opt not to correct the gas-phase metallicity for dust depletion, because 
(1) little depletion is expected from volatile elements such as O and C, 
(2) the SDSS spectra and photometry of the QSOs show no evidence of significant dust reddening, and
(3) the depletion factors are largely uncertain in external galaxies. 
As a reference, in the Milky Way's ISM, volatile elements (e.g., C, N, O, S, Zn) show depletions of ${\rm {\rm (X/H)}_{\rm ISM} - (X/H)}_{\rm gas} \lesssim 0.3$, while refractory elements (e.g., Mg, Al, Si, Fe, Ni) show $0.7 \lesssim {\rm (X/H)}_{\rm ISM} - {\rm (X/H)}_{\rm gas} \lesssim 2.0$ \citep{Savage96,Groves04a}. These local measurements from a metal-rich ISM provide strict upper limits on the level of depletion expected in the CGM of the SMG.

\section{Emission-Absorption Connection} \label{sec:connection}

\begin{figure}[!tb]
\epsscale{1.15}
\plotone{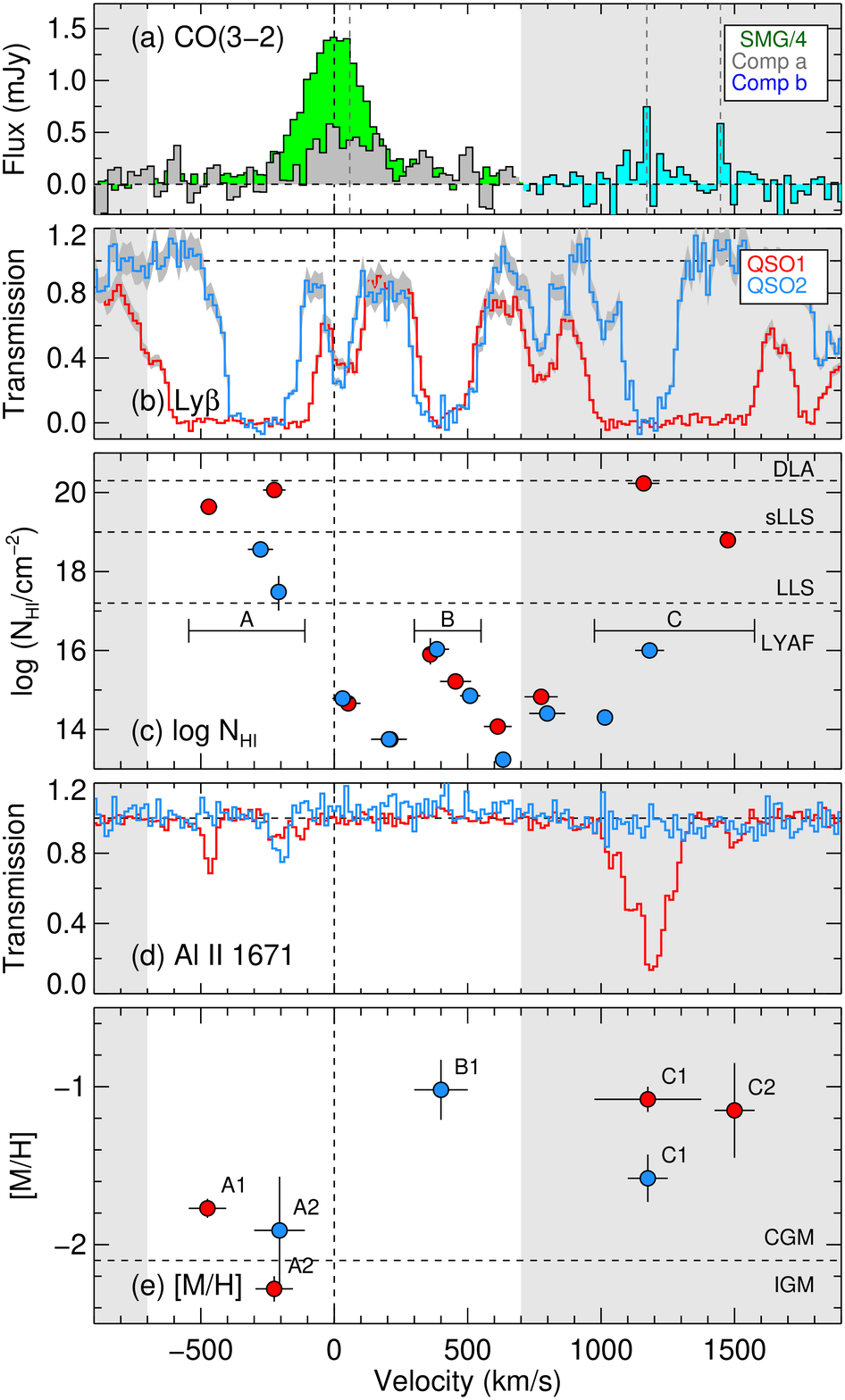}
\caption{Emission-absorption comparison. ({\it a}) \cothree\ spectra of the SMG and its companions, ({\it b}) \HI\ \Lyb\ absorption, ({\it c}) \HI\ column densities of Voigt components (Table\,\ref{tab:voigt}), ({\it d}) Al\,{\sc ii}$\lambda$1670.8 absorption, and ({\it e}) metallicities of metal-line-defined clouds (Table\,\ref{tab:aodm}). The absorbers toward \bgQSO\ and \fgQSO\ are color-coded in red and blue, respectively. In ({\it a}), the SMG spectrum has been divided by 4$\times$ to show it together with the spectra of its companions, and the vertical dashed lines indicate the centroid velocities of the CO emission lines at 0, 58, 1171, and 1447\,\kms.
All velocities are relative to the redshift of the SMG ($\zsmg = 2.674$) and the gray shaded regions indicate velocities beyond the escape velocity of a $10^{13}$\,\msun\ halo ($v_{\rm esc} \simeq 700$\,\kms). 
\label{fig:emabs}}
\epsscale{1.0}
\end{figure}

Identifying the emission counterparts of the intervening gas helps us compare the properties of the galaxies to those of their CGM. Having analyzed the ALMA CO emitters in \S~\ref{sec:smg} and the QSO absorption spectra in \S~\ref{sec:qsos}, we are now ready to draw connections between the emission and the absorption based on proximity in both spatial and redshift dimensions.

\begin{table}[!tb]
\begin{center}
\caption{Properties of the CGM around the SMG and \Chost}
\label{tab:subsys}
\begin{tabular}{ccccccc}
\hline
\hline
Name & $\delta v$ & Galaxy  & $R_\bot$ & $\logNHI$ & [M/H]  \\
          & (\kms)     &       & (kpc) &  ($\log {\rm cm}^{-2}$) &       \\
\hline
\bgQSO-A & $[-545,-110]$ & SMG    & 93.1 & $20.20_{-0.05}^{+0.07}$ & $-2.09\pm0.07$ \\
\fgQSO-A & ---           & ---    & 175.5& $18.59_{-0.43}^{+0.23}$ & $-1.91\pm0.34$ \\
\bgQSO-B & [300,550]     & SMG    & 93.1 & $15.98_{-0.23}^{+0.39}$ & \nodata \\
\fgQSO-B & ---           & ---    & 175.5& $16.06_{-0.15}^{+0.20}$ & $-1.02\pm0.19$ \\
\bgQSO-C & [975,1575]    & \Chost & 58.9 & $20.25_{-0.07}^{+0.06}$ & $-1.09\pm0.11$ \\ 
\fgQSO-C & ---           & ---    & 32.2 & $16.01_{-0.09}^{+0.10}$ & $-1.58\pm0.15$ \\
\hline
\end{tabular} 
\end{center}
\end{table}

Fig.~\ref{fig:emabs} directly compares the absorption profiles from the QSOs to the CO emission profiles from the SMG and its companions. The \HI\ \Lyb\ and Al\,{\sc ii}$\lambda$1670.8 profiles from the two QSOs are plotted together to illustrate the striking kinematic coherence. In addition, the figure illustrates $\logNHI$ from the Voigt profile solution in Table~\ref{tab:voigt} and the ionization-corrected metallicities of the metal-line-defined clouds in Table\,\ref{tab:aodm}. 

In \S~\ref{sec:qsos}, we have found that the $\zabs \approx 2.68$ \HI\ absorbers have total \HI\ column densities of $\logNHI = 20.53_{-0.06}^{+0.06}$ and $18.60_{-0.43}^{+0.23}$ toward \bgQSO\ and \fgQSO, respectively. Each absorber is resolved into three main subsystems (A, B, and C) with velocity spans of $\sim$1500-2000\,\kms. Although their \HI\ column densities vary significantly between the two QSO sightlines, their radial velocities show remarkable consistency, indicating that the QSOs are intercepting three expansive sheets/filaments of gas. At the same time, the extreme velocity widths of the absorption-line systems suggest that they probe merging systems \citep{Prochaska19}.

Results in Fig.~\ref{fig:emabs} show that subsystem C is unlikely to be in the same halo as subsystems A, for several reasons. First, subsystem C is 10$\times$ more metal-enriched than subsystem A (${\rm [M/H]} \simeq -1.1$ vs. $-2.1$). Secondly, the velocity spans of $\sim$1950\,\kms\ (\bgQSO) and $\sim$1460\,\kms\ (\fgQSO; Table\,\ref{tab:voigt}) and their asymmetric distributions around $\zsmg$ (absorption is centered at $\zabs = 2.68$) are inconsistent with gravitational motions inside even a $10^{13}$\,\msun\ halo centered on the SMG, because the escape velocity of such a halo following the NFW profile is flat at $\sim$700\,\kms\ between 60\,kpc and the virial radius of 186\,kpc. Lastly, we have detected CO emission at almost exactly the redshifts of the absorbing clouds in subsystem C in \Chost, which lies much closer to the QSOs than the SMG. Therefore, we consider subsystem A part of the CGM of the SMG at $z = 2.674$, and subsystem C part of the CGM of \Chost\ at $z = 2.6884$ and 2.6917.

As for subsystem B ($\logNHI \simeq 16$), although its velocity allows an association with the SMG, it is unimportant because its contribution to the CGM is negligible compared to subsystem A.

Once the emission counterparts are determined, the absorption-line measurements from the two QSO sightlines can be plotted against the impact parameter to show crude radial profiles of the CGM around the SMG and \Chost. We consolidate the results in Table\,\ref{tab:subsys}, where we have assigned a velocity window for each subsystem that captures its major clouds. We calculate the total \HI\ column densities from the \HI\ Voigt components within these velocity windows. When there are multiple metal-line clouds in a subsystem, the metallicity is calculated as the $N_{\rm H}$-weighted mean [$\alpha$/H], where $N_{\rm H}$ is the \HI\ + \HII\ column density based on the adopted {\sc cloudy} model. 

\begin{figure}[!tb]
\epsscale{1.2}
\plotone{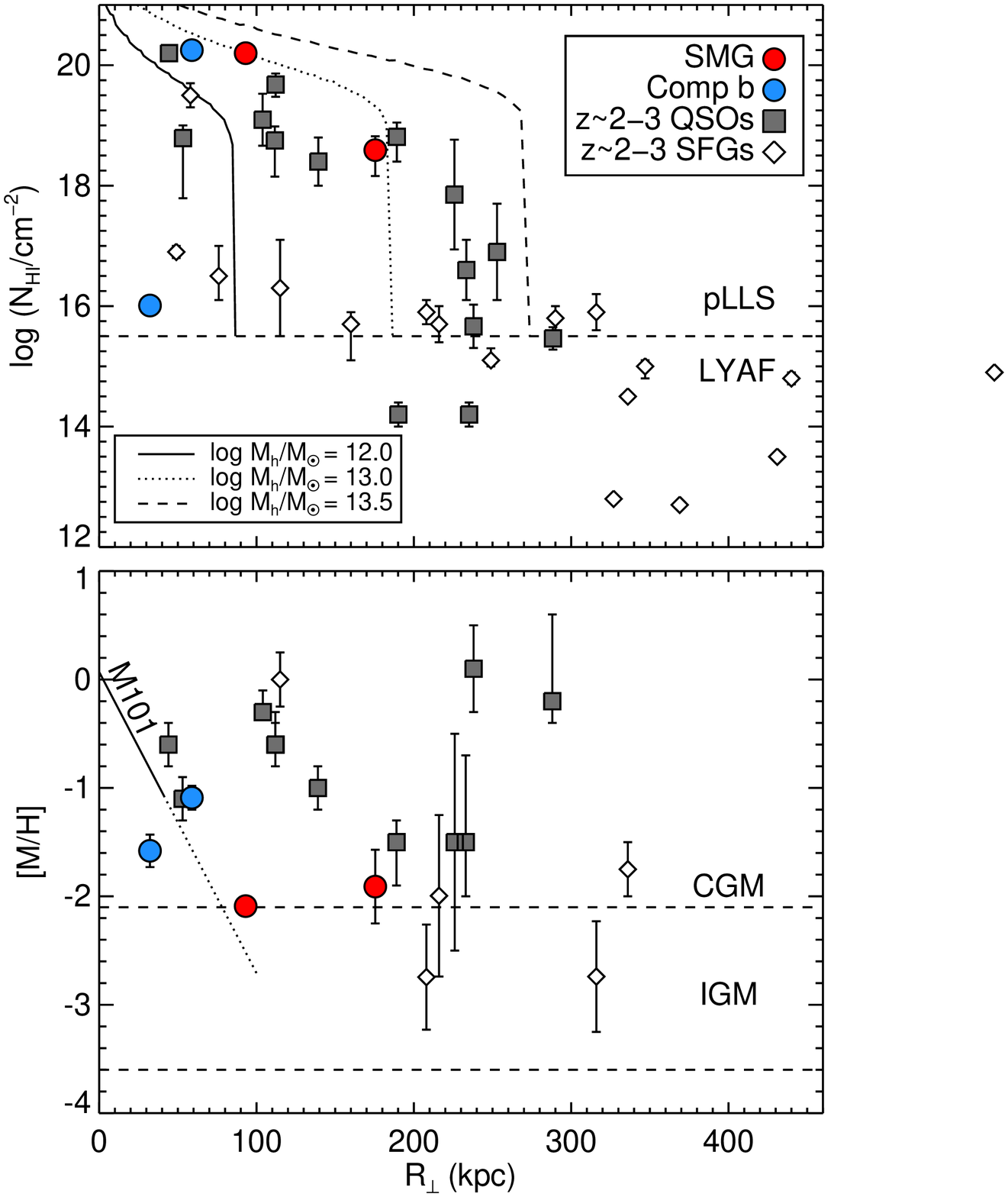}
\caption{Profiles of the CGM around the SMG ({\it red}) and \Chost\ ({\it blue}). {\it Top:} \HI\ column density vs. projected separation for the absorption-line clouds in \system\ and the literature. The curves show the projected \HI\ column densities of NFW halos, where the precipitous decline marks the virial radii. {\it Bottom:} Metallicity vs. projected separation for the absorption-line clouds in \system\ and the literature. Literature QSO CGM data are from \citet{Lau16}, SFG $\logNHI$ data are from \citet{Simcoe06}, \citet{Rudie12}, and \citet{Crighton13,Crighton15}, and SFG [M/H] data are from \citet{Simcoe06}. The horizontal dashed lines in the bottom panel indicate the range of IGM metallicities measured from the LYAF at $\zabs \sim 2.5$ \citep[${\rm [M/H]} = -2.85\pm0.75$;][]{Simcoe04}. The sloped line shows the oxygen abundance profile of the giant spiral galaxy M101 measured with an electron-temperature-based method \citep[Eq.~5 of][]{Kennicutt03}. The solid portion is covered by H\,{\sc ii} regions, while the dotted portion is an extrapolation.
\label{fig:CGMprof}}
\end{figure}

The \system\ system gives us the first glimpse into the CGM around an SMG. With high \HI\ column density and low metallicity at large impact parameters, the CGM of the SMG is distinct from the CGM of QSOs and normal star-forming galaxies at $z \sim 2-3$. 

The profile of \HI\ column density is shown in Fig.~\ref{fig:CGMprof}$a$. We compare our measurements with literature QSO absorption-line measurements in the surroundings of $z \sim 2-3$ QSOs \citep{Lau16} and Lyman Break Galaxies \citep[LBGs;][]{Simcoe06,Rudie12,Crighton13,Crighton15}. The \HI\ column densities of the SMG's CGM, similar to coeval QSOs, are significantly greater than those of star-forming galaxies at $R_\bot \gtrsim 70$\,kpc. The \HI\ column density declines as we move away from the SMG, with a gradient of $-2.0\pm0.4$\,dex per 100\,kpc. 
How do the observed column densities compare with the projected surface mass density $\Sigma_M(R)$ of NFW halos? To make this comparison, we first calculate $\Sigma_M(R)$ by integrating the NFW density profile $\rho(r)$ up to the virial radius:
\begin{equation}
\Sigma_M(R) = 2 \int_{R}^{R_{\rm vir}} \frac{r \rho(r)}{\sqrt{r^2-R^2}} dr
\end{equation}
We then convert it to \HI\ column densities by assuming a baryon$-$dark-matter density ratio of $f_b \equiv \Omega_b/\Omega_c = 0.187$ \citep{Planck-Collaboration20}, a Helium correction of $f_{\rm He} = M_{\rm H+He}/M_{\rm H} = 1.36$, and an arbitrary neutral fraction of 10\%\footnote{Comparable to the neutral fractions estimated by our photoionization models in Table\,\ref{tab:aodm}.}.
Fig.\,\ref{fig:CGMprof}$a$ shows that the \HI\ column density profiles of the SMG and the QSOs are consistent with the expectation from a $\sim$$10^{13}$\,\msun\ halo. This is in agreement with the detection of a high column of \HI\ ($\logNHI = 18.6$) at an impact parameter as far as 176\,kpc. The agreement also shows that the neutral gas in the absorption-line systems can account for $\sim$10\% of the total baryonic mass in the halo if they have a filling factor close to unity. On the other hand, the $\logNHI$ profile of LBGs is more consistent with a $\sim$$10^{12}$\,\msun\ halo.

The metallicity profile is shown in Fig.~\ref{fig:CGMprof}$b$. The literature data points show that,  metal-enriched gas with $-1 \lesssim {\rm [M/H]} \lesssim 0$ dominates the inner part ($R_\bot \lesssim 150$\,kpc) of the CGM around both QSOs and LBGs. By contrast, the CGM of the SMG is poor in metal, with almost a constant metallicity of [M/H]$ \approx -2$ across two sightlines separated by 86\,kpc (10.8\arcsec). Its metallicity is near the 1$\sigma$ upper bound of the metallicities measured in the LYAF \citep[i.e., the IGM;][]{Simcoe04}, and it is lower by $\sim$1.5 dex at $R_\bot = 93.1$ and by $\sim$0.5 dex at $R_\bot = 175.5$\,kpc than that of the CGM of QSOs. 

On the other hand, the CGM of \Chost\ show properties similar to that of normal star-forming galaxies at $z \sim 2-3$. Both line emitters in \Chost\ have orders of magnitude lower molecular gas mass than the SMG (see Table\,\ref{tab:coords}). Assuming a CO-to-molecular-mass correction factor of $\alpha_{\rm CO} = 4.3$ and a CO excitation correction of $r_{31} = 0.52$, the emitters at $z = 2.6884$ and 2.6917 have molecular gas masses of $M_{\rm mol} = 5.6\times10^9$ and $4.4\times10^9$\,\msun, respectively. The gas masses are comparable to those measured in the lensed normal star-forming galaxies that have stellar masses of $\sim10^{10}$\,\msun\ \citep{Saintonge13}. One would thus expect a CGM similar to that of LBGs. The corresponding subsystem C provides absorption-line measurements at $R_\bot = 32.3$ and 58.9\,kpc. It shows significantly metal-enriched gas (compared to the IGM level) with a large variation in \HI\ column density over just a difference of 27\,kpc in impact parameter. \Chost\ thus is surrounded by a metal-enriched clumpy medium extended to at least $\sim$60\,kpc.  

In the above, we have shown that the CGM of \SMG\ is distinctly different from the CGM of QSOs and normal star-forming galaxies. But how do our absorbers compare with other DLAs in terms of absorption-line properties only? The \HI\ column densities of subsystems A and C toward \bgQSO\ miss the DLA threshold of $\logNHI =  20.3$ by merely $\sim$0.1\,dex. Because such small differences are comparable to the measurement uncertainty, more liberal thresholds have been used to select DLAs \citep[e.g., $\logNHI > 20.1$ in][]{Rubin15}. Combined with the result that the two sub-DLAs have neutral fraction of $\sim$50\% from {\sc cloudy} models, we believe that it is appropriate to compare their properties with those of the general DLA population.

With high-resolution spectra of 100 DLAs at $\zabs \sim 2-4$, \citet{Rafelski12} found that their metallicity distribution is well fit by a Gaussian with a mean at ${\rm [M/H]} = -1.57$ and a dispersion of 0.57\,dex. The DLA metallicity only mildly decreases with redshift, but shows a strong correlation with the width of low-ion metal lines (e.g., $\Delta V_{90}$ or $W_{1526}$) \citep{Neeleman13}. This correlation between kinematics and metallicity is generally interpreted as a manifestation of the mass-metallicity relation \citep[e.g.,][]{Tremonti04,Erb06}, because the line width may reflect the halo mass (like in the Tully-Fisher relation), which in turn is proportional to the stellar mass \citep{Moller13}. 

Previous works have used higher-resolution spectra ($R \gtrsim 40000$) to measure $\Delta V_{90}$, the velocity interval including 90\% of the optical depth of an {\it unsaturated} line. And the alternative kinematic parameter $W_{1526}$, the rest-frame equivalent width of Si\,{\sc ii}$\lambda$1526.7, is unsuitable for our sub-DLAs because the line is {\it unsaturated} (see Fig.~\ref{fig:aodm}). The equivalent width only becomes a good kinematics indicator when the line is {\it saturated}, weakening its dependency on the Si$^+$ column density (and consequently, on metallicity). 
So to place our sub-DLAs on the relation, we adopt the velocity separation between kinematically resolved substructures as a surrogate of $\Delta V_{90}$. \bgQSO-A shows two velocity components of similar strength separated by $\sim$250\,\kms, and \bgQSO-C is dominated by the stronger C1 cloud, which appears to be a blend of four components with a velocity span of $\sim$200\,\kms. These estimates of the velocity width should be considered as lower limits on $\Delta V_{90}$, because they are the separations between peak optical depths. 

\begin{figure}[!tb]
\epsscale{1.2}
\plotone{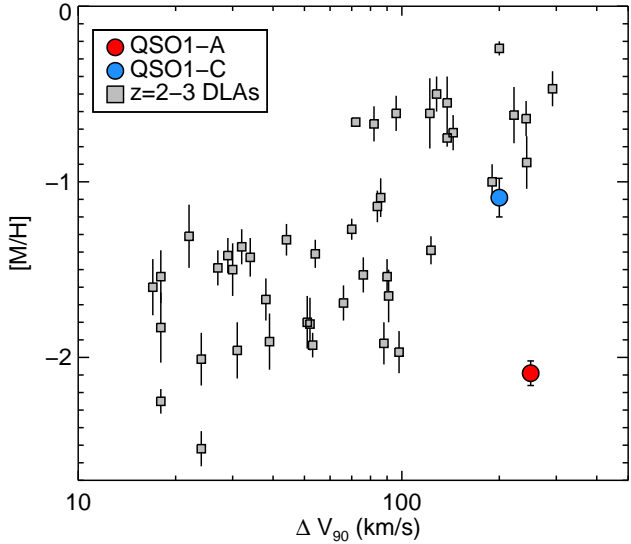}
\caption{Metallicity$-$line-width relation of (sub-)DLAs at $z \sim 2-3$. Gray squares are 44 DLAs between $2 < \zabs < 3$ compiled from the literature \citep{Neeleman13}. The red and blue circles show respectively subsystems A and C toward \bgQSO. For both sub-DLAs, we have estimated the velocity widths using the separation of resolved components in the $R \sim 11000$ X-shooter/VIS spectrum.
\label{fig:mh_v90}}
\end{figure}

Fig.\,\ref{fig:mh_v90} compares the two sub-DLAs against literature DLAs. To control the redshift evolution of the mass-metallicity relation, only the DLAs between $2 < \zabs < 3$ are plotted. While \bgQSO-C blends in the general trend established by DLAs, \bgQSO-A is a clear outlier. Firstly, very few DLAs have metallicities as low as \bgQSO-A: only 2 out of the 44 DLAs ($4.5_{-1.5}^{+5.5}$\%) have ${\rm [M/H]} \leq -2.1$. Secondly, its high velocity width places it significantly below the metallicity-line-width relation of DLAs, which would have predicted a velocity width of only $\sim$20\,\kms\ based on its metallicity at ${\rm [M/H]} = -2.1$. 

This finding suggests that most of the DLAs are likely closer to their hosts than our sub-DLA \bgQSO-A ($R_\bot = 93$\,kpc), consistent with previous observations of DLA host galaxies (see \S~\ref{sec:intro-connection}).   
More importantly, the unusual combination of high velocity-width and low metallicity provides a method -- based on absorption-line properties alone -- to potentially separate (sub-)DLAs associated with normal star-forming galaxies at low impact parameters with those associated with more massive galaxies like SMGs at large impact parameters.

\section{Summary \& Discussion} \label{sec:summary}

We have carried out a comprehensive study of the emission-absorption system \system\ at $z \sim 2.67$ with a multi-wavelength data set obtained primarily from \Herschel, ALMA, and VLT/X-shooter. The system consists of a bright SMG, its CO companion galaxies, and a number of optically thick \HI\ absorbers toward two background QSOs within 22\arcsec\ of the SMG. Our main results are: 

\begin{enumerate}

\item The \Herschel-selected SMG at $z = 2.674$, with an 870\,\um\ flux of 7.4\,mJy and an IR luminosity of $\sim10^{13}$\,\lsun, is one of the most luminous dusty star-forming galaxies. Its properties are similar to the general SMG population at $z \sim 2$, featuring a short gas depletion timescale of $\sim$0.1\,Gyr and compact (sub-arcsec) sizes in both dust emission and \cothree\ emission. The high S/N spectrum reveals two \cothree\ components at almost the same redshift: $\sim$1/4 of the line flux is in a broad component with a ${\rm FWHM} \simeq 900$\,\kms, while $\sim$3/4 of the flux in a narrow component with ${\rm FWHM} \simeq 250$\,\kms.     

\item Three companion CO emitters are identified within 30\arcsec\ and 1500\,\kms\ of the SMG. A comparison with the source counts from the ASPECS field survey indicates that the SMG lives in an over-dense environment. 

\item Two nearby QSOs provide background beacons to probe the CGM of the SMG. A DLA with a total \HI\ column density of $\logNHI = 20.5$ is identified at $\zabs \sim 2.68$ in the closer QSO sightline at $\theta = 11.7\arcsec$. The DLA is quite unusual, in terms of both the large impact parameter ($R_\bot = 93.1$\,kpc to the SMG) and the enormous velocity span ($\sim$2000\,\kms). X-shooter resolved the DLA into three major subsystems, including two sub-DLAs with distinctly different metallicities separated by $\sim$1600\kms. Remarkably, the same subsystems are also found in the farther QSO sightline at $\theta = 22.1\arcsec$: they have nearly the same velocities and metallicities as their counterparts at $\theta = 11.7\arcsec$, despite lower \HI\ column densities (total $\logNHI = 18.6$). 

\item We use the absorption-line systems to characterize the CGM of the SMG and its companion \Chost, and we compare their properties with the CGM of QSOs and normal star-forming galaxies. The CGM of the SMG forms its own category: while its high column densities at large impact parameters are similar to the massive halos inhabited by $z = 2-3$ QSOs, its metal content ($\sim$1\% solar) is an order of magnitude lower than the circum-QSO medium. On the other hand, the CGM of the much less luminous \Chost\ is more consistent with that of normal star-forming galaxies at $z = 2-3$: showing significant metal enrichment ($\sim$10\% solar) within $R_\bot \lesssim 60$\,kpc.

\end{enumerate}

The detection of high-column density, mostly neutral, metal-poor gas in the CGM of a massive dusty starburst galaxy at $z = 2.674$ has powerful implications to theories of galaxy formation and evolution. The remarkable consistency of the \HI\ absorbers in both kinematics and metallicity across two sightlines separated by 86\,kpc is at odds with CGM models that assume randomly floating \HI\ clouds in pressure equilibrium with hot X-ray gas. Instead, it is logical to assume that the background QSOs have intercepted a single filament of cool gas permeating in the halo. 

Narrow filaments of cool gas and satellite galaxies can penetrate the hot CGM of massive halos without ever being shock-heated to the virial temperature, because (1) massive halos are rare and tend to form at the intersections of the cosmic web and (2) the cooling time is shorter in the filaments than in the halo \citep{Dekel06,Dekel09,Keres09}. At $z \gtrsim 2.5$, such cold streams can survive even in halos more massive than $\sim$$10^{13}$\,\msun\ (although note that this mass limit is highly uncertain). In particular, stream-feeding is likely important in the bright SMGs with $S_{850} > 5$\,mJy because their comoving space density exceeds the expectation from minor and major mergers \citep{Dekel09}. 

The observed properties of the absorption-line systems match the simulation-predicted properties of cold streams. First, the simulations of \citet{Dekel09} show that the inflow velocity is comparable to the virial velocity and is roughly constant along the filament. This is consistent with the velocity shift ($\delta v = -300$\,\kms) and the kinematic coherence we saw between the clouds in both QSO sightlines. Secondly, by post-processing gas in seven simulated halos with $\mh = 10^{10}-10^{12}$\,\msun\ between $1.5 < z < 4.0$, \citet{Fumagalli11a} found that the absorption-line systems associated with the smooth stream component have systematically lower metallicity ($\sim$1\% solar). This is exactly the level of the gas metallicity we measured in the absorbers associated with the SMG. 

Radially inflowing on nearly a free-fall timescale, the cold streams may account for the bulk of the baryonic accretion rate and become the dominant mechanism to feed the growth of galaxies \citep{Keres09}. We can crudely estimate the gas accretion rate from the filament that the QSOs intercepted. The filament has a length $>$176\,kpc and a depth on the order of 10\,kpc at $R_\bot = 93$\,kpc. The former is estimated from the impact parameter of \fgQSO, and the latter is estimated from the ratio between the column density and the volume density of Hydrogen, $l = N_{\rm H}/n_{\rm H}$, inferred from the photoionization model of the sub-DLA \bgQSO-A2. The depth-to-distance ratio is 0.11\,radian or $6^\circ$, comparable to the opening angles of $20-30^\circ$ seen in simulations \citep{Dekel09}. Our photoionization models also indicate similar \HI+\HII\ column densities for \bgQSO-A2 ($\log N_{\rm H} = 20.4$) and \fgQSO-A2 ($\log N_{\rm H} = 20.1$), the two main clouds associated with the SMG at $R_\bot = 93, 176$\,kpc. By assuming an opening angle of $\beta = 25^\circ$, an average hydrogen column density of $\log N_{\rm H} = 20.2$, and a $10^{13}$\,\msun\ NFW halo at $z = 2.674$, we estimate that the mass of the filament is $M_{\rm fil} = f_{\rm He} m_p N_{\rm H} (\beta~R_{\rm vir}^2/2) \sim 1.3\times10^{10}$\,\msun. Given an accretion timescale of $\tau_{\rm acc} = R_{\rm vir}/V_{\rm vir} = 4\times10^8$\,yr, the gas accretion rate from this single filament is $\sim$33\,\msunyr. Typically three main filaments are seen in the simulations, so our estimate shows that cold-mode accretion can supply gas at a rate of $\sim$100\,\msunyr. Although accounting for only $\sim$10\% of the current SFR, our estimated cool gas accretion rate is in fact comparable to the rate of total gas supply to the central galaxies in $10^{13}$\,\msun\ halos at $z = 2$ from cosmological simulations \citep[see Fig.~9 of][]{Keres09}, and at this rate the molecular gas reservoir of $10^{11}$\,\msun\ can be acquired in just $\sim$1\,Gyr. On the other hand, star formation at the current intensity seems unsustainable despite the efficient gas supply from cold streams. 

In this work, we have presented the first observational evidence that supports the existence of cold streams in the CGM of a massive starburst galaxy. The \system\ system has an excellent data set and the results are highly informative, but larger samples are clearly desired to draw conclusions on the general properties of the CGM. We hope that this will serve as a springboard for upcoming statistical studies of the CGM in similar galaxies. As an attempt to inform these future studies, it is worth discussing the major technical challenges we had faced in this program:
\begin{enumerate} 

\item The large beam of \Herschel\ (17.8\arcsec\ at 250\,\um) makes it inefficient to identify SMG-QSO pairs with small angular separations ($\theta \lesssim 10\arcsec$). As a result, \bgQSO\ in \system\ is the only one that probes below 100\,kpc; and despite intercepting a sub-DLA it has yet to expose the chemically enriched area of the CGM of the SMG.  

\item High S/N spectra with moderately high spectral resolution are needed to unambiguously detect optically thick absorbers with $17.2 < \logNHI \lesssim 19$. For example, with the $R \sim 2000$ SDSS spectrum of \fgQSO, we couldn't have identified the LLS associated with the SMG (i.e., \fgQSO-A2 with $\logNHI = 18.6$). But the LLS is unambiguously detected in the X-shooter spectrum because of its resolved \HI\ Lyman profiles and the clear detection of low-ion metal lines. Similarly low column densities are expected in most of our sample, because all of the other spectroscopically confirmed SMG-QSO pairs we have so far have impact parameters between $100 < R_\bot < 300$\,kpc \citep[][and unpublished data]{Fu16} and none of them show obvious (sub-)DLA features (which may be expected given the large impact parameters).

\item It has been difficult to obtain spectroscopic redshifts of the \Herschel\ sources because (1) they require sub-arcsec positions from interferometers to place spectroscopic slits, (2) the near-IR spectral range suffers from heavy telluric absorption and OH emission, and (3) SMGs tend to be weak in rest-frame optical lines. The latter two points are the main reasons why our redshift success rate is only $\sim$60\%. 

\end{enumerate}

Possible solutions to these issues may be already on the horizon.
To address the first challenge, we need to design an efficient interferometer imaging survey, because a sample of \Herschel\ sources with less than 10\arcsec\ apparent offsets from optical QSOs is likely overwhelmed by contaminating sources. Two third of the \Herschel-SDSS-selected pairs with apparent separations between 5\arcsec\ and 10\arcsec\ turned out to be IR-luminous QSOs instead of projected SMG-QSO pairs. A good strategy is to observe multiple sources located within a $\sim$10$^\circ$ diameter circle in a single ALMA scheduling block (SB). In \citet{Fu17}, we managed to observe $\sim$10 targets in a single $\sim$50\,min SB, achieving an on-source integration time of 200\,s per source and an rms of 0.12\,mJy~beam$^{-1}$. Do we have enough such pairs to populate a 50-min SB? The surface density of \Herschel-QSO pairs with offsets less than 10\arcsec\ is 0.16\,deg$^2$ (26 pairs in the 161.6 deg$^2$ H-ATLAS GAMA fields), which gives $\sim$13 targets for a $\sim$10$^\circ$ diameter circle.

To address the second challenge, we need QSO spectra with quality similar to the X-shooter spectra presented here to better sample the spatial profile of the CGM. The QSOs in our sample are selected to have $g < 22$, the majority of which requires $\sim$1.5 hours' exposure time with an Echellette spectrograph on a 10-m-class telescope to reach a sufficient S/N at $R \sim 8000$ \citep[e.g., see the Keck/ESI survey of DLAs by][]{Rafelski12}.

To address the last challenge, we need a more efficient method to obtain SMG redshifts. One potential approach is to exploit the frequency scan mode offered by modern millimeter interferometers. This method has the additional advantage of bypassing the interferometer imaging step because the primary beam is usually larger than the \Herschel\ positional uncertainty and the line detection also provides positional information. For instance, with NOEMA scans of the 2\,mm and 3\,mm bands (only two spectral configurations per band), \citet{Neri20} obtained 12 secure redshifts for 13 sources. The average telescope time spent is $\sim$105\,min per source, including $\sim$40\,min overhead. Although the targets in this Pilot study have 500\,\um\ flux densities greater than 80~mJy (many are strongly lensed), this technique could be applied to fainter sources (like ours with $S_{500} > 20$\,mJy) as the instrument sensitivity and overheads continue to improve. 
 
\acknowledgments
We thank D.~Kere\v{s} and J.~Hennawi for discussions. 
This work is partially supported by the National Science Foundation (NSF) grant AST-1614326.
D.~N. acknowledges support from NSF AST-1909153.
The National Radio Astronomy Observatory is a facility of the NSF operated under cooperative agreement by Associated Universities, Inc.
This paper makes use of the following ALMA data: ADS/JAO.ALMA\#2015.1.00131.S, ADS/JAO. ALMA\#2018.1.00548.S. ALMA is a partnership of ESO (representing its member states), NSF (USA) and NINS (Japan), together with NRC (Canada), MOST and ASIAA (Taiwan), and KASI (Republic of Korea), in cooperation with the Republic of Chile. The Joint ALMA Observatory is operated by ESO, AUI/NRAO and NAOJ.

{\it Facilities}: Herschel, ALMA, Gemini/GNIRS, VLT/X-shooter, Keck/LRIS, Sloan, KiDS, VISTA


\newpage
\appendix

\section{Lensing of the SMG} \label{sec:fakecp}

In the KiDS $gri$ pseudo-color image in Fig.\,\ref{fig:fchart}, there is an extended optical source just 0.8\arcsec\ to the NW of the ALMA source. This offset cannot be an astrometry error, because the ALMA \cothree\ emission of \fgQSO\ agrees with its KiDS optical position within 0.1\arcsec. The optical source is in the KiDS-VISTA 9-band photometric catalog \citep[$ugriZYJHK_s$;][]{Kuijken19} with a designation of KiDSDR4\,J091339.496$-$010656.17 and a photometric redshift of $z_p = 0.07_{-0.04}^{+0.05}$. We obtained an optical spectrum with the Low Resolution Imaging Spectrometer \citep[LRIS;][]{Oke95} on the Keck I telescope on 2017 Mar 23. Strong emission lines, such as [O\,{\sc ii}]$\lambda$3728, [O\,{\sc iii}]$\lambda\lambda$4960,5008, were detected at high significance in the 20\,min exposure, placing its redshift at 0.055. 

The source is detected in all of the nine photometric bands included in the KiDS$+$VISTA photometry catalog, with $r = 21.58\pm0.02$ and $K_s = 20.84\pm0.18$. Our best-fit stellar population synthesis model of the photometry reveals a stellar mass of $\sim3\times10^7$\,\msun\ and an SFR of $\sim0.006$\,\msunyr. We have used the \citet{Bruzual03} models assuming exponentially declining star-forming histories and the \citet{Chabrier03} initial mass function. 

Could the SMG be gravitationally magnified by this foreground dwarf galaxy? Using the halo-mass$-$stellar-mass relation from abundance matching \citep[Fig.\,6 of][]{Bullock17}, we estimate a halo mass of $\sim3\times10^{10}$\,\msun. The halo mass corresponds to a line-of-sight velocity dispersion ($\sigma$) of just $\sim$29\,\kms, assuming a singular isothermal sphere (SIS; $\rho(r) = \sigma^2/2\pi G r^2$) and the fitting function of the halo overdensity $\Delta_c(z)$ from \citet{Bryan98}. The Einstein radius of the SIS can be calculated as \citep[e.g.,][]{Kneib11}: 
\begin{equation}
r_E = 4\pi \frac{\sigma^2}{c^2} \frac{D_{ds}}{D_s}
\end{equation}
where $c$ is the speed of light, $D_{ds}$ is the angular diameter distance between the deflector and the background source, and $D_s$ is the angular diameter distance between the observer and the background source. For our system consisting of a foreground lens at $z_d = 0.055$ with $\sigma = 29$\,\kms\ and a background source at $z_s = 2.674$, the Einstein radius is $r_E \sim 0.024\arcsec$. Because $r_E$ is 33$\times$ smaller than the 0.8\arcsec\ offset between the SMG and the foreground galaxy, we conclude that the foreground galaxy is unlikely to have any measurable lensing effect on the SMG.

\clearpage
\section{Blind Line Detection in the ALMA Band-3 Data} \label{sec:blindCO}

To search for faint emission-line sources with unknown line widths in 3D data cubes, a matched-filtering algorithm is commonly used: e.g., in SoFiA \citep{Serra15}, FindClump \citep{Walter16}, LineSeeker \citep{Gonzalez-Lopez19}, and MF3D \citep{Pavesi18}. We wrote an IDL program to implement this simple algorithm. The convolution kernel we chose to filter the data in the spectral dimension is a top-hat function with a variable half-width between $n = 1$ and $n = 9$. For each channel, the data in the neighboring $\pm n$ channels are stacked with equal weighting. Given the average channel spacing of $\sim$25\,\kms, the convolved channel widths range between $\sim$75\,\kms\ and $\sim$475\,\kms. As shown in \citet{Gonzalez-Lopez19}, the simple top-hat function is as effective as the more sophisticated Gaussian kernels in detecting low S/N line emission. 

In each convoluted channel map, we measure the rms noise level with a robust sigma routine and detect unresolved sources near the highest S/N pixel. An elliptical Gaussian fixed to the shape of the core of the dirty beam is fit to the 8\arcsec$\times$8\arcsec\ subregion centered on the pixel and subtracted from the image. The parameters of the best-fit Gaussians are saved at each iteration to form the raw source catalog. The iterative process continues until the image contains no pixels above the S/N threshold of ${\rm S/N}_{\rm pix,th} = 4.5$. It is worth noting that this source-detection algorithm is similar to the minor cycles of the {\sc clean} deconvolution algorithm, but here we subtract only the core of the dirty beam to save computing time. This simplified approach is justified by the low S/N of the sources other than the SMG. 

Given that a single source can be detected in multiple channels and with multiple convolution kernels, we remove duplicated detections by iteratively looping through the raw source list from the highest to the lowest S/N and discard all detections within 2\arcsec\ and 0.2\,GHz from the highest S/N source remaining in the list.

Because our search is restricted to point sources, the source S/N simply scales with the ratio between the peak of the best-fit Gaussian ($S_{\rm peak}$) and the rms noise of the convolved channel map:
\begin{equation}
{\rm S/N} = 0.77 \frac{S_{\rm peak}}{{\rm rms}},
\end{equation}
where a scaling factor is used to account for fitting errors (Eq.\,9 of \citealt{Rengelink97}, see also \citealt{Condon97}). 

To estimate the fidelity of the detected sources, we search for sources in simulated noise-only interferometer data instead of using negative ``sources'' in the actual data \citep[e.g.,][]{Gonzalez-Lopez19}. First, we introduce random thermal noise by replacing the calibrated MS's visibilities with a normally distributed random array of complex numbers generated with {\it numpy.random}. This is equivalent to the CASA {\tt simulator} function {\tt setnoise} in the ``simplenoise'' mode, but our approach is faster because it only writes the DATA column once and does not add MODEL\_DATA and CORRECTED\_DATA columns to the MS. Unlike the fixed random number seed (11111) adopted in the CASA {\tt simulator}, each noise realization uses a different seed in our code. We set the widths of the normal distributions for the real and imaginary parts to the standard deviations of the original visibilities measured with {\tt visstat} ($\sigma \sim 250$\,mJy\,visibility$^{-1}$ with a slight dependence on the spectral window). Then, we use {\tt tclean} to image the noise-only visibilities into spectral data cubes with natural weighting. The same {\tt tclean} parameters are adopted except that we turn off de-convolution by setting {\tt niter} = 0, because the simulated data contain no sources. For each spectral window, we generate a set of 10 simulated data cubes to provide enough source counts for the source fidelity calculation. 

We run the same line search code on the simulated data with the same detection parameters. We compare the normalized cumulative distribution functions (CDF) of the detected sources in the actual data and in the simulated data to estimate the source fidelity. We use the following equation, similar to LineSeeker used in the ASPECS-LP survey \citep{Gonzalez-Lopez19}: 
\begin{equation} \label{eq:fidelity}
{\rm Fidelity} = 1-\frac{F_{\rm sim}(\ge {\rm S/N}~|~n)}{F_{\rm data}(\ge {\rm S/N}~|~n)},
\end{equation}
where $F(\ge {\rm S/N}~|~n)$ is the fraction of sources detected at or above the source S/N, with its detection kernel width $n$, and at any frequency channel in the spectral window. The subscript indicates whether it is from the actual data or the simulated noise-only data. $F_{\rm data}$ is measured directly from the CDF, while $F_{\rm sim}$ is obtained from the best-fit error function of the CDF to mitigate noise at high S/N due to low source counts. The fidelity is thus the probability that the detected source is not due to random noise. A source has a high fidelity near unity when $F_{\rm sim} \ll F_{\rm data} $, and a low fidelity near zero when $F_{\rm sim} \approx F_{\rm data}$.

\begin{figure}[!tb]
\epsscale{0.7}
\plotone{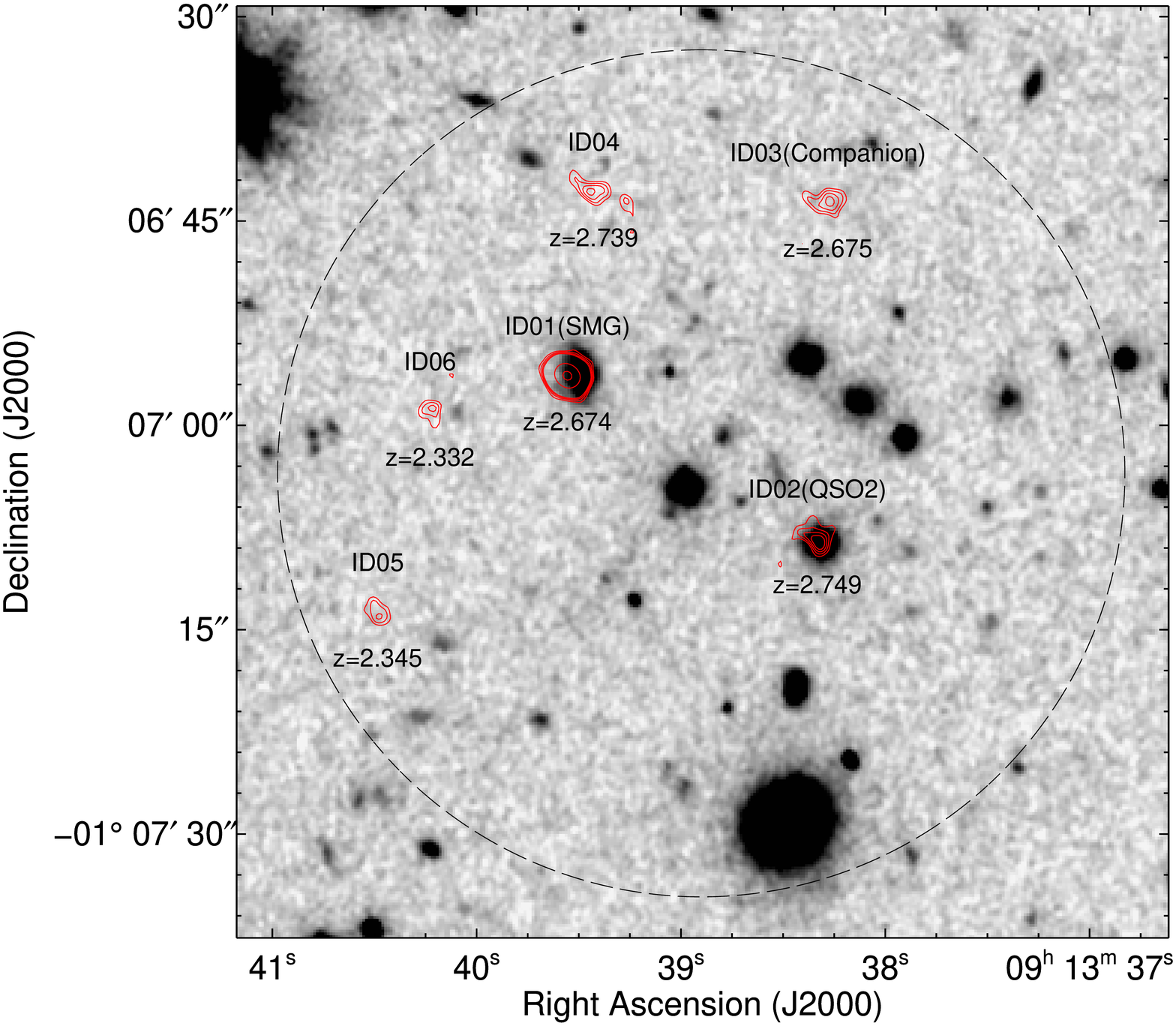}
\caption{ALMA emission-line detections in the \system\ field. The KiDS $r$-band image is overlaid with contours from the ALMA band-3 emission line sources. The emission line sources are labeled with its ID number and redshift (when interpreted as \cothree). The long-dashed circle shows the ALMA primary beam at 94\,GHz.
\label{fig:codist}} 
\epsscale{1.0}
\end{figure}

\begin{figure*}[!tb]
\epsscale{0.58}
\plotone{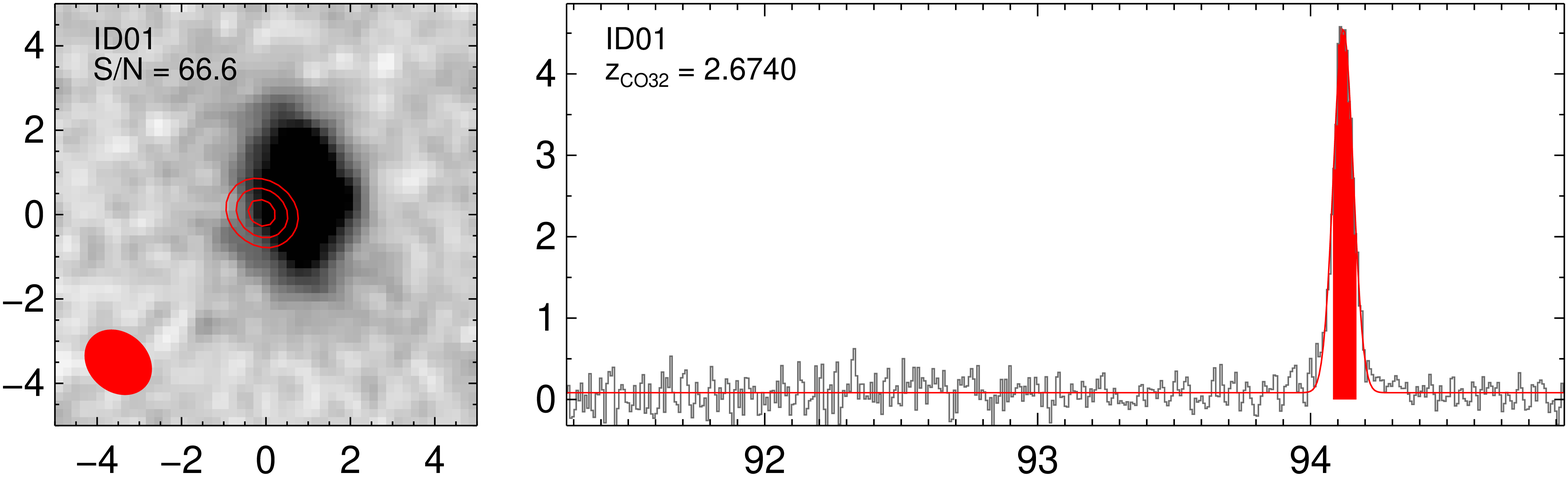}
\plotone{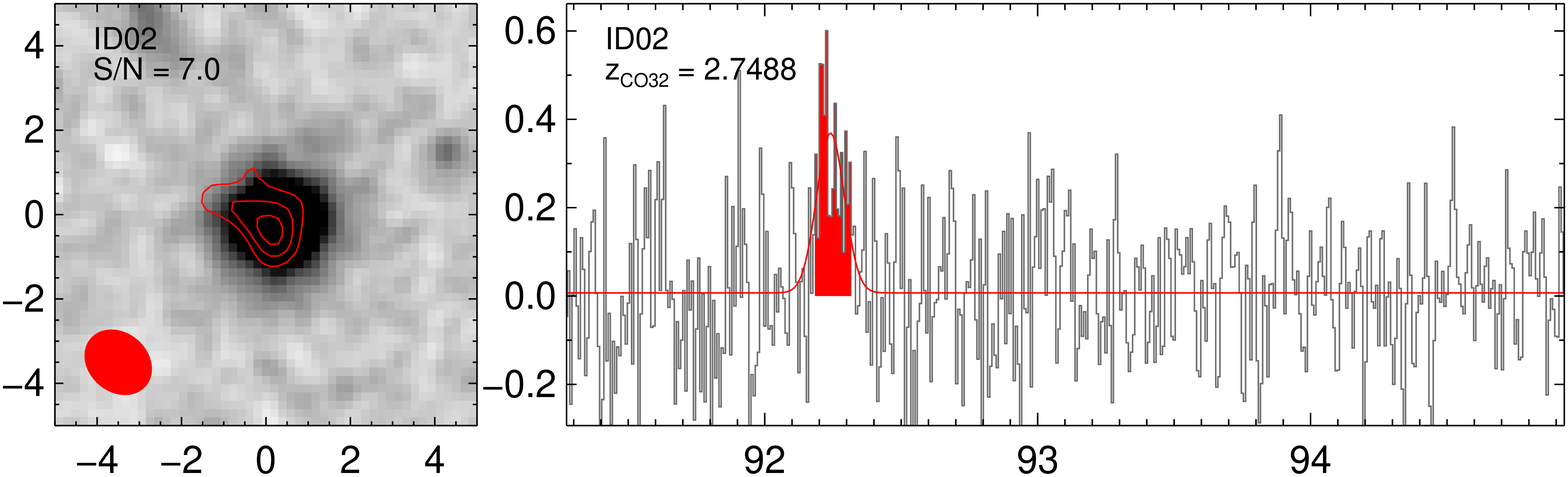}
\plotone{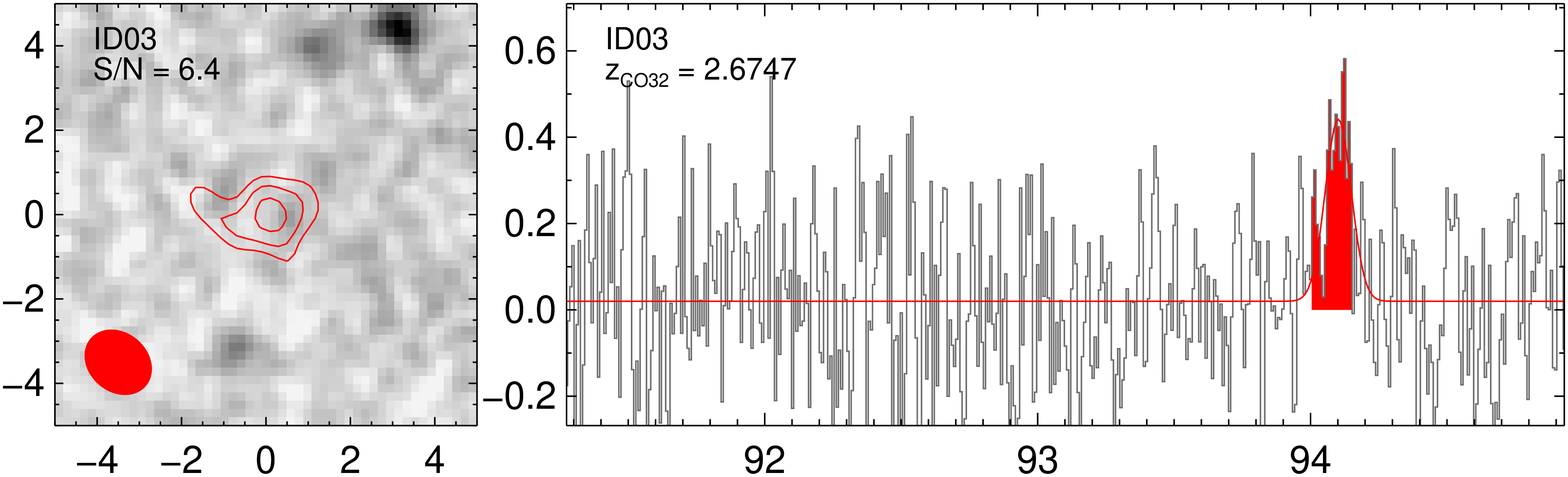}
\plotone{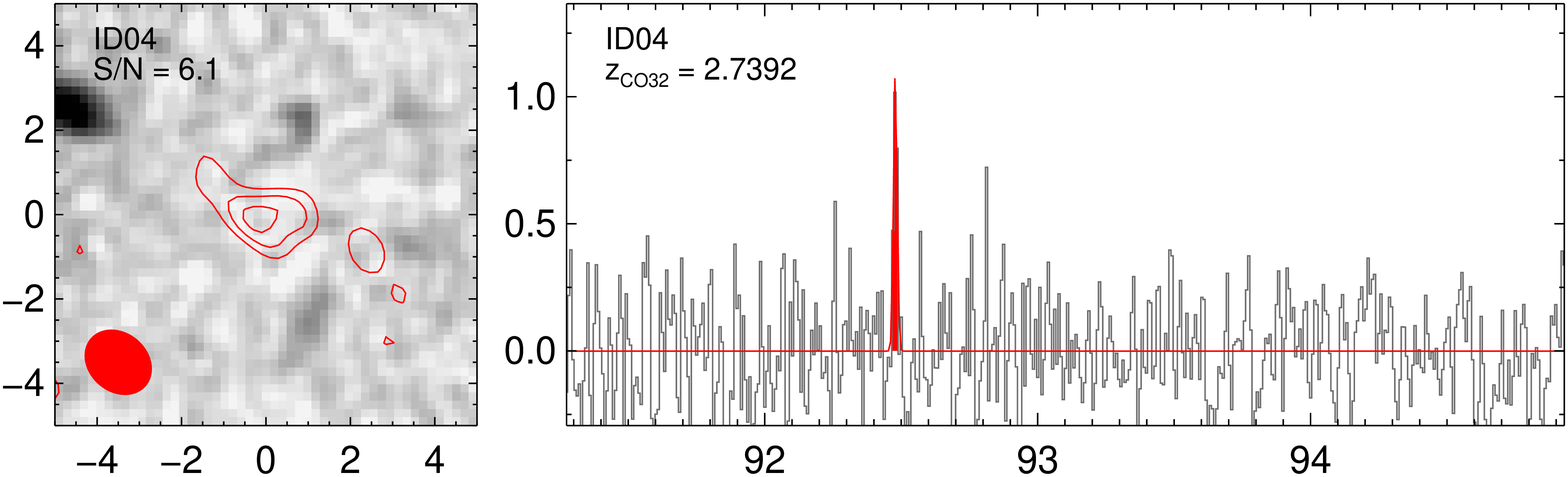}
\plotone{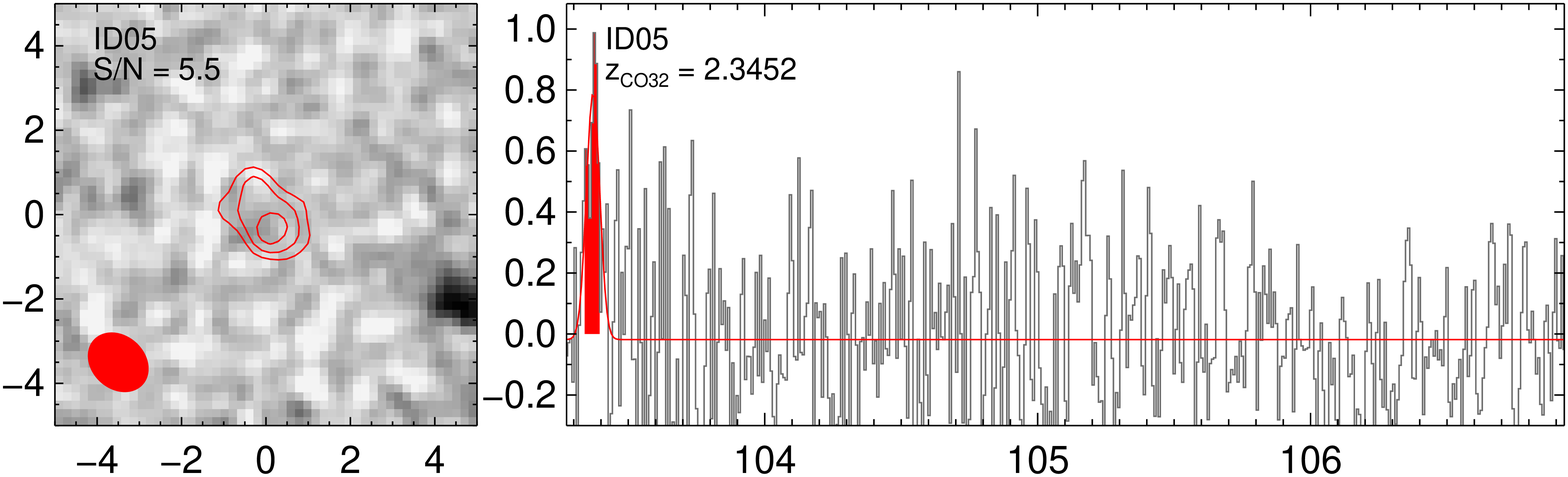}
\plotone{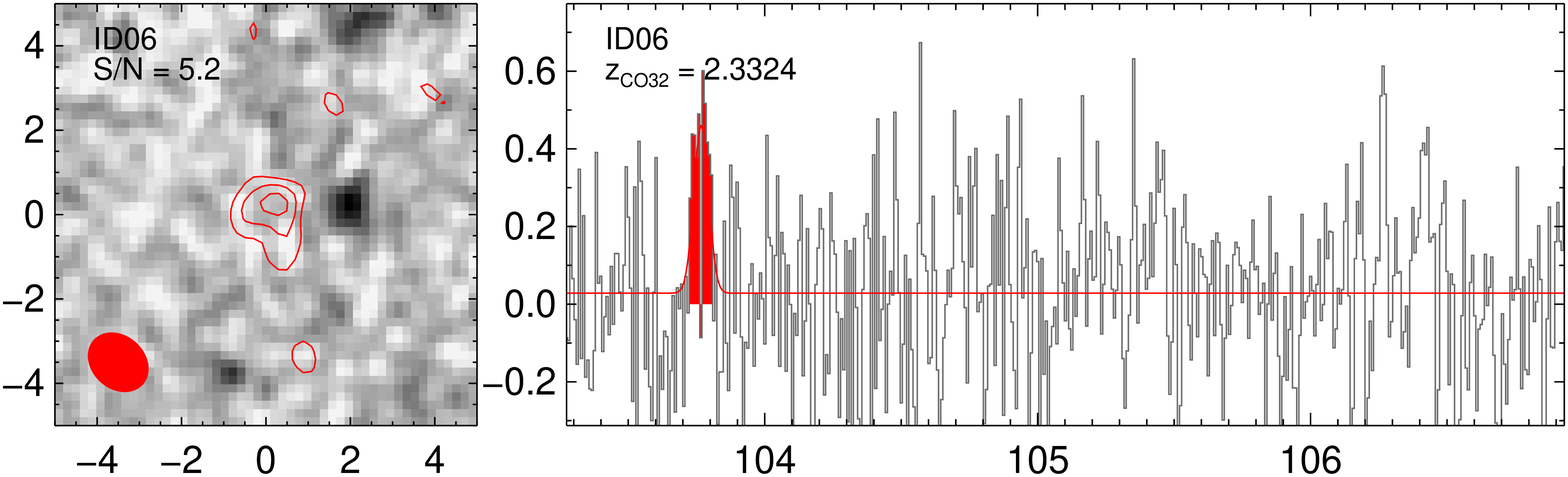}
\caption{For each line emitter, we show a 10\arcsec$\times$10\arcsec\ KiDS cutout image overlaid with its ALMA line emission map as contours, along with the source integrated spectrum (Flux Density in mJy vs. Observed Frequency in GHz). The ALMA maps are created by integrating line emission over narrow spectral windows, highlighted in red in their corresponding spectra. The red filled ellipses illustrate the synthesized beam size. 
\label{fig:coemitters}}
\epsscale{1.0}
\end{figure*}

We identified a total of six emission-line sources with ${\rm fidelity} > 0.9$ within the primary beam from the non-interpolated datacubes of all four spectral windows. We extracted the spectrum for each detection from the linear-interpolated and primary-beam-corrected data cubes with an elliptical aperture matched to the sizes of the restoring beams; i.e., we had assumed that the sources are unresolved. We measured the central frequency ($\nu_0$), FWHM, and line flux with a single-Gaussian model and list the results in Table\,\ref{tab:codetect}. Fig.\,\ref{fig:codist} illustrates the distribution of the detected sources in the field, and Fig.\,\ref{fig:coemitters} shows the zoomed-in version of the integrated intensity maps and their ALMA spectra.

\begin{table*}[!tb]
\begin{center}
\caption{Line Emitters Found in the ALMA Band-3 Data (Sorted in S/N)}
\label{tab:codetect}
\begin{tabular}{lcccc rrccc cc}
\hline
\hline
ID & R.A. (J2000) & Decl. (J2000) & $n$ & S/N & Fidelity & $\nu_{\rm obs}$ & FWHM &  Line Flux & $z_{\rm CO32}$ & $L'_{\rm CO32}$ & PB \\
   & (hms)& (dms) &     &     &          & (GHz)           &(\kms)& (Jy\,\kms) &                & (K\,\kms\,pc$^2$)& \\
\hline
 1&09:13:39.55&$-$01:06:56.5&  5&66.6&1.00& 94.120&$271\pm  5$ &$1.3435\pm0.0264$ &$2.6740\pm0.0001$ &$10.69\pm0.01$ &0.92\\
 2&09:13:38.33&$-$01:07:08.4&  8& 7.0&1.00& 92.241&$388\pm 69$ &$0.1482\pm0.0282$ &$2.7488\pm0.0004$ &$9.75\pm0.08$ &0.95\\
 3&09:13:38.28&$-$01:06:43.8&  9& 6.4&1.00& 94.102&$358\pm 67$ &$0.1724\pm0.0318$ &$2.6747\pm0.0003$ &$9.80\pm0.08$ &0.73\\
 4&09:13:39.42&$-$01:06:43.0&  1& 6.1&1.00& 92.479&$ 51\pm 10$ &$0.0590\pm0.0134$ &$2.7392\pm0.0001$ &$9.35\pm0.10$ &0.74\\
 5&09:13:40.47&$-$01:07:13.8&  3& 5.5&1.00&103.371&$166\pm 32$ &$0.1509\pm0.0293$ &$2.3452\pm0.0002$ &$9.64\pm0.08$ &0.58\\
 6&09:13:40.22&$-$01:06:59.1&  5& 5.2&0.97&103.767&$184\pm 53$ &$0.0488\pm0.0260$ &$2.3324\pm0.0002$ &$9.15\pm0.23$ &0.72\\
\hline
\end{tabular}
\end{center}
\tablecomments{The Column $n$ lists the half-width of the convolution window that yielded the highest S/N detection. The Columns Line Flux and $L'_{\rm CO32}$ have been corrected for the primary beam pattern, and the corresponding correction factors are listed in the Column PB.}
\end{table*}

\clearpage
\section{Other Absorbers in the QSO Spectra} \label{sec:otherabs}

\begin{figure}[!b]
\epsscale{0.8}
\plotone{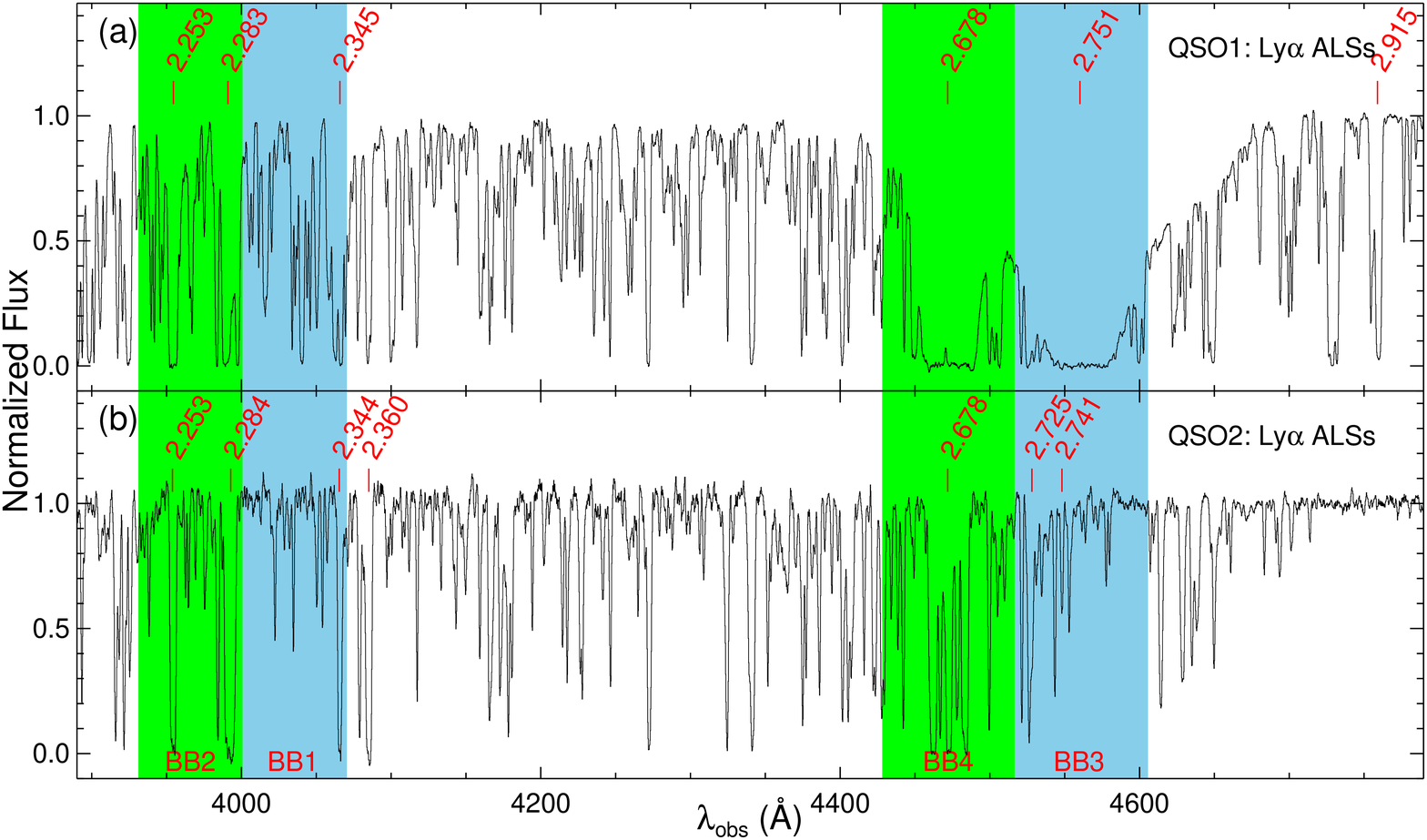}
\plotone{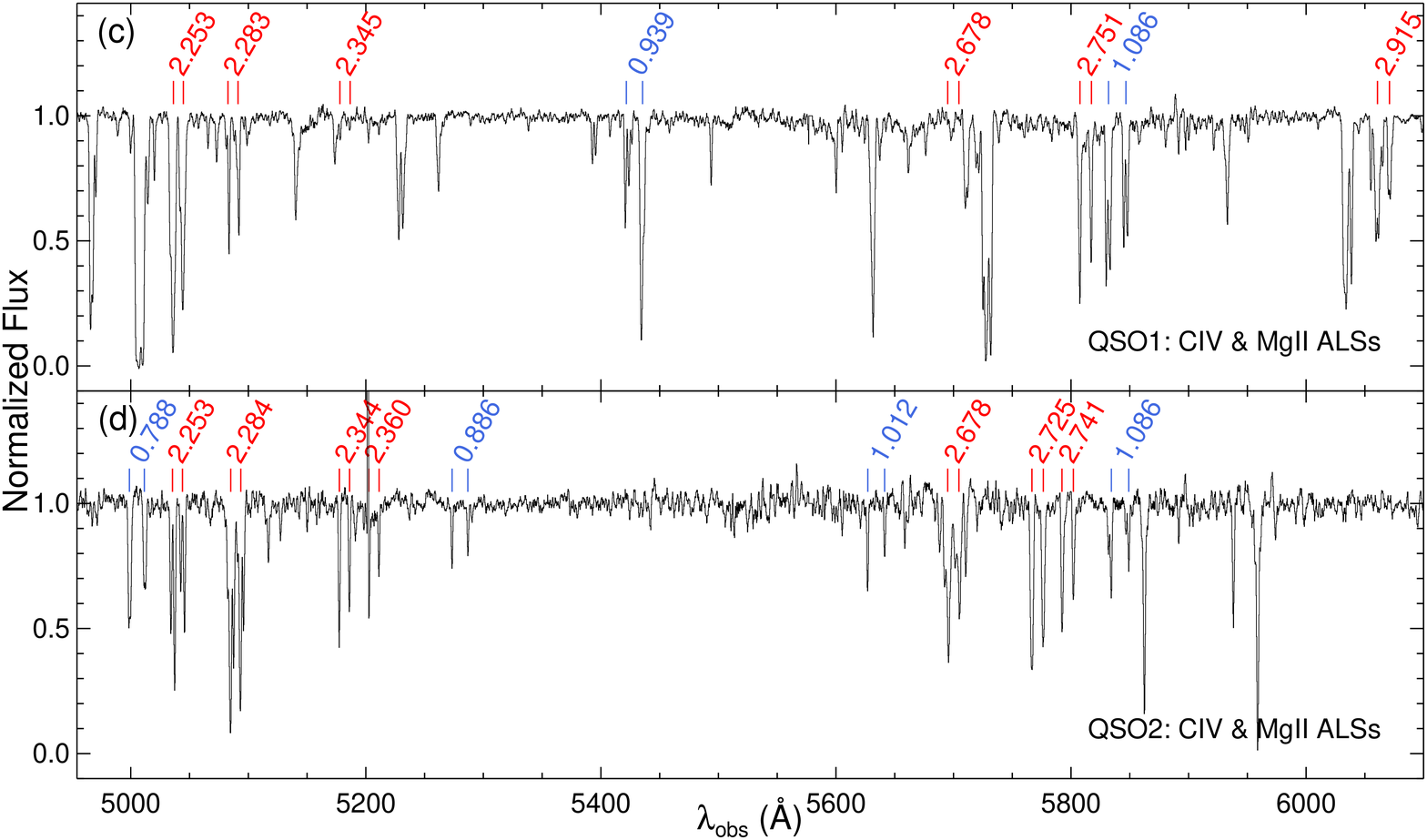}
\caption{Continuum-normalized QSO spectra from X-shooter. ({\it a,b}): Labeled are the main \HI\ \Lya\ absorbers between $2.20 < \zabs < 2.94$. The redshift ranges covered by the ALMA observations for \cothree\ are highlighted. ({\it c,d}): Labeled in {\it red} are the main \CIV\,$\lambda\lambda$1548.2,1550.8 absorbers between $2.20 < \zabs < 2.94$ and in {\it blue} the main \MgII\,$\lambda\lambda$2796.4,2803.5 absorbers between $0.77 < \zabs < 1.18$.
\label{fig:absorbers}} 
\epsscale{1.0}
\end{figure}
 
Using the low-resolution ($R \sim 2000$) BOSS spectrum of \bgQSO, \citet{Noterdaeme12} identified two DLA candidates at $\zabs = 2.680$ and 2.751. Subsequently, a number of \MgII\ and \CIV\ absorbers were identified toward both QSOs using the BOSS spectra: $\zabs = 0.9388, 2.2530, 2.7512$ toward \bgQSO\ and $\zabs = 0.7876, 1.0126, 1.5010, 2.7248, 2.7418$ toward \fgQSO\ \citep{Chen15,Chen16}.
The X-shooter spectra confirm all of the previously known absorbers and reveal several additional absorbers toward both QSOs: $\zabs = 1.0855, 2.283, 2.9147$ toward \bgQSO\ and $\zabs = 0.886, 1.0865, 2.2525,2.2845,2.3065,2.3445,2.3595$ toward \fgQSO. Fig.\,\ref{fig:absorbers} shows portions of the X-shooter spectra to illustrate all of the major absorbers we have identified (8 toward \bgQSO\ and 12 toward \fgQSO), omitting only the \MgII\ absorber at $z = 1.501$ toward \fgQSO.

\begin{figure}[!tb]
\epsscale{0.5}
\plotone{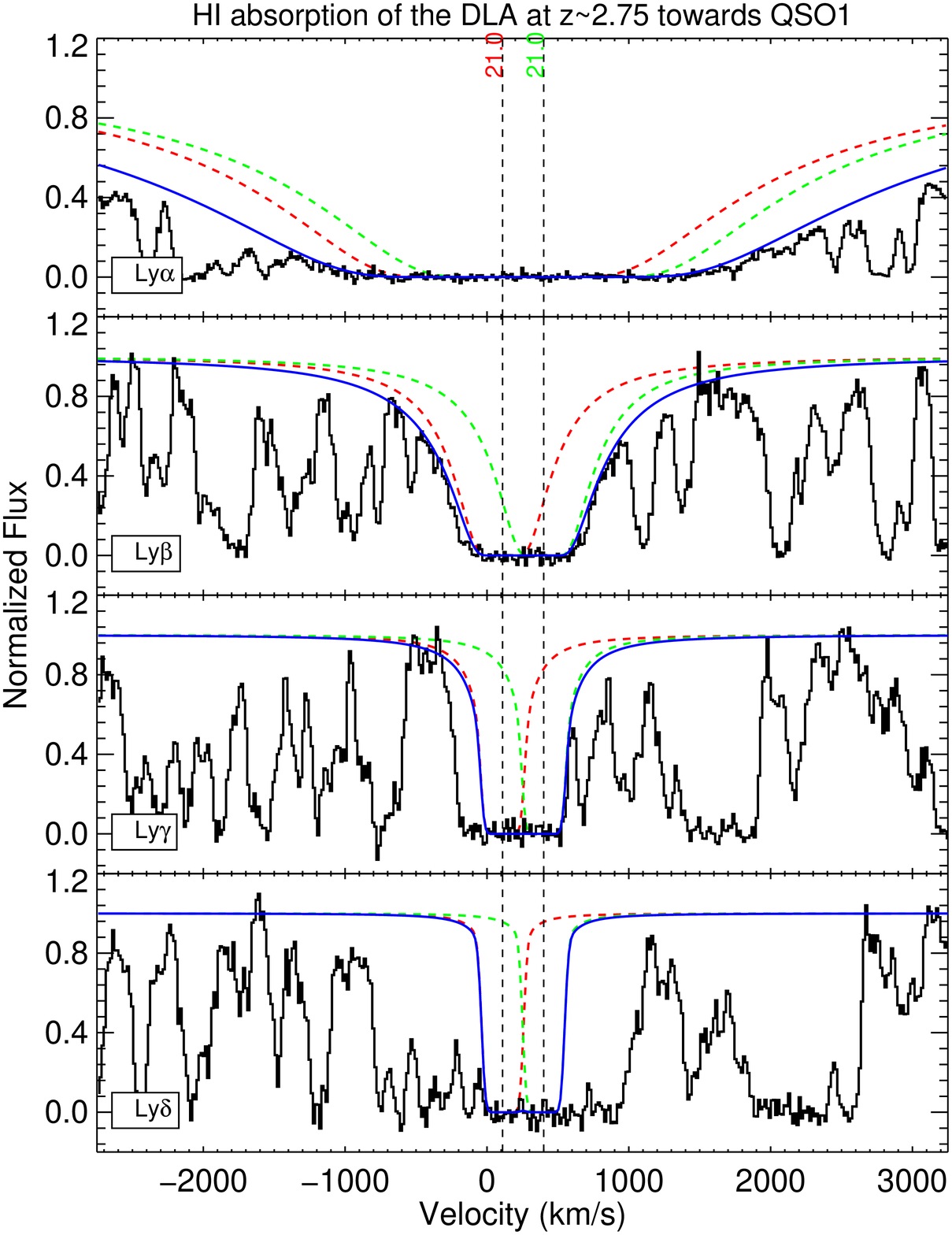}
\plotone{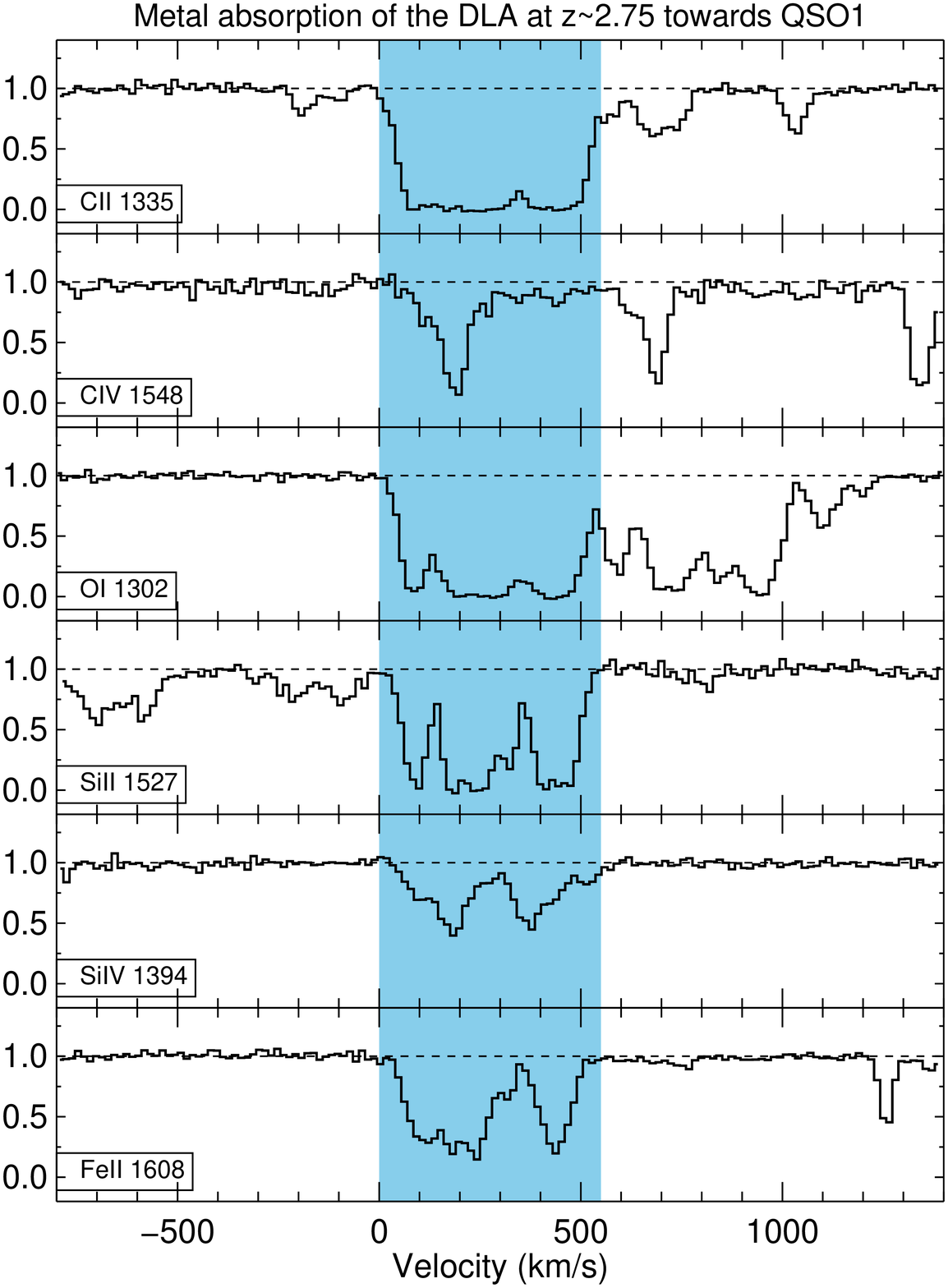}
\caption{The DLA at $z \approx 2.75$ toward \bgQSO. {\it Left}: \HI\ Lyman series and Voigt profile fit ({\it blue}). {\it Right}: selected metal transitions and the adopted AODM velocity integration windows. All velocities are relative to the systemic redshift defined by the \cothree\ line of \fgQSO\ at $z = 2.7488$. The DLA is at an impact parameter of $R_\bot = 85$\,kpc.}
\label{fig:cgm_qso2}
\epsscale{1.0}
\end{figure}

The $\zabs \approx 2.75$ DLA toward \bgQSO\ is apparently associated to \fgQSO\ at $z = 2.7488$. The DLA has been previously analyzed as part of the Quasar Probing Quasar (QPQ) project using a lower-resolution GMOS spectrum, from which they measured $\logNHI = 21.3\pm0.15$ \citep{Prochaska13a} and rest-frame equivalent widths of $2.60\pm0.05$\,\AA\ for C\,{\sc ii}$\lambda$1334.5 and $0.51\pm0.05$\AA\ for C\,{\sc iv}$\lambda$1548.2 \citep{Prochaska14}. We obtained similar results using the X-shooter spectrum (Fig.\,\ref{fig:cgm_qso2}). With Voigt profile fitting, we find that the \HI\ Lyman series is adequately fit with two components separated by 290\,\kms\ ($\zabs = 2.7502, 2.7538$), each with $\logNHI = 21.0$ and $b = 40$\,\kms. We thus obtain a total column density of $\logNHI = 21.3$ with an estimated systematic uncertainty of $\sim$0.1\,dex. With the AODM method and ICs from a {\sc cloudy} model with $\logNHI = 21.3$, $\log U = -2.5$, and the HM12 radiation background, we measure an ionization-corrected $\alpha$ metallicity of [C/H] = $-1.2\pm0.1$ from C\,{\sc iv}$\lambda$1550.8 and an iron metallicity of [Fe/H] = $-1.6\pm0.2$ from Fe\,{\sc ii}$\lambda$1608.5 (see Table\,\ref{tab:qso1dla}). Given the impact parameter of $R_\bot = 85$\,kpc, this DLA fits nicely with the CGM profiles of $z = 2-3$ QSOs in Fig.\,\ref{fig:CGMprof} \citep{Lau16}. 

\begin{table*}[!tbh]
\begin{center}
\caption{Metal Line Measurements of the $\zabs \approx 2.75$ DLA toward \bgQSO}
\label{tab:qso1dla}
\begin{tabular}{lc rrrrrrrr}
\hline
\hline
Ion & $\lambda_{\rm rest}$ & EW & $\log N$ & [X/H]$^\prime$ & [X/H] \\
    & (\AA)                & (\AA) & (cm$^{-2}$) & & \\
\hline
 C~\sc{ii}&1334.5323& $ 2.13\pm0.01$	&       $>15.48$&       $>-2.25$&       $>-2.26$	\\
 C~\sc{iv}&1548.2040& $ 0.63\pm0.02$	& $14.40\pm0.03$& $-3.33\pm0.10$& $-1.31\pm0.10$	\\
 \nodata  &1550.7776& $ 0.45\pm0.02$	& $14.50\pm0.04$& $-3.23\pm0.11$& $-1.21\pm0.11$	\\
 O~\sc{i}&1302.1685& $ 1.94\pm0.01$	&       $>15.86$&       $>-2.13$&       $>-2.14$	\\
Mg~\sc{ii}&2796.3543& $ 4.37\pm0.04$	&       $>14.44$&       $>-2.46$&       $>-2.26$	\\
 \nodata  &2803.5315& $ 3.40\pm0.08$	&       $>14.59$&       $>-2.31$&       $>-2.11$	\\
Al~\sc{ii}&1670.7886& $ 1.88\pm0.01$	&       $>13.99$&       $>-1.76$&       $>-1.52$	\\
Al~\sc{iii}&1854.7183& $ 0.43\pm0.01$	& $13.45\pm0.05$& $-2.30\pm0.11$& $-1.59\pm0.11$	\\
Si~\sc{ii}&1260.4221& $ 2.07\pm0.00$	&       $>14.54$&       $>-2.27$&       $>-2.31$	\\
 \nodata  &1304.3702& $ 1.69\pm0.01$	&       $>15.46$&       $>-1.35$&       $>-1.40$	\\
 \nodata  &1526.7070& $ 1.89\pm0.02$	&       $>15.23$&       $>-1.58$&       $>-1.63$	\\
Si~\sc{iv}&1393.7602& $ 0.67\pm0.01$	& $13.98\pm0.02$& $-2.83\pm0.10$& $-1.48\pm0.10$	\\
 \nodata  &1402.7729& $ 0.24\pm0.01$	& $13.79\pm0.07$& $-3.02\pm0.12$& $-1.68\pm0.12$	\\
Fe~\sc{ii}&1608.4508& $ 1.33\pm0.01$	& $15.22\pm0.01$& $-1.58\pm0.10$& $-1.62\pm0.10$	\\
 \nodata  &2382.7642& $ 3.01\pm0.03$	&       $>14.67$&       $>-2.13$&       $>-2.16$	\\
 \nodata  &2600.1725& $ 3.16\pm0.03$	&       $>14.73$&       $>-2.07$&       $>-2.10$	\\
\hline
\end{tabular}
\end{center}
\tablecomments{All measurements were made using a velocity integration window between 0 and 550\,\kms\ relative to $\zsys = 2.7488$. We use ``$<$'' for upper limits due to non-detections, ``$>$'' for lower limits due to line saturation, and ``$\lesssim$'' for upper limits due to line blending.}
\end{table*}

\clearpage
\section{The AODM Method and Results} \label{sec:aodmtables}

This Appendix gives a brief overview of the AODM method \citep{Savage91,Prochaska01} and provides tables of ionic column densities, ionization corrections, and ionization-corrected metallicities for all of the selected metal transitions (Tables\,\ref{tab:logN}, \ref{tab:IC}, and \ref{tab:XH}). Equations are all in SI units. The corresponding expressions in cgs units can be obtained by setting $\epsilon_0 = 1/4\pi$.

The velocity-dependent scattering cross section of resonance line photons is \citep{Meiksin09}:
\begin{equation}
\sigma(u) = \left( \frac{\pi e^2}{m_e c} \right) \left[ \frac{1}{4\pi \epsilon_0} \right] f \lambda_0 \phi_u, \label{eq:sigma}
\end{equation}
where the constants $e$, $m_e$, $\epsilon_0$, $c$, $f$, and $\lambda_0$ are respectively the electron charge, electron mass, the permeability of vacuum, the speed of light, the oscillator strength, and the rest-frame wavelength of the transition, and the function $\phi_u$ is the probability density function per unit velocity due to line broadening (i.e., a Voigt profile). 
For QSO absorption lines, the optical depth [$\tau(u)$] at velocity $u$ is the product of this velocity-dependent cross section and the column density ($N_a$):
\begin{equation}
\tau(u) = \frac{e^2}{4 \epsilon_0 m_e c} f \lambda_0 N_a \phi_u \equiv \frac{e^2}{4 \epsilon_0 m_e c} f \lambda_0 N_{a,u} \label{eq:Nau}
\end{equation}
where we have defined the column density per unit velocity, $N_{a,u} \equiv N_a \phi_u$, by shifting the velocity dependency from the cross section to the column density.

The above relation provides a method to measure column densities from observed line profiles, because the optical depth is the natural logarithmic of the ratio between the incident continuum intensity ($I_0$) and the observed attenuated intensity ($I_{\rm obs}$): 
\begin{equation}
\tau(u) = \ln \frac{I_0(u)}{I_{\rm obs}(u)}.
\end{equation}

The total column density can then be calculated by integrating the apparent optical depth over the velocity integration window:
\begin{equation}
N_a = \sum N_{a,u} \Delta u = \sum \frac{4 \epsilon_0 m_e c}{e^2 f \lambda_0} \tau(u) \Delta u
\end{equation} 
and the 1$\sigma$ {\it statistical} variance on the column density through standard error propagation is:
\begin{equation}
\sigma^2_{\rm sta}(N_a) = \sum \left( \frac{4 \epsilon_0 m_e c}{e^2 f \lambda_0} \right)^2 \sigma^2[\tau(u)] \Delta u^2
\end{equation}
where the statistical uncertainty of optical depth is estimated from the noise spectrum: 
\begin{equation}
\sigma_{\rm sta}[\tau(u)] = \sigma[I_{\rm obs}(u)]/I_{\rm obs}(u)
\end{equation}

Similar to the \HI\ Voigt profile fitting (but to a less extent because of the narrower velocity range), the ionic column density is also affected by the systematic uncertainty in our empirical model of the QSO continuum. We again adopt a $\pm$10\% error in the QSO continuum ($I_0$), which directly leads to $\sigma_{\rm sys}[\tau(u)] = 0.1$ and the equation for the systematic error of the ionic column density:
\begin{equation}
\sigma^2_{\rm sys}(N_a) = \sum \left( \frac{4 \epsilon_0 m_e c}{e^2 f \lambda_0} \right)^2 0.1^2 \Delta u^2.
\end{equation}

\begin{table*}[!tbh]
\begin{center}
\caption{AODM Ionic Column Densities}
\label{tab:logN}
\begin{tabular}{lc rrrrrrrr}
\hline
\hline
Ion & $\lambda_{\rm rest}$ & \multicolumn{8}{c}{$\log N$} \\
    & (\AA)                & QSO1-A1 & QSO1-A2 & QSO1-B1 & QSO1-C1 & QSO1-C2 & QSO2-A2 & QSO2-B1 & QSO2-C1 \\
\hline
 C~\sc{ii}&1334.5323	&$\lesssim14.74$	& $13.94\pm0.05$	&       $<12.90$	&       $>15.21$	& $14.18\pm0.03$	& $14.01\pm0.06$	&       $<13.33$	&       $<13.25$	\\
 C~\sc{iv}&1548.2040	&       $<12.88$	&       $<12.88$	&       $<12.91$	& $14.08\pm0.04$	&       $<12.91$	&       $<13.20$	& $14.23\pm0.03$	& $13.61\pm0.08$	\\
 \nodata  &1550.7776	&       $<13.21$	&       $<13.19$	&       $<13.19$	& $14.15\pm0.07$	&$\lesssim14.83$	&$\lesssim14.30$	& $14.30\pm0.04$	& $13.72\pm0.13$	\\
 O~\sc{i}&1302.1685	& $14.57\pm0.03$	& $14.48\pm0.04$	&       $<13.16$	&       $>15.55$	& $14.36\pm0.06$	&       $<13.78$	&       $<13.79$	&       $<13.75$	\\
Mg~\sc{ii}&2796.3543	& $13.24\pm0.07$	& $12.99\pm0.08$	       & \nodata	&       $>14.09$	& $13.21\pm0.05$	       & \nodata	       & \nodata	&       $<12.60$	\\
 \nodata  &2803.5315	& $13.49\pm0.06$	& $13.15\pm0.09$	&$\lesssim14.11$	&       $>14.32$	& $13.08\pm0.09$	       & \nodata	&       $<12.93$	&       $<12.76$	\\
Al~\sc{ii}&1670.7886	& $12.37\pm0.11$	& $11.97\pm0.28$	&       $<11.70$	& $13.49\pm0.02$	& $12.15\pm0.20$	& $12.28\pm0.19$	&       $<12.12$	&       $<12.06$	\\
Al~\sc{iii}&1854.7183	&       $<12.17$	&       $<12.14$	&       $<12.15$	& $12.91\pm0.16$	&       $<12.12$	&       $<12.50$	&       $<12.53$	&       $<12.49$	\\
Si~\sc{ii}&1304.3702	& $13.76\pm0.12$	& $13.22\pm0.40$	&       $<12.93$	& $14.81\pm0.02$	& $13.32\pm0.34$	&       $<13.52$	&       $<13.55$	&       $<13.48$	\\
 \nodata  &1526.7070	& $13.78\pm0.08$	&       $<13.23$	&       $<13.15$	& $14.76\pm0.02$	& $13.55\pm0.12$	&       $<13.58$	&       $<13.50$	&       $<13.38$	\\
Si~\sc{iv}&1393.7602	&       $<12.30$	&       $<12.30$	&       $<12.36$	&$\lesssim13.70$	&       $<12.56$	& $13.15\pm0.11$	&       $<12.76$	&       $<12.77$	\\
 \nodata  &1402.7729	&       $<12.82$	&       $<12.84$	&       $<12.79$	& $13.54\pm0.11$	&       $<12.80$	&       $<13.22$	&       $<13.42$	&       $<13.12$	\\
Fe~\sc{ii}&1608.4508	& $13.55\pm0.26$	&       $<13.39$	&       $<13.44$	& $14.52\pm0.05$	&       $<13.36$	&       $<13.64$	&       $<13.74$	&       $<13.71$	\\
 \nodata  &2382.7642	& $13.34\pm0.05$	& $12.90\pm0.13$	&       $<12.58$	&$\lesssim14.47$	&$\lesssim13.90$	&       $<12.79$	&       $<12.66$	&       $<12.71$	\\
 \nodata  &2600.1725	&$\lesssim13.64$	& $13.13\pm0.10$	&       $<12.58$	&       $>14.25$	& $12.91\pm0.17$	&       $<13.18$	&       $<12.92$	&       $<12.94$	\\
\hline
\end{tabular}
\end{center}
\tablecomments{This table lists the measured column densities from each transition, where we use ``$<$'' for upper limits due to non-detections, ``$>$'' for lower limits due to line saturation, ``$\lesssim$'' for upper limits due to line blending, and ``...'' for no measurement due to poor data quality.}
\end{table*}

\begin{table*}[!tbh]
\begin{center}
\caption{Ionization Correction}
\label{tab:IC}
\begin{tabular}{l rrrrrrrr}
\hline
\hline
Ion & 
\multicolumn{7}{c}{${\rm IC} \equiv \log f_{\rm HI} - \log f_{\rm X}$ }\\
    & QSO1-A1 & QSO1-A2 & QSO1-B1 & QSO1-C1 & QSO1-C2 & QSO2-A2 & QSO2-B1 & QSO2-C1 \\
\hline
C~\sc{ii}	&$-0.18$	&$-0.10$	&$-1.74$	&$-0.08$	&$-0.58$	&$-0.90$	&$-1.74$	&$-1.74$	\\
C~\sc{iv}	&$ 2.78$	&$ 1.90$	&$-2.82$	&$ 2.00$	&$ 1.98$	&$ 0.40$	&$-2.82$	&$-2.82$	\\
O~\sc{ i}	&$-0.01$	&$-0.01$	&$ 2.56$	&$-0.01$	&$-0.04$	&$ 0.02$	&$ 2.56$	&$ 2.56$	\\
Mg~\sc{ii}	&$ 0.14$	&$ 0.25$	&$ 0.04$	&$ 0.27$	&$-0.36$	&$-0.71$	&$ 0.04$	&$ 0.04$	\\
Al~\sc{ii}	&$-0.05$	&$ 0.06$	&$-1.08$	&$ 0.12$	&$-0.68$	&$-1.20$	&$-1.08$	&$-1.08$	\\
Al~\sc{ii}	&$ 0.61$	&$ 0.55$	&$-1.18$	&$ 0.59$	&$ 0.21$	&$-0.51$	&$-1.18$	&$-1.18$	\\
Si~\sc{ii}	&$-0.21$	&$-0.14$	&$-1.17$	&$-0.12$	&$-0.62$	&$-0.95$	&$-1.17$	&$-1.17$	\\
Si~\sc{iv}	&$ 1.61$	&$ 1.03$	&$-2.37$	&$ 1.13$	&$ 0.83$	&$-0.46$	&$-2.37$	&$-2.37$	\\
Fe~\sc{ii}	&$-0.12$	&$-0.08$	&$ 1.58$	&$-0.07$	&$-0.37$	&$-0.49$	&$ 1.58$	&$ 1.58$	\\
\hline
\end{tabular}
\end{center}
\end{table*}

\begin{table*}[!tbh]
\begin{center}
\caption{Ionization-Corrected Metallicities}
\label{tab:XH}
\begin{tabular}{lc rrrrrrrr}
\hline
\hline
Ion & $\lambda_{\rm rest}$ & 
\multicolumn{8}{c}{${\rm [X/H]} \equiv {\rm [X/H]}^\prime + {\rm IC}$ }\\
    & (\AA) & QSO1-A1 & QSO1-A2 & QSO1-B1 & QSO1-C1 & QSO1-C2 & QSO2-A2 & QSO2-B1 & QSO2-C1 \\
\hline
 C~\sc{ii}&1334.5323	&$\lesssim-1.51$	& $-2.66\pm0.08$	&       $<-1.25$	&       $>-1.53$	& $-1.62\pm0.29$	& $-1.91\pm0.34$	&       $<-0.87$	&       $<-0.92$	\\
 C~\sc{iv}&1548.2040	&       $<-0.41$	&       $<-1.72$	&       $<-2.32$	& $-0.59\pm0.09$	&       $<-0.33$	&       $<-1.42$	& $-1.05\pm0.19$	& $-1.63\pm0.13$	\\
 \nodata  &1550.7776	&       $<-0.08$	&       $<-1.41$	&       $<-2.04$	& $-0.52\pm0.10$	&$\lesssim 1.59$	&$\lesssim-0.32$	& $-0.98\pm0.19$	& $-1.53\pm0.16$	\\
 O~\sc{i}&1302.1685	& $-1.77\pm0.06$	& $-2.28\pm0.08$	&       $<3.05 $	&       $>-1.38$	& $-1.15\pm0.30$	&       $<-1.48$	&       $<3.63 $	&       $<3.62 $	\\
Mg~\sc{ii}&2796.3543	& $-1.87\pm0.09$	& $-2.42\pm0.10$	       & \nodata	&       $>-1.47$	& $-1.53\pm0.29$	       & \nodata	       & \nodata	&       $<1.04 $	\\
 \nodata  &2803.5315	& $-1.61\pm0.08$	& $-2.26\pm0.11$	&$\lesssim 2.57$	&       $>-1.24$	& $-1.66\pm0.30$	       & \nodata	&       $<1.34 $	&       $<1.20 $	\\
Al~\sc{ii}&1670.7886	& $-1.77\pm0.12$	& $-2.47\pm0.29$	&       $<0.19 $	& $-1.08\pm0.08$	& $-1.77\pm0.35$	& $-1.96\pm0.38$	&       $<0.56 $	&       $<0.52 $	\\
Al~\sc{iii}&1854.7183	&       $<-1.31$	&       $<-1.82$	&       $<0.54 $	& $-1.18\pm0.17$	&       $<-0.91$	&       $<-1.05$	&       $<0.87 $	&       $<0.87 $	\\
Si~\sc{ii}&1304.3702	& $-1.60\pm0.13$	& $-2.49\pm0.40$	&       $<0.27 $	& $-1.05\pm0.08$	& $-1.60\pm0.45$	&       $<-1.53$	&       $<0.84 $	&       $<0.80 $	\\
 \nodata  &1526.7070	& $-1.57\pm0.09$	&       $<-2.48$	&       $<0.48 $	& $-1.10\pm0.08$	& $-1.37\pm0.31$	&       $<-1.48$	&       $<0.79 $	&       $<0.70 $	\\
Si~\sc{iv}&1393.7602	&       $<-1.23$	&       $<-2.24$	&       $<-1.50$	&$\lesssim-0.91$	&       $<-0.91$	& $-1.41\pm0.35$	&       $<-1.14$	&       $<-1.11$	\\
 \nodata  &1402.7729	&       $<-0.72$	&       $<-1.70$	&       $<-1.07$	& $-1.07\pm0.13$	&       $<-0.68$	&       $<-1.34$	&       $<-0.49$	&       $<-0.76$	\\
Fe~\sc{ii}&1608.4508	& $-1.71\pm0.27$	&       $<-2.24$	&       $<3.54 $	& $-1.27\pm0.09$	&       $<-1.30$	&       $<-0.94$	&       $<3.79 $	&       $<3.79 $	\\
 \nodata  &2382.7642	& $-1.92\pm0.07$	& $-2.73\pm0.15$	&       $<2.68 $	&$\lesssim-1.33$	&$\lesssim-0.76$	&       $<-1.79$	&       $<2.71 $	&       $<2.80 $	\\
 \nodata  &2600.1725	&$\lesssim-1.62$	& $-2.51\pm0.12$	&       $<2.69 $	&       $>-1.55$	& $-1.76\pm0.34$	&       $<-1.39$	&       $<2.98 $	&       $<3.02 $	\\
\hline
\end{tabular}
\end{center}
\end{table*}

\clearpage
\section{The Identification of The Emission Counterpart of Subsystem C} \label{sec:optCO}

\begin{figure*}[!b]
\epsscale{1.18}
\plotone{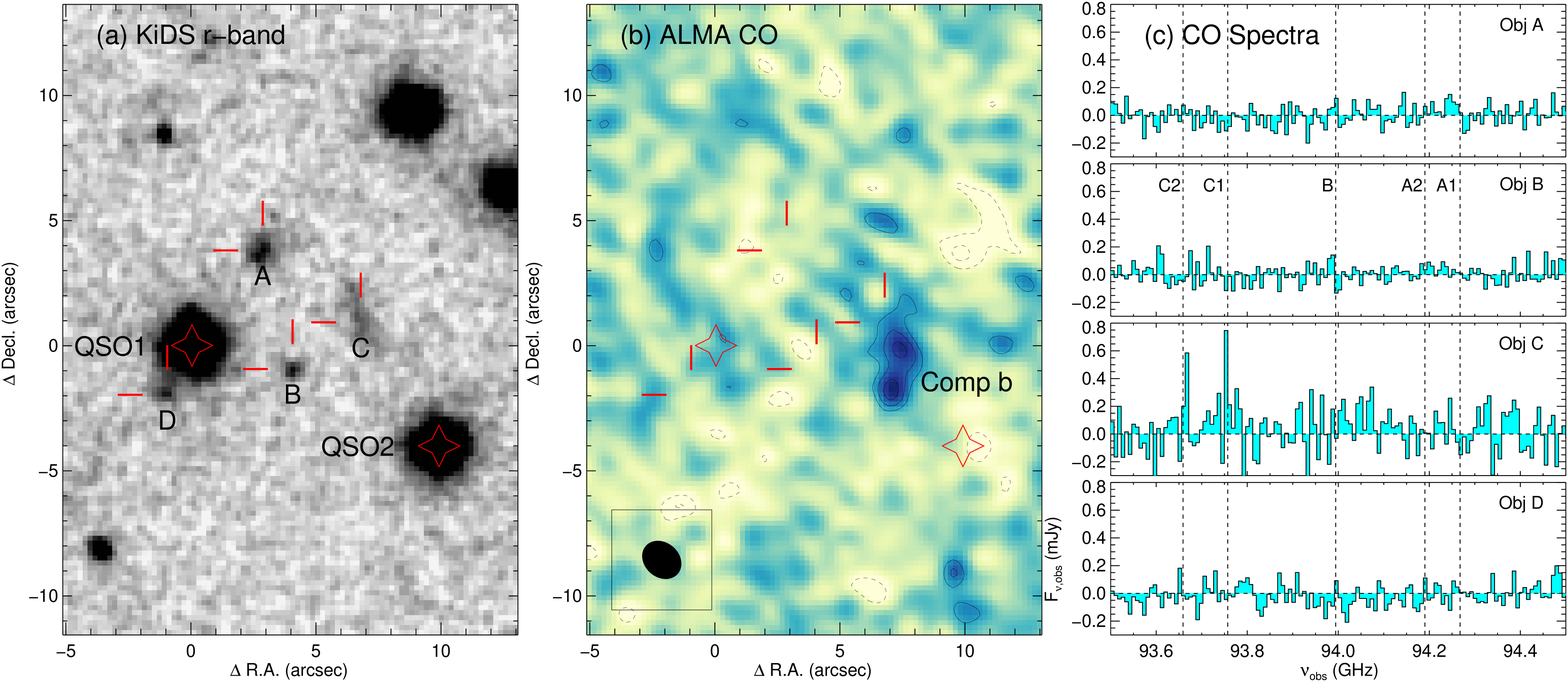}
\caption{($a$) The four faint ($r > 23$) optical sources within 7\arcsec\ of \bgQSO. ($b$) An ALMA CO map constructed by combining the channels at 93.7535 and 93.6676\,GHz, where CO emission from Object C is detected. The contours are drawn at $-3$, $-2$ ({\it dashed}), 2, 3, and 4$\sigma$ ({\it solid}). In both images, \bgQSO\ sets the origin of the coordinates. In the four panels in $c$, we show the ALMA spectra of the objects. The vertical dashed lines indicate the \cothree\ frequencies that correspond to the redshifts of the major absorption-line clouds toward \bgQSO.   
\label{fig:fchartC}}
\epsscale{1.0}
\end{figure*}

In the deep $r$-band image from KiDS (5$\sigma$ limit at $\sim$25\,mag) shown in Fig.~\ref{fig:fchartC}$a$, we labeled four faint optical sources within 7\arcsec\ of \bgQSO. Here we explore whether any of these sources is connected to the DLA at $\zabs \approx 2.68$ by examining their photometric redshifts and their ALMA spectra. 

Three of the sources (A, B, C) are listed in the joint KiDS-VISTA 9-band photometric catalog \citep{Kuijken19}: 
\begin{itemize}
\item KiDSDR4\,J091338.791$-$010700.71 (A): $r = 23.6\pm0.1$, $H = 22.3\pm0.4$, $z_{\rm p} = 1.09^{+0.09}_{-0.15}$; 
\item KiDSDR4\,J091338.711$-$010705.48 (B): $r = 24.5\pm0.2$, $H = 22.9\pm0.7$, $z_{\rm p} = 0.45^{+0.78}_{-0.12}$.
\item KiDSDR4\,J091338.527$-$010703.60 (C): $r = 23.8\pm0.1$, $H = 22.0\pm0.3$, $z_{\rm p} = 0.79^{+0.45}_{-0.06}$.
\end{itemize}
where the $r-$ and $H$-band magnitudes are from the homogenized ``Gaussian Aperture and PSF (GAaP)'' photometry and $z_{\rm p}$ are the 9-band photometric redshift estimates from the Bayesian photometric redshift code \texttt{BPZ} \citep{Benitez00}. The 68\% confidence intervals of the photometric redshifts suggest that both sources are in the far foreground of the SMG \SMG\ ($\zsmg = 2.674$). 
The fourth object (D) is not in the catalog likely because its proximity to the bright \bgQSO. We measured its position directly from the image: R.A. = $09^{\rm h}13^{\rm m}39.05^{\rm s}$, Decl. = $-01^\circ07\arcmin06.5\arcsec$. 

We then extracted spectra at their optical positions from the ALMA band-3 datacube. For objects A, B, and D, we adopted elliptical apertures matching the synthesized beam size (1.7\arcsec$\times$1.3\arcsec\ at PA = 49$^\circ$). We show these spectra in Fig.~\ref{fig:fchartC}$c$. Even at the depth of our ALMA data (rms$ = 0.155$\,mJy~beam$^{-1}$~channel$^{-1}$ in BB4), none of the sources show emission lines at a detectable level.  

For Object C, we initially used an aperture centered on the optical position and detected hints of emission lines at the expected frequencies of absorption-line clouds C1 and C2. We then made a CO map by combining the two channels that show the most significant emission. The CO image in Fig.~\ref{fig:fchartC}$b$ led to the discovery of \Chost\ as it reveals a highly significant source $\sim$3\arcsec\ to the SSW of the optical position, which we have designated as \Chost. Guided by the CO image, we re-extracted a spectrum from a 3.4\arcsec$\times$1.8\arcsec\ elliptical aperture that matches the geometry of \Chost\ to optimize the line detection. This spectrum is shown in the Object C panel of Fig.~\ref{fig:fchartC}$c$. Line emission is clearly detected at the expected frequencies of absorption-line clouds C1 and C2 toward \bgQSO. Through this exercise, we have identified \Chost\ as the most likely emission counterpart of absorption subsystem C toward both QSOs. 

\end{document}